\documentclass[proof]{WileyASNA-v1}
\articletype{Article Type}
\usepackage{graphicx}
\usepackage{tabularx}	
\usepackage{chemformula}
\let\ce\ch
\usepackage{dblfloatfix}
\usepackage{placeins}
\usepackage{float}
\usepackage[utf8]{inputenc}
\usepackage[english]{babel}
\usepackage{xcolor}
\usepackage{pifont}

\definecolor{darkgreen}{rgb}{0.0, 0.5, 0.0}

\def\tick{\textcolor{green}{\ding{52}}}
\def\cross{\textcolor{red}{\ding{56}}}

\received{4 May 2023}
\revised{21 Aug 2023}
\accepted{31 Aug 2023}
\begin{document} 

\title{The sulphur species in hot rocky exoplanet atmospheres}

\author[1,2,6,7]{L. J. Janssen$^*$}
\author[3]{P. Woitke}
\author[4]{O. Herbort}
\author[5]{M. Min}
\author[6,7]{K. L. Chubb}
\author[3,8]{Ch. Helling}
\author[3]{L. Carone}

\authormark{JANSSEN \textsc{et al.}}

\address[1]{\orgname{Leiden Observatory}, \orgaddress{\state{Niels Bohrweg 2, 2333 CA Leiden}, \country{The Netherlands}}}

\address[2]{\orgname{Anton Pannekoek Institute for Astronomy}, \orgaddress{\state{Science Park 904, 1098 XH Amsterdam}, \country{The Netherlands}}}

\address[3]{\orgname{Institut f\"ur Weltraumforschung, \"Osterreichische Akademie der Wissenschaften}, \orgaddress{\state{Schmiedlstr. 6, A-8042, Graz}, \country{Austria}}}

\address[4]{\orgname{Department of Astrophysics, University of Vienna}, \orgaddress{\state{T\"urkenschanzstr. 17, A-1180, Wien}, \country{Austria}}}

\address[5]{\orgname{SRON Netherlands Institute for Space Research}, \orgaddress{\state{Sorbonnelaan 2, 3584 CA Utrecht}, \country{The Netherlands}}}

\address[6]{\orgname{Centre for Exoplanet Science, University of St Andrews}, \orgaddress{\state{St. Andrews}, \country{United Kingdom}}}

\address[7]{\orgname{SUPA, School of Physics \& Astronomy, University of St Andrews}, \orgaddress{\state{North Haugh, St Andrews, KY16 9SS}, \country{United Kingdom}}}

\address[8]{\orgname{Fakult\"at für Mathematik, Physik und Geod\"asie, TU Graz}, \orgaddress{\state{Petersgasse 16, Graz}, \country{Austria}}}

\corres{*\email{ljanssen@strw.leidenuniv.nl}}
%\presentaddress{*test address}
 
\abstract{The first JWST observations of hot Jupiters showed an
    unexpected detection of \ce{SO2} in their hydrogen-rich
  atmospheres.  We investigate how much sulphur can be expected in the
  atmospheres of rocky exoplanets and which sulphur molecules can be
  expected to be most abundant and detectable by transmission
  spectroscopy.  We run thermo-chemical equilibrium models at
  the crust-atmosphere interface, considering surface temperatures
  $500-5000$\,K, surface pressures $1-100$\,bar, and various sets of
  element abundances based on common rock compositions.  Between
  1000\,K and 2000\,K, we find gaseous sulphur concentrations of
  up to 25\% above the rock in our models. \ce{SO2}, \ce{SO},
  \ce{H2S} and \ce{S2} are by far the most abundant sulphur
  molecules. \ce{SO2} shows potentially detectable features in
  transmission spectra at about 4\,$\mu$m, between 7 and 8\,$\mu$m,
  and beyond 15\,$\mu$m.  In contrast, the sometimes abundant
  \ce{H2S} molecule is difficult to detect in these spectra, which
  are mostly dominated by \ce{H2O} and \ce{CO2}. Although
  the molecule PS only occurs with concentrations $<\!300$\,ppm, it
  can cause a strong absorption feature between 0.3 and
  0.65\,$\mu$m in some of our models for high surface
  pressures. The detection of sulphur molecules would enable a
  better characterisation of the planetary surface.}


%\jnlcitation{\cname{%
%\author{Janssen, L.J.},
%\author{Woitke, P.}, 
%\author{Herbort, O.}, 
%\author{Min, M.}, 
%\author{Chubb, K.L.},
%\author{Helling, Ch.}, and 
%\author{Carone, L.}} (\cyear{2023}), 
%\cjournal{Atsron.Nachr}} %\ceprint{e20230075}, %\curl{https://doi.org/10.1002/asna.20230075}}

\maketitle

\section{Introduction}

In our Solar System, sulphur is the fifth most abundant volatile
element \cite{2009ARA&A..47..481A}, \cite{Gao2017},
\cite{2021A&A...653A.141A}, with a sulphur/hydrogen ratio of
$\ce{[S/H]}\!=\!10^{-4.85}$ \citep{2019arXiv191200844L}. This is small
compared to carbon $[\ce{C/H]}\!=\!10^{-3.53}$, nitrogen
$[\ce{N/H]}\!=\!10^{-4.15}$, and oxygen
$[\ce{O/H]}\!=\!10^{-3.29}$. 
{It is therefore remarkable that the sulphur carrying species 
\ce{SO2} was among the first new molecules discovered by the James Webb Space Telescope
  (JWST) in an exoplanet atmosphere. Sulphur-dioxide
  \ce{SO2} was observed in the gas giant WASP\,39\,b jointly with
  \ce{H2O}, \ce{CO2}, CO, and Na\,I \citep{2023Natur.614..659R}.
  \cite{Tsai2023} suggest that the observed \ce{SO2} is
  photochemically produced from \ce{H2S}, which actually is the main
  sulphur carrier in this atmosphere, but remained undetected. Their
  model requires a large metallicity of $10\times$ solar.
  \cite{2023arXiv230108492C} argue that sulphur may be overabundant in
  the inner atmosphere in WASP\,39\,b compared to Mg, Si and Fe,
  making it a candidate for tracing planet formation.
  \cite{2023arXiv230317622C} elaborates this idea by considering
  gas-phase kinetics for C/N/O/H/S species without the feedback of
  condensation processes. Earlier studies by \cite{Zahnle2016} for
  51\,Eri\,b did not predict large amounts of \ce{SO2}, but these
  authors used low metallicities and low irradiations in their
  exoplanet models.}

%\lj{Especially in our solar system the research on the sulphur
%  chemistry has been extensive.}
The element sulphur plays a key role in the atmospheric chemistry of
Venus \citep{2012Icar..218..230K, Rimmer2021}, Jupiter
\citep{Visscher2006}, { and its moon Io \citep{Zolotov1999},
  \cite{Spencer2000}, \cite{Moses2002}, \cite{2007iag..book..231L}}.
Mars does not show any traces of sulphur in its atmospheric spectrum
\citep{2011A&A...530A..37E}. { Several sulphur-bearing species have been observed in the coma of comets such as 67P/Churyumov–Gerasimenko~\citep{16CaAlBa}.}

Venus is known to contain about 150\,ppm \ce{SO2} in its \ce{CO2}-rich
atmosphere at $\sim\!90$ bar, with $[\ce{S/H]}\!\approx\!2$, and
\ce{H2SO4} cloud particles at $\sim\!1$ bar \citep[][{ see
    references of previous modelling work on the Venusian atmosphere
    in this article}]{Rimmer2021}.

\citet{Visscher2006} modeled Jupiter's atmospheric conditions and
found sulphur being predominantly present in the form of \ce{H2S}.
Photolysis of \ce{H2S} could be a step in the chemical reaction chain
towards sulphur hazes \citep{Gao2017}.  These aerosols are expected to
have a smoothing effect on infrared transit spectra of exoplanets
\citep[e.g.][]{Gao2021, Dymont2021}.  Both laboratory and space
research have shown that sulphur can stimulate haze formation.
\citet{2020DPS....5240304H} conducted laboratory experiments and
confirm this for warm, \ce{CO2}-rich atmospheres.  { Haze particles
  can form, e.g.\ from \ce{S8}-molecules \citep{Zahnle2016}, and act
  as a catalyser, triggering reactions of species such as \ce{NH3} or
  \ce{CH4}, which can change the optical appearance of the respective
  spectral features.}

{ During the SL-9 impact on Jupiter, large amounts of CS, \ce{CS2},
  and OCS were detected in the impact region, but SO and \ce{SO2} were
  not \citep{Zahnle1994}. The authors argue, that these detections are
  an indicator for a local carbon to oxygen ratio of $\rm C/O\!>\!1$,
  in which case sulphur forms CS, \ce{CS2} and OCS, whereas in the
  case $\rm C/O\!<\!1$, sulphur would form SO and \ce{SO2}.
  In protoplanetary disks, high CS and low SO and SO$_2$ abundances
  have been suggested as indicators for a high C/O ratio \citep{LeGal2021}.}

The simplest and most straightforward approach to model the near-crust
composition of exoplanet atmospheres is to assume thermo-chemical
equilibrium at the interface between planet crust and atmosphere.
While the computation of gas phase equilibrium concentrations has
been long established, for example by \cite{Tsuji1973},
\cite{Gail1984}, \cite{Allard1995}, \cite{Woitke2004}, it is the
inclusion of solid and liquid species in phase equilibrium that
allows us to discuss the composition of gases above silicate
materials, for example \cite{Lodders2002}, \cite{Hashimoto2007},
\cite{Schaefer2012}, \cite{Ito2015}, \cite{Woitke2018},
\cite{Wood2019}, \cite{Fegley2020}, \cite{Herbort2020}, and
\cite{Fegley2023}.  In particular, \cite{Fegley2016} have
studied the solubility of rock in steam atmospheres.  
\cite{Schaefer2012} and \cite{Herbort2020} have
calculated the chemical composition of the near-crust gas above common
rock materials as function of surface temperature and pressure,
showing a large variety of results for different element mixtures.  
Recently, \cite{Timmermann2023} have developed an open-source
python code for equilibrium condensation, and compared their results
to those of {\sc GGchem} \citep{Woitke2018}.
%\chh{ \cite{Herbort2020} also address element abundances that may 
%indicate the potential non-solar composition of extrasolar planets 
%derived from white dwarf (PWD) spectral analysis.}
{\cite{Fegley_2023} have published an extensive study of 69
elements in thermo-chemical equilibrium, using dry and wet Bulk Silicate
Earth (BSE) abundances. Considering temperature $1000-4500\,$K, they 
conclude that the silicate vapour behaves ideally at least up to 100\,bars,
and discuss the effects of treating the silicate melt as a non-ideal solution.}
%Themical equilibrium distribution of 69 elements between gas and melt is modeled for bulk silicate Earth also recently by \cite{Fegley_2023} based on the fugacity concept.

The validity of such an equilibrium approach can be questioned.
For example, \citet{2021arXiv210108327H} used a combination of {\sc
  Levi}, a photo-chemical kinetics code, and FastChem
\citep{Stock2018}, a thermo-chemical equilibrium solver to study the
sulphur chemistry on hot Jupiters. The authors compared their
modelling results to a model proposed by \citet{2017ApJ...850..199W}
and showed that thermo-chemistry dominates in most parts -- except for
the uppermost layers -- for the example of a $T_{\rm eq}\!=\!2000\,$K
atmosphere with solar metallicity. \citet{Rimmer2021} showed that
thermo-chemical equilibrium can simultaneously explain the crust
composition and near-crust gas composition of Venus.

Arriving at similar conclusions, \citet{Shulyak2020} modeled hot
Jupiter atmospheres around A0, F0, G2 and K0 stars for temperatures
around 2000\,K. They demonstrated that disequilibrium processes such
as vertical mixing and stellar XUV radiation do not change the
spectral results much at these temperatures.  However, for lower
temperatures around 1000\,K the picture changes. Shulyak et
al.\ predicted that the molecular mixing ratios in such UV-dominated
atmospheres are strongly affected by disequilibrium processes, and
possibly detectable with the James Webb Space Telescope (JWST).

The main JWST targets will be hot Jupiters and close in rocky
exoplanets with equilibrium temperatures between about 900\,K and
2200\,K around stars of types FKGM
\citep{Stevenson2016,2021arXiv211202031K}.  { For these objects, an
  application of thermo-chemical models can at least provide a good
  starting point for the composition of their atmospheres.} To classify the chemistry in these atmospheres, it is an important
question whether sulphur molecules might be detectable.

{ The observable concentrations of sulphur molecules in the
  atmospheres of rocky exoplanets are linked to the sulphur
  abundance (and those of the other elements) in the crust, which is
  ultimately set by planet formation and evolution.  These conditions
  include, among other factors, the temperature conditions and
  availability of elements during planet formation and the proximity
  to the star, for example \cite{2018ApJ...869L..41A},
  \cite{2019ApJ...885..114K}, \cite{Khorshid2021},
  \cite{2022arXiv220202483C}, \cite{2022MNRAS5106059S}} Depending on
the element abundances present in the planet's crust and surface, the
atmospheric conditions can vary from oxidising to reducing
\citep{Ortenzi2020}, and evolve with time.  Observational data for the
atmospheres of rocky planets are still very sparse even if utilizing
JWST \citep{Wordsworth2021,2023arXiv230314849G}.

In Sect. 2, we run thermo-chemical equilibrium models to predict
the abundances of sulphur molecules in the near-crust atmosphere, and
determine which are the most abundant sulphur molecules to be expected
in the atmosphere of hot rocky planets.  In Sect.~3, we run
simple radiative transfer models based on hydrostatic model
atmospheres with constant temperature and chemical composition to
find out which sulphur molecules might be detectable.  We use ARCiS
\citep{2020A&A...642A..28M} to generate transmission spectra, and
then simulate JWST/NIRSpec and JWST/MIRI observations in Sect.~4,
to discuss which spectral fingerprints of sulphur molecules might be
detectable.  We conclude and discuss our findings in
Sect.~\ref{sec:Summary}.

\begin{figure*}[!t]
\centering
\begin{tabular}{cc}
\hspace*{-5mm}
\includegraphics[width=90mm,trim=16 10 11 10,clip]{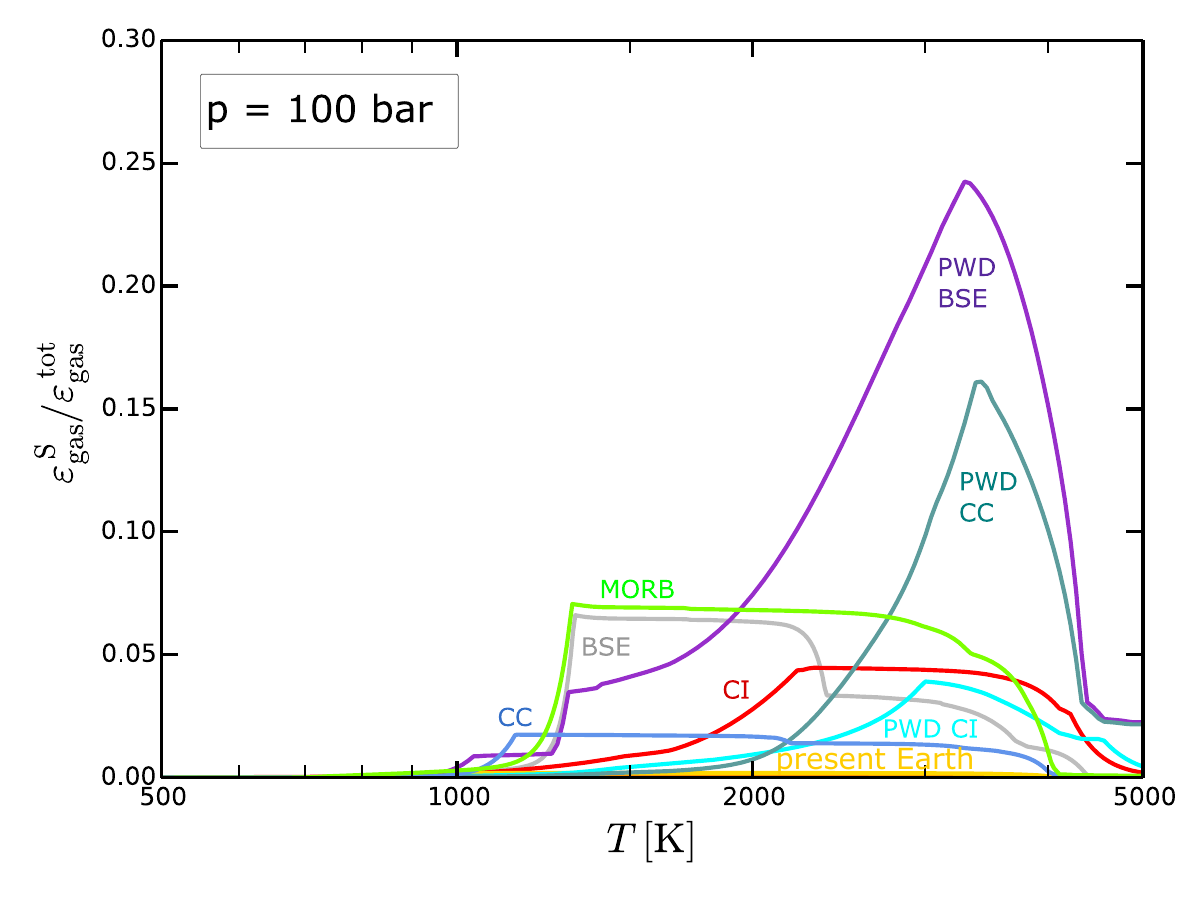} &
\hspace*{-5mm}
\includegraphics[width=90mm,trim=16 10 11 10,clip]{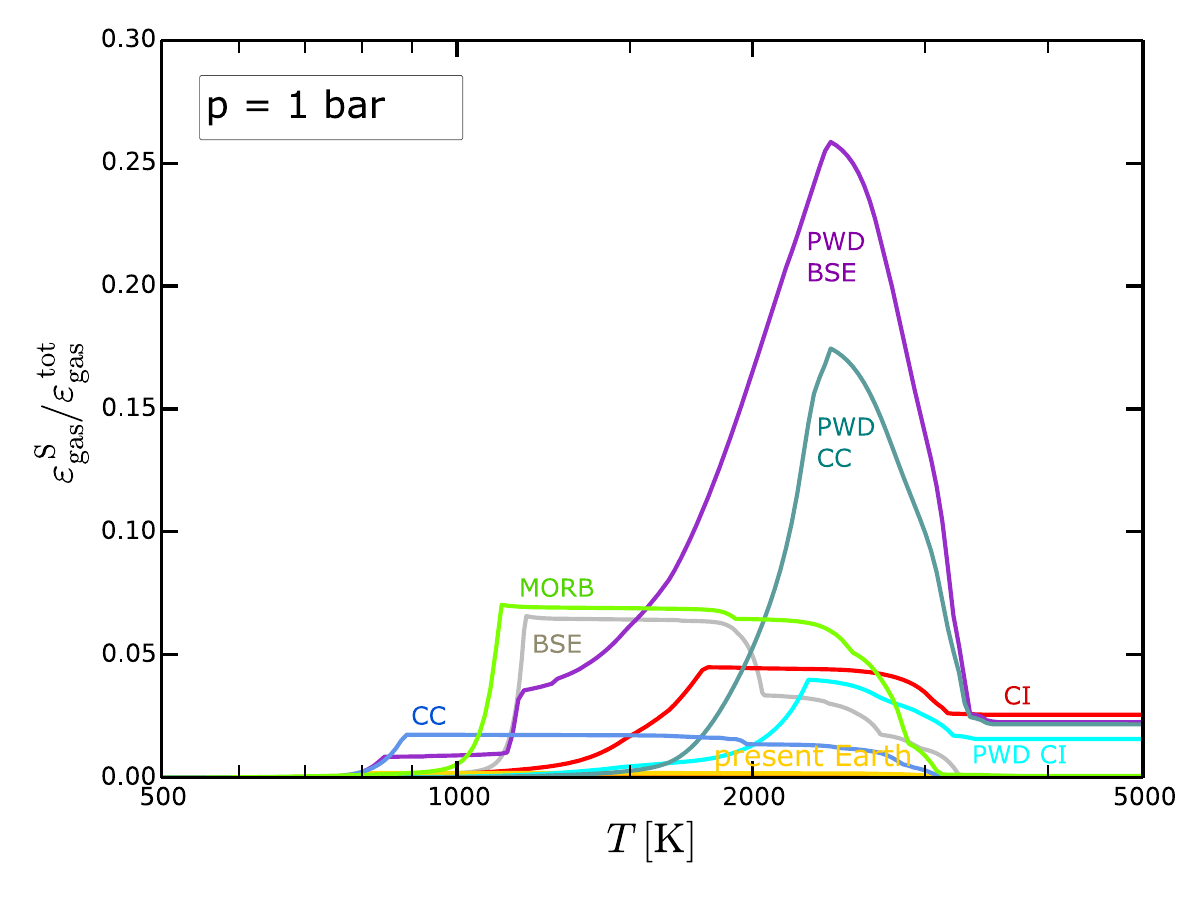} \\[-7mm]
\hspace*{-5mm}
\includegraphics[width=90mm,trim=5  10 11 10,clip]{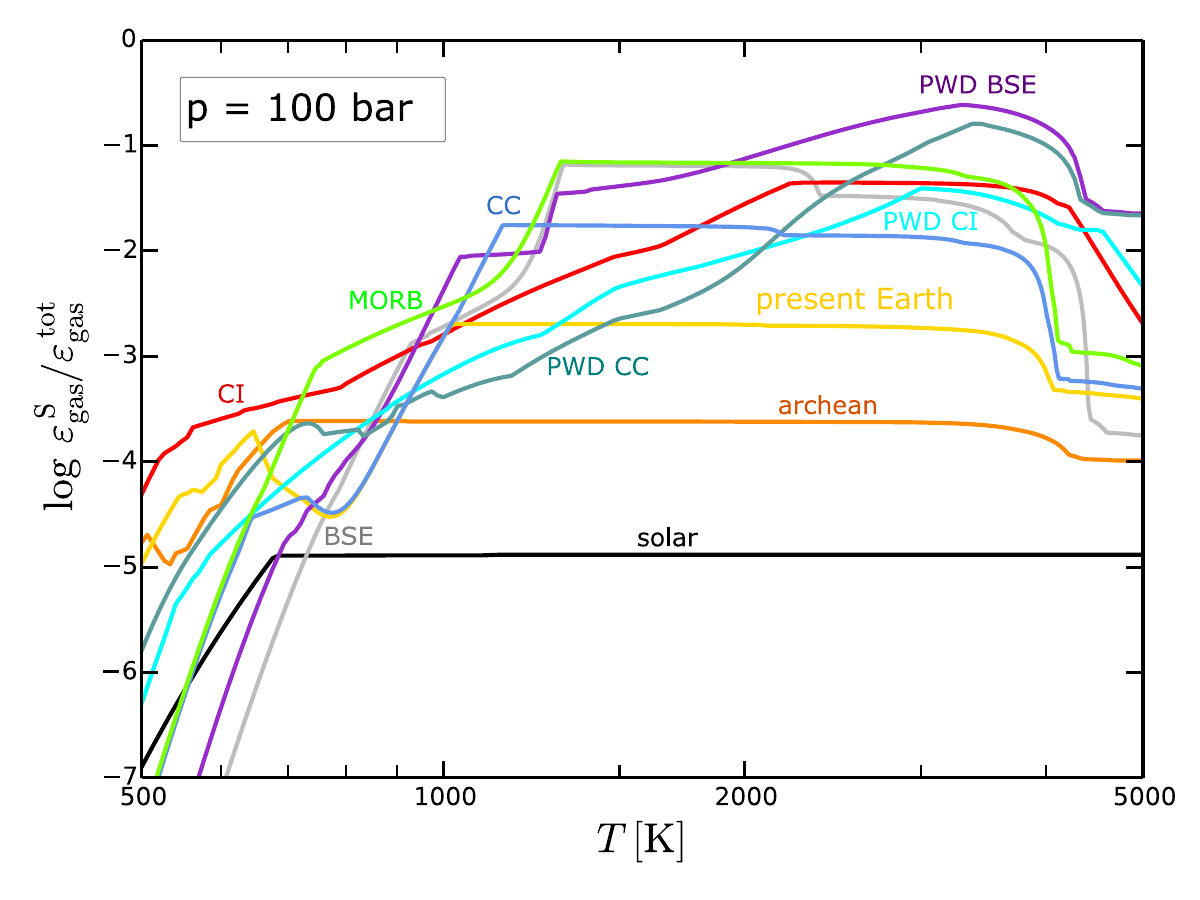} &
\hspace*{-5mm}
\includegraphics[width=90mm,trim=5  10 11 10,clip]{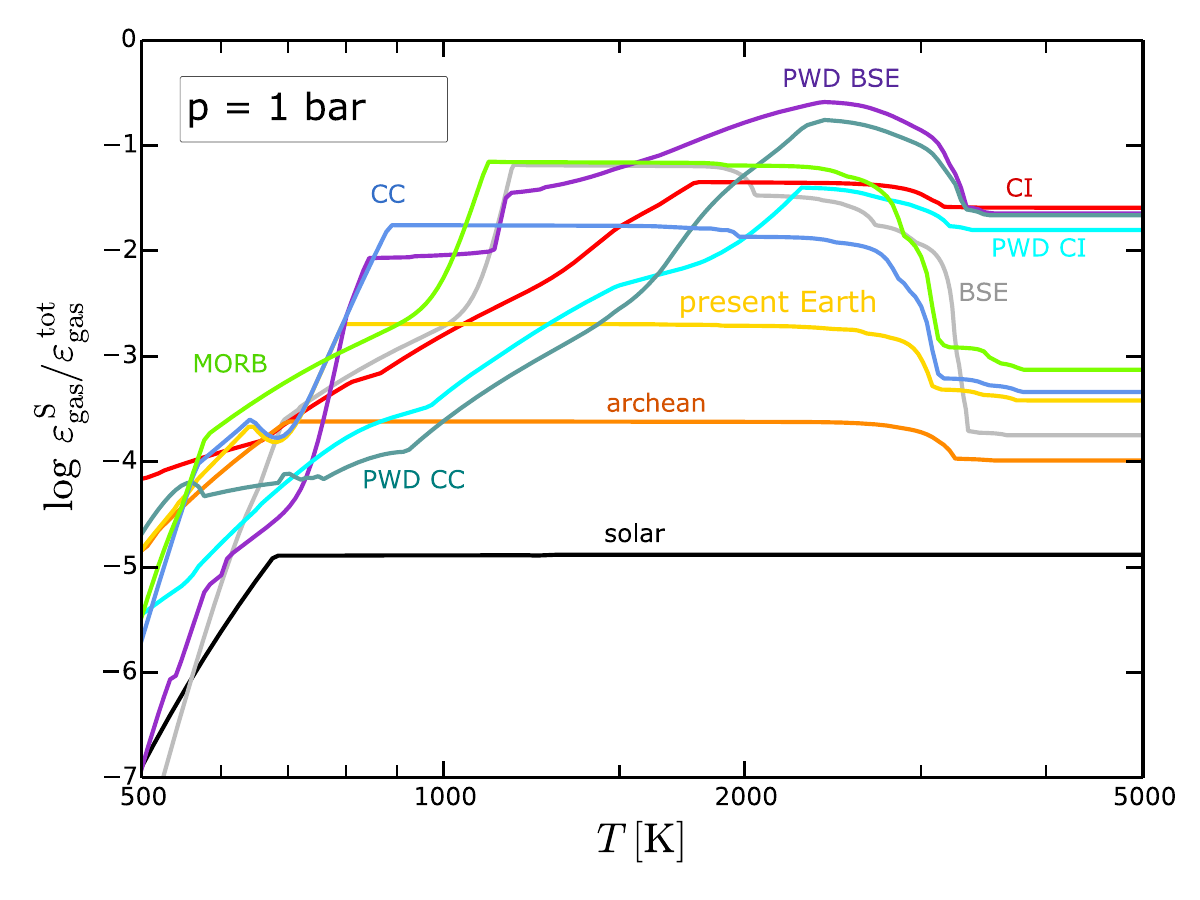}\\[-3mm]\end{tabular}
\caption{Sulphur abundances in the gas over rocky surfaces of
  different materials (labelled) as function of temperature.
  $\epsilon^{\rm S}_{\rm gas}$ is the element abundance of S in the
  gas phase and $\epsilon^{\rm tot}_{\rm gas}$ the sum of all element
  abundances in the gas phase.  Results are shown on a linear axis on
  the top, and on a logarithmic axis at the bottom. The left and right
  figures show the results for 100\,bar and 1\,bar, respectively. The
  material labels are explained in Sect.~\ref{sec:model}.}
\label{sulphur_gasform}
\vspace*{-2mm}
\end{figure*}

\section{Phase equilibrium models}
\label{sec:sulphur}

\subsection{Modelling approach}
\label{sec:model}

We use the principle of minimisation of Gibbs free energy to determine both the chemical composition of the gas and the material composition of the crust at the surface of a rocky planet.  We use the Fortran code {\sc GGchem} \citep{Woitke2018} in this work.  Similar phase equilibrium models have been developed e.g.\ by \cite{Lodders2002}, \cite{Schaefer2012}, \cite{Ito2015}, \cite{Fegley2016}, \cite{Wood2019}, and \cite{Timmermann2023}.  Our modelling approach is visualised by Fig.~1 in \citet{Herbort2020}.  A given set of total element abundances is considered in chemical and phase equilibrium at pressure $p$ and temperature $T$, assuming an equilibrium between outgassing and deposition. The code determines which condensates are stable, how much of these condensates are deposited, and calculates the composition of the remaining gas phase in contact with them.  The results include the chemical concentrations of all considered sulphur molecules.  The resulting fractions of the condensed species are interpreted as the surface mineral composition of the rocky exoplanet. The {\sc GGchem} model assumes a mixture of ideal gases, so we use partial pressures instead of the more general concept of fugacities.

Our model includes 18 elements (H, C, N, O, F, Na, Mg, Al, Si, P, S,
Cl, K, Ca, Ti, Cr, Mn, and Fe).  According to this selection of
elements, {\sc GGchem} finds 471 molecules (71 of them contain
sulphur) and 208 liquid and solid condensates in its database. We do not include silicic acid as a gaseous species \citep[as
    proposed by][]{Fegley2016}.  Concerning the assumed total element
abundances prior to condensation, we consider the following 10
datasets as described in \citet{Herbort2020} and \citet{Herbort2022}:
Bulk Silicate Earth (BSE), Continental Crust (CC), CI chondrite (CI),
Mid Oceanic Ridge Basalt (MORB), Archean, present Earth 
(see Appendix ~A in \citealt{Herbort2022}), and solar abundances.  We also
consider Polluted White Dwarf (PWD) compositions.  The spectroscopic measurements of PWDs \citep[e.g.][]{Bonsor2020} do not
  allow for certain key element abundances to be determined, in
  particular hydrogen, but also nitrogen, flourine and chlorine, to
  mention a few.  We therefore completed one basic PWD dataset of
  element mass fractions \citep{Melis2016} with the mass fractions of
  the missing elements from either BSE, CC or CI, see table~2 in
  \citet{Herbort2022} for more details. For example, the notation
PWD-BSE means that we consider base PWD abundances, completed by BSE
abundances.

\subsection{Resulting gas compositions}

Similar to \cite{Hashimoto2007}, \cite{Schaefer2012} and
\cite{Fegley2016}, we studied the gas phase composition above rocks
in models for $p\!=\!1\,$bar and 100\,bar.
Table~\ref{Tab_crusttypes_100bar} lists our results for the 10
different sets of total element abundances in form of the main
molecules and condensates that occur at different temperatures between
100\,K and 5000\,K for a constant surface pressure of
$p\!=\!100$\,bar. This pressure is similar to Venus' surface pressure.
We also list the sulphur concentrations that result to remain in the
gas phase, i.e.\ in the atmosphere.

Figure~\ref{sulphur_gasform} plots the same results as function of
temperature. According to these results, rocky exoplanet atmospheres
can contain up to $25\%$ of gaseous sulphur. This
maximum value is found at 3000\,K and 100\,bar in the PWD-BSE
model. This gaseous elemental sulphur concentration is significantly
higher, for example, than in our model with Earth-like element
abundances, where it only reaches a maximum of about $1.7\%$ at about
1000\,K.

Both Tables~\ref{Tab_crusttypes_100bar} and
\ref{Tab_crusttypes_1bar} show that all models (except for the
hydrogen-rich CI chondrite model) are featured by an \ce{N2}-rich
atmosphere at the lowest temperatures. As the hydrogen content in CI
chondrite atmospheres is high, \ce{NH3} replaces \ce{N2} at these
temperatures. Ammonia is further discussed in
  \cite{Hashimoto2007} and \cite{Schaefer2007}.

Similar to gas giants such compositions would be classified as an
A-type atmosphere following the scheme of \citet{Woitke2020}
based on H, C, N, O element abundances.  {We have extended this
  classification to include sulphur in \ref{A1}.  It provides a very
  helpful scheme to understand the very complex sulphur chemistry and
  the condensations that can occur therein
  (Table~\ref{Tab_condens}). The sulphur chemistry is not only
  characterised by the redox-state (see $y$-axis in
  Fig.\,\ref{atmos_types}), but we have a two-dimensional problem, where
  a suitable second axis is found to be a relative carbon content in
  the gas (see $x$-axis in Fig.\,\ref{atmos_types})}.  The molecule
\ce{CO2}, which is predicted to be abundant in type~B and type~C
atmospheres, cannot form in A-type atmospheres. And the molecule
\ce{CH4}, which is found to be abundant in type~A and C atmospheres,
cannot form in B-type atmospheres.
%This atmospheric type sets the red-ox state in the atmosphere, and
%the sulphur molecules found to be abundant change accordingly, see \ref{A1} for more details.

\begin{figure}[!t]
\hspace*{-1.6mm}
\includegraphics[width=87.6mm,trim=0 33 0 0,clip]{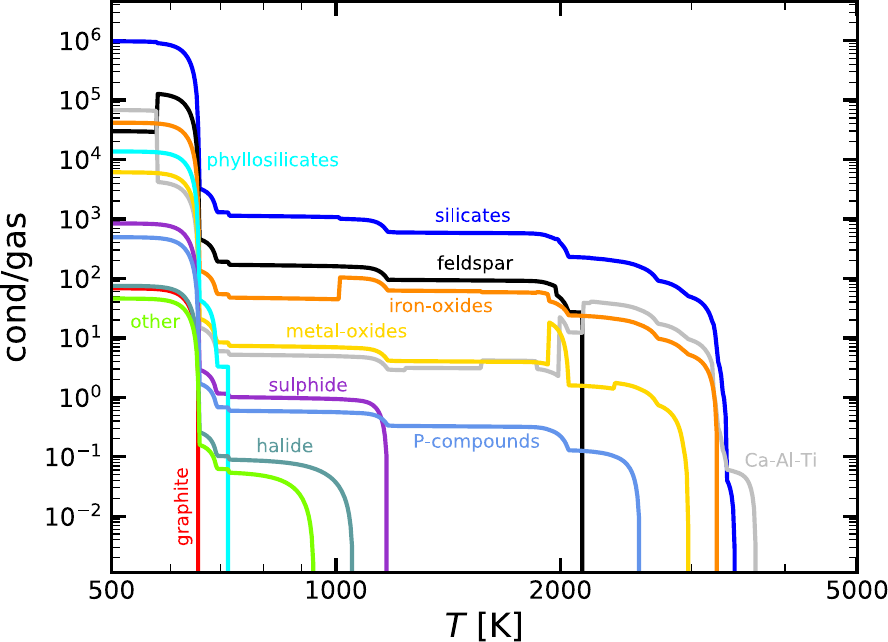}\\
\hspace*{0.0mm}
\includegraphics[width=86.0mm,trim=0  0 0 0,clip]{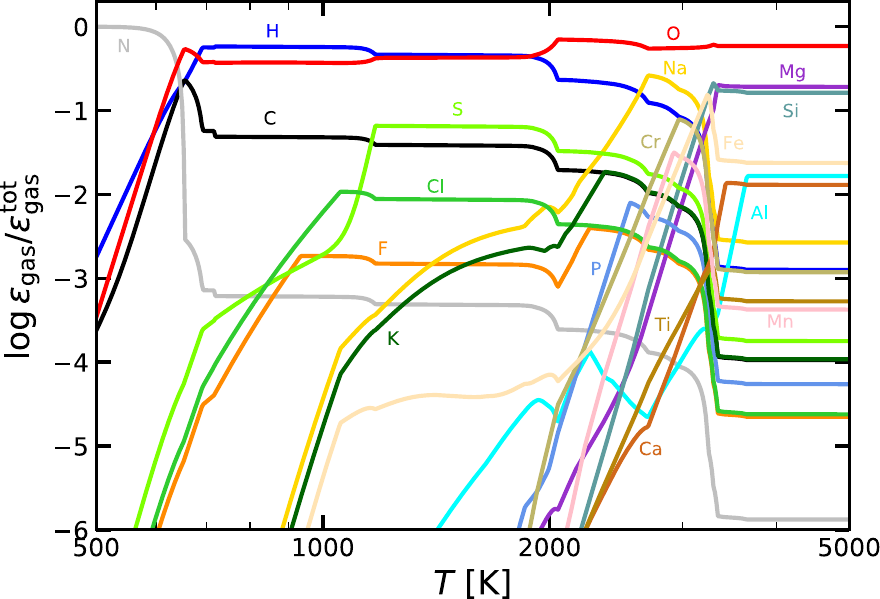}\\*[-6mm]
\caption{Equilibrium condensation model at constant pressure
  $p\!=\!1$\,bar for Bulk Silicate Earth (BSE) total element
  abundances. The {upper plot} shows the condensate-to-gas mass
  ratio, i.e.\ the mass of families of condensed species with respect
  to the mass of the gas. The {lower plot} shows the relative
  element abundances in the gas phase { on a $\log_{10}$ axis}.}
\label{one_model}
\vspace*{-2mm}
\end{figure}

Figure~\ref{one_model} shows the results of one of our equilibrium
condensation models in more detail.  We selected the BSE model at
$p\!=\!1\,$bar for this plot, see Table~\ref{Tab_crusttypes_1bar} for
details.  The relative sulphur abundance in the gas phase peaks around
$1000-2000$\,K (green line, labeled \ch{S}), reaching a maximum value
of about 6.5\%.  At these temperatures, sulphur becomes the third most
abundant element in the gas phase after hydrogen and oxygen, more
abundant than carbon and nitrogen.  The family names of condensates
used in the upper part of Fig.\,\ref{one_model} are explained in the
appendix in Table~\ref{condensates}.

\subsection{Why so much sulphur?}

The peaking sulphur abundance as function of temperature is a consequence of condensation.
%is a consequence of (i) the assumed
%total (gas\,$+$\,condensate) element abundances and (ii) 
%{ the subsequent formation of condensates, which remove certain
%  elements from the gas phase}, where sulphur is among the more volatile
%elements.
%This finding is supported by our kinetic cloud model where
%the effect of increasing sulfur element abundances did not change the
%relative amount if the non-sulfur metal-oxide condensates in
%oxygen-rich atmospheres \citep{Helling2019}.
From hot to cold
temperatures, the relative sulphur abundance first increases, by about
two orders of magnitude in the BSE model, as the first Ca-Al-Ti
compounds form, as well as some of the most stable liquid silicates
and iron-oxides, which drastically reduce the Mg, Si, Fe, Ca and Al
abundances in the gas phase (see Fig.\,\ref{one_model}).  At
temperatures $\lesssim\!1170$\,K in this model, sulphur starts to condense
as well, here in form of \ce{FeS}[s] {\sl (troilite)}, below which the
sulphur abundance in the gas phase falls quickly. Thus, although
sulphur is only the 12$^{\rm th}$ most abundant element at high
temperatures, it becomes the 3$^{\rm rd}$ most abundant element in
the gas between about 1000\,K and 2000\,K, due to condensation.

At 1500\,K, the composition of the atmospheric gas (molar mixing
ratios) is 66\% \ce{H2O}, 19\% \ce{SO2}, 11\% \ce{CO2}, 1.5\%
\ce{HCl}, 0.64\% \ce{NaCl}, 0.43\% \ce{HF}, and 0.37\% \ce{KCl},
followed by \ce{H2}, \ce{N2}, \ce{CO} and \ce{FeCl2}, with all other
molecules having concentrations $<\!100\,$ppm. { We note that
  results like these depend on the completeness of molecular and
  condensed species included in the model. For example,
  \cite{Fegley2016} claim that silicic acids is among the
  species present in a hot steam atmosphere.}

At $T\!\lesssim\!650$\,K, phyllosilicates and graphite start to become
stable in this model, which successively removes all remaining oxygen,
hydrogen and carbon from the gas phase, leaving behind a pure \ce{N2}
atmosphere. A condensate to gas mass ratio of $10^6$ is reached below
about 600\,K, which means that one gram of gas corresponds to one ton
of condensates.

The first idea to explain the large range of gaseous sulphur
concentrations is to consider the total sulphur (gas and condensed)
mass fraction that we use as input. However, we do not see a clear
correlation here, see Table~\ref{Tab_crusttypes_100bar}.  For example,
the model PWD-BSE with the highest gaseous sulphur abundance has an
input sulphur mass fraction of 3.3\%, whereas the Earth-model, which
results in one of the lowest gaseous sulphur concentrations, uses
4.8\%.
%the BSE model has the lowest of all considered input sulphur
%mass fractions (0.027\%), yet the gaseous sulphur concentration is
%almost as high as in the MORB model, which shows the highest gaseous
%sulphur concentration but uses an input sulphur mass
%fraction of 0.11\%.
%The input sulphur mass fraction alone is hence insufficient to predict
%the sulphur abundance in the gas phase.
%We need to consider the abundances of the other elements, too.

The amount of sulphur remaining in the gas phase is controlled by the
thermal stability of the condensates. { In fact, it is the
  determination of the stable condensates, not the amount of them,
  that sets the gas phase results.  Each stable condensate
  provides one auxiliary condition to the solution of the gas phase
  equilibrium, namely that a certain combination of partial pressures
  must equal the given saturation pressure of that condensate.
  The amount of each stable condensate can hence be arbitrarily
  increased without changing the gas phase results, see
  Appendix B of \cite{Woitke2018}.}
%The dimension $N$ cannot be larger
%than the number of elements included in phase equilibrium.

The most frequent sulphur condensates are found to be \ce{FeS}[l,s] {\sl
  (trolite)}, \ce{FeS2}[s] {\sl (pyrite)}, \ce{CaSO4}[s] {\sl
  (anhydrite)} and \ce{MnS}[s] {\sl (alabandite)}, see
Table~\ref{Tab_crusttypes_100bar}.  This would suggest that the
gaseous S concentration should depend on the availability of Fe and Ca
to form those sulphur compounds. The more Fe and Ca was available, the 
more sulphur condensates should form, which would consume more S and
hence lower the S gas concentration.  However, this idea also fails to
explain our results. All PWD models, for example, use a large input Fe
mass fraction of about 10\%, whereas the MORB and BSE models use less,
7.3\% and 6.3\%, respectively; yet the PWD-BSE and PWD-CC models
have more gaseous sulphur.
%Similar for calcium. Whereas the PWD models have a Ca mass fraction
%of 6.9\%, the BSE composition only has 2.7\%.

The best explanation we can provide is that it depends on the
hierarchy of condensation, i.e.\ how well the most refractory
elements, such as Si, Mg, Fe, Al and Ca, can be put together into
highly refractory condensates, which depends on their element
stoichiometry.  If there is a leftover of Fe or Ca (or, precisely
speaking, if it is necessary to put these leftovers into less stable
condensates), then sulphur condensates start to form, and the S
concentration in the gas phase drops.

More insight into the regulating processes can be obtained by studying
the $T$-dependence shown in Fig.~\ref{sulphur_gasform}. We can
distinguish between two different cases.
\begin{itemize}
\item[(1)] The PWD-BSE, PWD-CC and PWD-CI models show clear peaks of
  the sulphur concentration that remains in the gas phase as function
  of $T$, and a positive slope in the $1500-3000$\,K region at 1\,bar
  (the peak value in the PWD-CI model is lower by about a factor of
  2-4). At these temperatures, the models show a gradual transition
  from \ce{CO2} to CO on a $\approx\!40$\%-level (see
  Table~\ref{Tab_main_mols}), an about constant \ce{SO2}
  concentration, a rising \ce{S2} concentration, and a gradual
  evaporation of \ce{FeS}[l] while the \ce{FeO}[l] concentration rises
  with increasing temperature.  Our conclusion is that the
  dissociation $\ce{CO2} \to \ce{CO} + \ce{O}$ is followed by
  $\ce{FeS[s]} + \ce{O} \to \ce{FeO[l]} +
  \frac{1}{2}\ce{S2}$. \\*[-3.5ex]
\item[(2)] The MORB and BSE models show a constant gaseous sulphur
  abundance between about 1200\,K and 2000\,K at 1\,bar.  These models
  are featured by a steam atmosphere with high and about constant
  \ce{H2O}-concentration (see Table~\ref{Tab_main_mols}), followed by
  constant \ce{SO2}, \ce{CO2} and \ce{O2}
  concentrations. There are no stable sulphur condensates at these
  temperatures, and no \ce{CO2}-dissociation, so the gaseous S
  abundance stays about constant.
\end{itemize}
Other models show a combination of these two patterns.  For example,
the CI model at 1\,bar first follows the pattern of case (1) up to
$\sim\!1800$\,K, above which \ce{FeS}[l] is no longer stable
and the gaseous S concentration becomes constant. The CI model has
more hydrogen, so \ce{H2O} remains the most abundant molecule
throughout, and \ce{H2} is present as well. Thus, the release of
oxygen as described in case (1) leads to a gradual change of the redox
character, causing the \ce{H2S} concentration to drop, whereas
\ce{SO2} increases, while \ce{S2} is also present.

Changing the gas pressure from 1\,bar to 100\,bar (see right and
left parts of Fig.~\ref{sulphur_gasform}) does not change the
qualitative behaviour of the mixture of gas and condensates.
It mainly causes a shift of the thermal stability to
higher temperatures such that the temperature thresholds, at which
certain phase transitions occur are raised by a factor of roughly
1.3-1.4.

The CC model looses its last sulphur condensate \ce{CaSO4}[s] at about
900\,K at 1\,bar, above which there is a constant steam atmosphere
made of 44\% \ce{H2O} and 34\% \ce{CO2} with about 5\% \ce{SO2}, see
Table~\ref{Tab_main_mols}.  The model with solar input abundances
shows a straightforward behaviour, with \ce{FeS}[s] as the main
sulphur condensate up to about 650\,K at 1\,bar, above which there is
a constant but low concentration of \ce{H2S} which changes to HS at
higher temperatures. Gas giants as investigated by \citet{Gao2017},
\citet{2021arXiv210108327H} and \citet{2021A&A...649A..44H} are
relevant examples for such A-type atmospheres, see \ref{A1}. However,
the majority of our rocky exoplanet models favours the production of
\ce{SO2}, which is classified as BC1, BC2 or BC3-type atmospheres in
\ref{A1}.

In the PWD-CC and PWD-BSE models, the molecule \ce{S2} ({\sl
  di-sulphur}) also reaches concentrations of a few percent beside the
molecule \ce{SO2}. Tables \ref{Tab_crusttypes_100bar} and
\ref{Tab_crusttypes_1bar} show that these high \ce{S2} concentrations
coincide with the peak of the gaseous sulphur abundance around
2500-3500\,K shown in Fig.~\ref{sulphur_gasform}.

Beside the expected \ce{SO2}, \ce{H2S} and \ce{S2} molecules, we find
smaller concentrations of the molecule PS ({\sl phosphorus sulfide})
in some of our models, see Table~\ref{Tab_PS}. In certain cases, this
molecule shows a surprisingly strong absorption feature at optical
wavelengths, see Sect.~\ref{PS_features}.  In the models using polluted 
white dwarf abundances, and in the Archean and CI models, the 
concentrations of PS are between about 1 and several 100\,ppm at 2000\,K 
and 100\,bar. 

\begin{table}
\centering
\caption{The concentration of phosphorus sulphide (PS) at 2000\,K and 10\,bar. "$\ll$" means a concentration $<10^{-15}$.}
\label{Tab_PS}
\vspace*{-1mm}
\begin{tabular}{c|c} 
   \hline
   model & PS concentration [ppm]\\ 
   \hline
   CC & $\ll$ \\
   BSE & $\ll$\\
   CI & $2.7$ \\
   MORB &  $\ll$ \\
   Archean & $1.6$ \\
   present Earth & $\ll$ \\
   PWD-CC & $4.3$ \\
   PWD-BSE & $2.9$ \\
   PWD-CI & $260$ \\
   solar & $3.4\times 10^{-3}$\\
   \hline
\end{tabular}
\end{table}

In the following section, we explore whether it might be possible to
detect sulphur molecules -- in particular \ce{SO2}, \ce{H2S} and
\ce{PS} -- by transmission spectroscopy using the James Webb Space
Telescope (JWST), given the high sulphur abundances that we predict
for hot rocky exoplanets.
%As the diatomic molecule \ch{S2} has no permanent dipole moment we do
%not expect detectable features caused by \ce{S2} in the wavelength
%regime that JWST operates in.

\section{Transmission spectra of sulphur-rich exoplanets}

Current and future observational missions like JWST \citep{2021ExA....51..109B}, 
%CHEOPS \citep{2021ExA....51..109B}, 
LUVOIR, HabEx and HABITATS \citep{Wang2020} can be expected to provide an unprecedented quality and variety of transit spectra of diverse rocky exoplanetary worlds. 
%we use the software ARCiS {\sl (ARtful modelling Code for exoplanet Science)} to explore which spectral features one can expect. 
As shown in Tables \ref{Tab_crusttypes_100bar} and \ref{Tab_crusttypes_1bar}, our models suggest that some of the atmospheres of these rocky planets might be very rich in sulphur. Therefore, the question arises whether we can detect sulphur-molecules by transmission spectroscopy with present and future observational facilities also for rocky planets.  We will not focus on any particular targets, but instead provide a first guidance of where to look for detectable spectroscopic signatures of sulphur-bearing molecules.

\subsection{The model atmosphere}\label{ss: slab}

% CHH: I suggest to keep this as simple as possible. Our aim here is to derive dominating processes, and not to get entangles with complex atmosphere simualtions. All text is still there and can be re-included or, if  at all, used for a discussion later in the paper. I am trying to concentrate here on what was done and what is possible.

We use simple 1D hydrostatic models with an isothermal structure and 
constant molecular concentrations.  The molecular concentrations are 
taken from the {\sc GGchem} phase equilibrium models for the surface 
as described in Sect.~2.  Constant molecular concentrations were 
previously adopted for atmospheres of magma planets, see e.g.\ 
\citet{Barth2021} and \citet{Graham2021}.  Strong vertical mixing is 
expected to occur in such atmospheres, keeping the molecular mixing 
ratios sustained throughout. We neglect cloud formation.

\subsubsection{The set of opacities in ARCiS} \label{ARCiS_code}

\begin{table}
\begin{center}
\caption{Atoms and molecules that can be fairly abundant in our
  atmosphere models without opacity data in ARCiS$^{(1)}$.}
\vspace*{0mm}
\label{missing_mols}
\vspace*{-2mm}\hspace*{-1mm}
\resizebox{87mm}{!}{\begin{tabular}{c|c|c|cc} 
   \hline
   &&&&\\[-2.4ex]
    $3\%>x>1\%$ & 
    $x>1\%$ & 
    $x>100\,$ppm & 
    \multicolumn{2}{c}{$x<100\,{\rm ppm}$}\\
   &&&&\\[-2.4ex]
   \hline
   &&&&\\[-2.4ex]
  \ch{NaOH} & \ch{AlF2O} & \ch{Na2SO4} & \ch{Cl} & \ch{P2}\\
  \ch{KOH} &\ch{COS} & \ch{H} & \ch{Mn} & \ch{S4} \\
  \ch{PO2} & \ch{P4O6} & \ch{KO} &\ch{CS2} & \ch{SiO2} \\
  \ch{S2O} & \ch{NaO} & \ch{S3} & \ch{K+} & \ch{TiO2} \\
   && \ch{HO2} & \ch{(KCl)2} & \ch{FeCl2}\\
   && \ch{CrO3} & \ch{MnH} & \ch{Mg(OH)2} \\
   && \ch{Fe(OH)2} & \ch{K2SO4} & \ch{Ca(OH)2}\\
   && \ch{CrO2} & \ch{MnO} & \ch{Cl-}\\
   && \ch{(NaCl)2} & \ch{Fe} & \ch{Mg}\\
   \hline
\end{tabular}}
\end{center}
\vspace*{-2mm} \resizebox{87mm}{!}{\parbox{95mm}{$^{(1)}$: Maximum gas
    particle concentrations $x\!=\!n_{\rm mol}/n_{\rm tot}$ of atoms
    and molecules across all our atmospheric models with no opacity
    data in ARCiS, ordered by concentration.  We note that KOH and
    NaOH line lists are now available from \citet{21OwTeYu}.}}
\vspace*{-2mm}
\end{table}

{ We used ARCiS \citep{2020A&A...642A..28M} to generate transmission
spectra.  The molecular opacities in these models are determined from
standardised pre-computed k-correlations from
\citet{2021A&A...646A..21C}, completed with a few species such as
H$^-$~\citep{88John}. The line list data in ARCiS are collected from
HITRAN~\citep{HITRAN_2016}, HITEMP~\citep{HITEMP},
MoLLIST~\citep{MOLLIST}, ExoMol~\citep{jt631,ExoMol2020} and
NIST~\citep{NISTWebsite}.  The ARCiS code has a list of molecules
for which opacity data is available, which are:}

\medskip\noindent
\ch{H-}, \ch{Na}, \ch{K}, \ch{CH4}, \ch{CO2}, \ch{H2O}, \ch{H3+}, \ch{H2}, \ch{VO}, \ch{TiH}, \ch{SiH4}, \ch{SiH}, \ch{ScH}, \ch{SO3}, \ch{HS}, \ch{PS}, \ch{PH3}, \ch{P2H2}, \ch{OH}, \ch{OH+}, \ch{NaF}, \ch{NaCl}, \ch{NO}, \ch{NH}, \ch{MgO}, \ch{MgH}, \ch{MgF}, \ch{LiH}, \ch{LiH+}, \ch{LiF}, \ch{LiCl}, \ch{KCl}, \ch{H2S}, \ch{H2CO}, \ch{H2+}, \ch{FeH}, \ch{CrH}, \ch{CaO}, \ch{CaH}, \ch{CaF}, \ch{CS}, \ch{CO}, \ch{CN}, \ch{CH3F}, \ch{CH3Cl}, \ch{AlO}, \ch{AlH}, \ch{BeH}, \ch{AsH3}, \ch{AlF}, \ch{AlCl}, \ch{TiO}, \ch{SiS}, \ch{SO2}, \ch{PO}, \ch{H2O2}, \ch{PN}, \ch{O3}, \ch{O2}, \ch{O}, \ch{C2}, \ch{NO2}, \ch{NH3}, \ch{N2O}, \ch{N2}, \ch{HCN}, \ch{HCl}, \ch{SiO}, \ch{NS}, \ch{HNO3}, \ch{CH}, \ch{HF}, \ch{KF}, \ch{CP}, \ch{C2H4}, \ch{C2H2}.

\medskip\noindent The line list data of particular importance 
for this paper can be found in \citet{jt635} for \ch{SO2}, in
\citet{jt640} for \ch{H2S}, in \citet{jt734} for \ch{H2O}, in
\citet{15LiGoRo.CO} for CO, in \citet{jt703} for PS, and in
\citet{20YuMeFr.co2} for \ch{CO2}. { Table~\ref{missing_mols}
lists some molecules with missing opacity data that could be interesting
for hot rocky exoplanet atmospheres.}

\begin{table}
\begin{center}
\caption{Input parameters for our ARCiS spectral models$^{(1)}$.}
\label{ARCIS_parms}
\vspace*{-1mm}
\begin{tabular}{c|c} 
   \hline
   parameter &setting\\ 
   \hline
   &\\[-2.2ex]
   effective temperature (star) & 5196\,K\\
   stellar radius & 0.98\,$R_\odot$ \\
   planetary system & 55\,Cnc\,e\\
   atmospheric temperature & 2000\,K \\
   surface pressure & 10\,bar\\
   planetary radius & $1.875\,R_{\rm Earth}$ \\
   planetary mass  & $7.99\,M_{\rm Earth}$ \\
   highest atmospheric layer & $10^{-8}$\,bar \\
   nb. of atmospheric layers & $50$ \\
   wavelength range & $0.3 - 29\,\mu$m\\
   spectral resolution & $100$ \\
   \hline
\end{tabular}
\end{center}
\vspace*{-2mm} \resizebox{87mm}{!}{\parbox{95mm}{$^{(1)}$:
    %The table summarises the stellar and planetary parameters used for
    %our spectroscopic models with ARCiS. The molecular concentrations
    %are assumed constant in our models and set by the {\sc GGchem}
    %simulations for the surface conditions.
    Stellar temperature and radius, as well as planetary radius and
    mass, are taken from estimates for 55\,Cnc\,e, see Table~1 in
    \citet{Tabernero2020}. By setting the name of the planetary
    system, these values are automatically adopted by ARCiS.
    %The surface pressure is adopted from \citet{Hammond2017}, but
    %rounded to 10\,bar.
    The models use 50 pressure layers for the hydrostatic atmospheric
    structure. The spectral resolution is set according to the
    resolution of the MIRI LRS spectrograph on JWST which is tailored
    to mid infrared spectroscopy.}}
\end{table}

\subsubsection{Input parameters}

{ The physical input parameters for the ARCiS models are listed in Table
  \ref{ARCIS_parms}.}  We use the hot Super Earth 55\,Cnc\,e as a
reference for the stellar and planetary parameters. 55\,Cnc\,e orbits
a GV\,8 star \citep{von_Braun_2011,Folsom2020,Tabernero2020}, has an
equilibrium temperature of 2350\,K and a surface pressure of
$5\!-\!10$\,bar \citep{Hammond2017}.  { We set the
  surface pressure to 10\,bar and the atmospheric temperature to
  2000\,K for these simple explorative models, except for those
  models where we investigate the dependencies on pressure and
  temperature.}
%In all cases, we assume an isothermal atmosphere with
%an exponentially decreasing pressure profile down to $10^{-8}$ bar.
%The pressure range of the layers as defined in ARCiS gives an
%indication of the atmospheric extent.

\subsubsection{Atmospheric composition and extent}

The molecular concentrations, taken from the simulations of the near-crust atmospheric composition at $p_{\rm surf}$, $T_{\rm surf}$ (Sect.~\ref{sec:sulphur}), result in very different mean molecular weights, see Table ~\ref{Tab_main_mols}.   which has an important impact on the observability of the spectral features, because large mean molecular weights translate into compact atmospheres with a small atmospheric extent.

\begin{figure*}
\centering
\begin{tabular}{cc}
\hspace*{-3mm}  
\includegraphics[width=91mm,height=79mm,trim=13 10 0 10,clip]{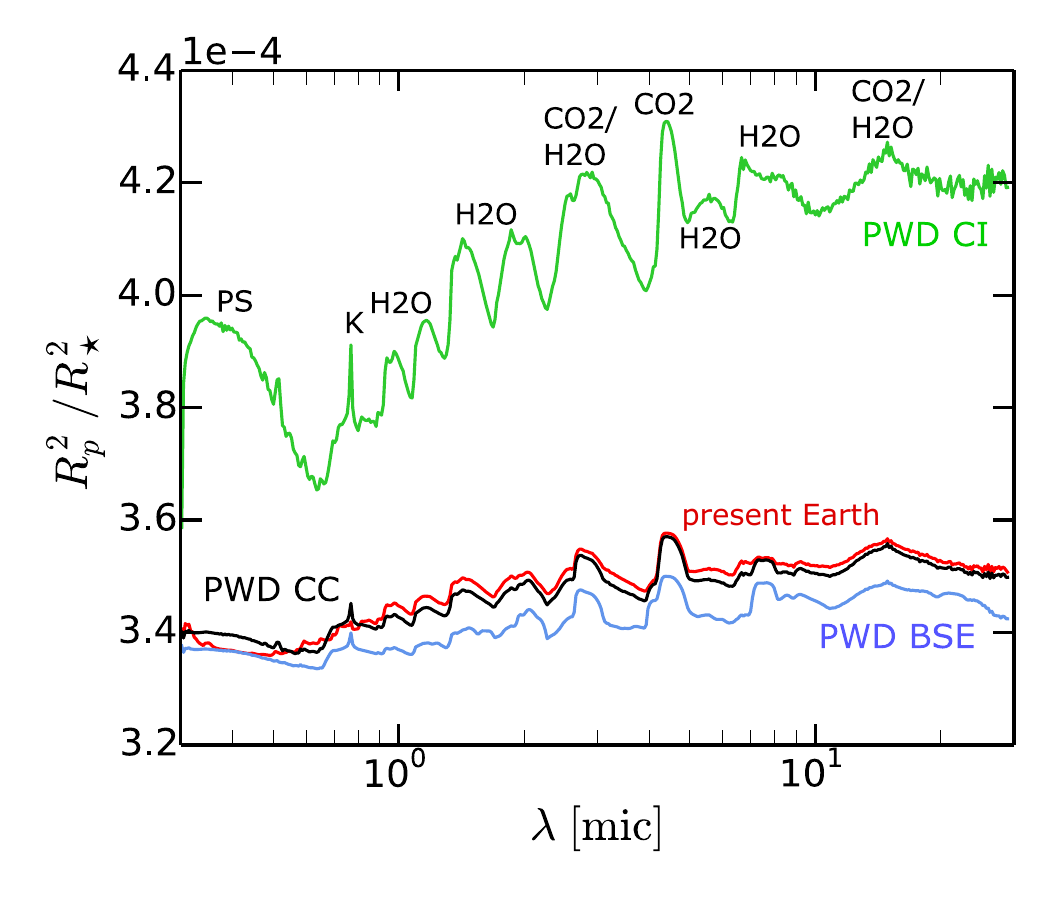} &
\hspace*{-7mm}
\includegraphics[width=91mm,height=79mm,trim=13 10 0 10,clip]{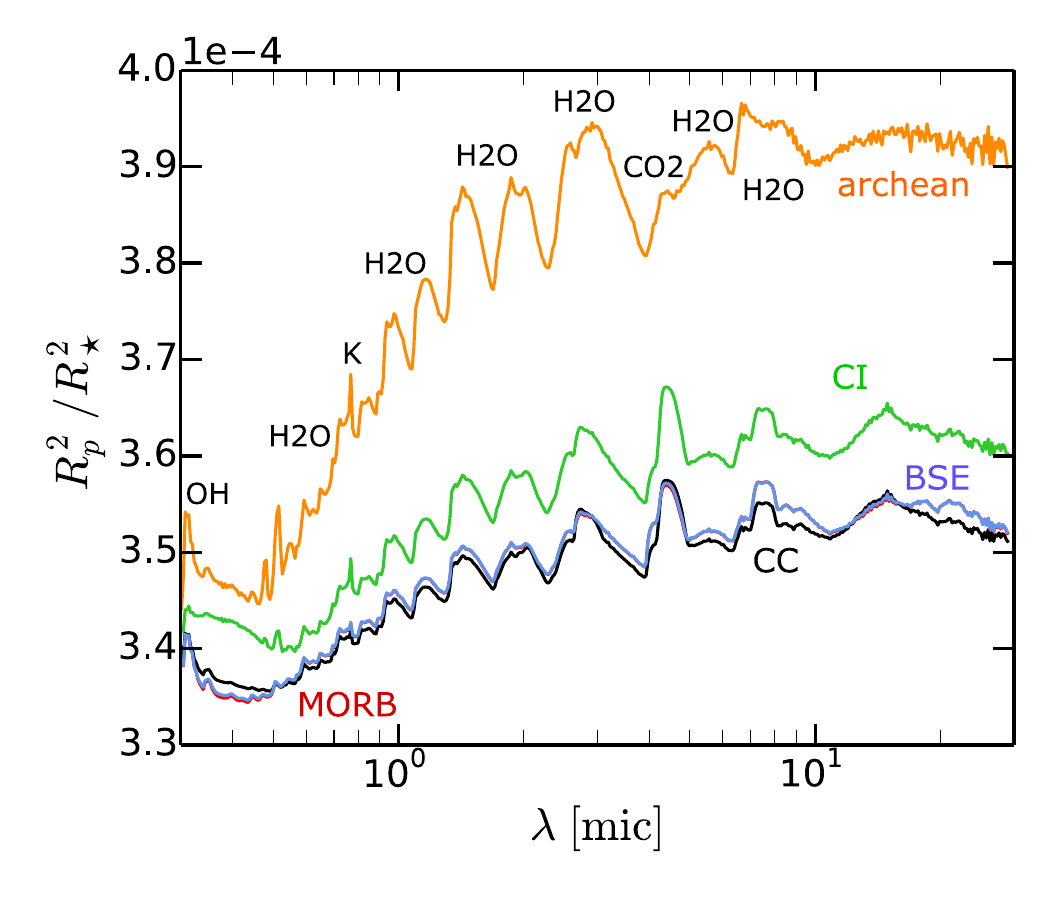} \\[-6mm]
\end{tabular}
\caption{Transmission spectra calculated with ARCiS for input
  parameters listed in Table~\ref{ARCIS_parms}. Nine models with
  different total element abundances are shown on the left and
  right. The transmission depths vary according to the different mean
  molecular weights $\mu$. We note the different scaling of the
  $y$-axis on the left and right. All spectra have been computed for a
  surface pressure of 10\,bar and an isothermal atmosphere of
  2000\,K. The solar model is not shown here because its mean
  molecular weight $\mu\!\approx\!2.05$, see Table~\ref{Tab_main_mols},
  is so small that the transit depth is significantly larger than
  all other models.}
\label{composition_spectra}
\vspace*{-1mm}
\end{figure*}

Some of the molecules and atoms, which are included in {\sc GGchem}, are not implemented in ARCiS yet.  These molecules are not passed to ARCiS for spectra generation, and are listed in Table~\ref{missing_mols}.  Other species are implemented in ARCiS, but have no opacity data there, such as \ch{SO} and \ch{S2}. The absorption features of these molecules cannot be predicted.  Ignoring some of the abundant molecules in the ARCiS simulations might have a small impact on the calculation of the mean molecular weight in ARCiS.  However, as Table~\ref{missing_mols} shows, none of these molecules exceeds a concentration of 3\%, and the molecules $>\!1\%$ only occur in our hottest models $T\!\ge\!2500$\,K or in the artificially sulphur-enriched models.  Among these molecules is \ch{NaOH} in the Archean and BSE models, for which opacity data has recently been published by \cite{21OwTeYu} like for the SO molecule.  The impact of the mean molecular weight of our models is discussed further in Sect.~3.2 and Sect.~3.3.

\subsubsection{Pressure broadening and CIA}

In our models, the atmospheres of hot rocky exoplanets are mostly
constituted of \ce{CO2}, \ce{H2O}, \ce{CO}, \ce{H2}, and \ce{SO2} (see
Tab. \ref{Tab_main_mols}). { These atmospheres are hence very
  different from the atmospheres of gas giants where \ce{H2} and He
  dominate}.  The opacities we use in this work assume pressure
broadening by \ch{H2} and \ch{He}.
%However, we would expect the pressure broadening to be different
%in atmospheres made up of heavier molecules, as is the case with
%our models.
The available broadening data of HITRAN2020~\citep{HITRAN2020} for
other species indicate that pressure broadening from \ch{H2O} in
particular is expected to be relevant. Works such as
\cite{19GhLi.broad} and \cite{22AnChCh} illustrate the effect that
self pressure broadening can have for \ch{H2O} in steam atmospheres.
However, even if we were able to take these pressure broadening effects into
account, we do not expect them to affect the main conclusions of the
present work.  We compute our models considering only collision
induced absorption (CIA) from H$_2$-H$_2$ and H$_2$-He.
%We do not expect CIA between heavier species
%to have a significant impact on our models.
Future work will show whether updated CIA opacities have an influence 
on these results.
%to update the modelling tool and to compute
%rocky atmosphere spectra with taking into account pressure broadening
%and CIA due to these molecules.

\subsection{Resulting transmission spectra}

{ The following sections show our calculated transmission spectra
  for 10 different rocky element mixtures and different atmospheric
  temperatures and surface pressures, see Fig.~\ref{composition_spectra}
  for an overview. In each of these cases, we are
  looking out for observable spectral features of the various sulphur
  molecules.}
%In order to investigate the contribution of a
%particular molecule to the transit spectrum, we run the same model
%again with zero opacity data for the molecule under investigation.

\begin{figure*}
\centering
\begin{tabular}{cc}
\hspace*{-7mm}
\includegraphics[width=85mm,height=65mm,trim=4 0 0 18,clip]{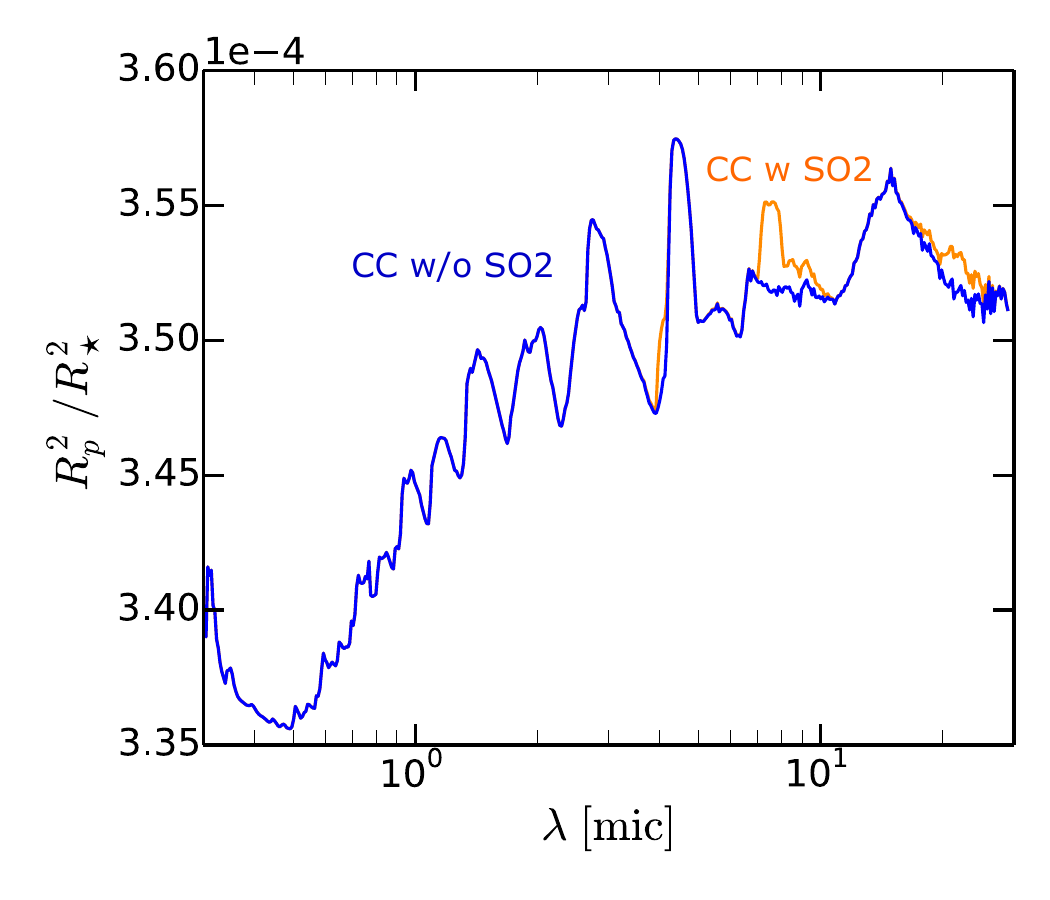} &
\hspace*{-7mm}
\includegraphics[width=85mm,height=65mm,trim=4 0 0 18,clip]{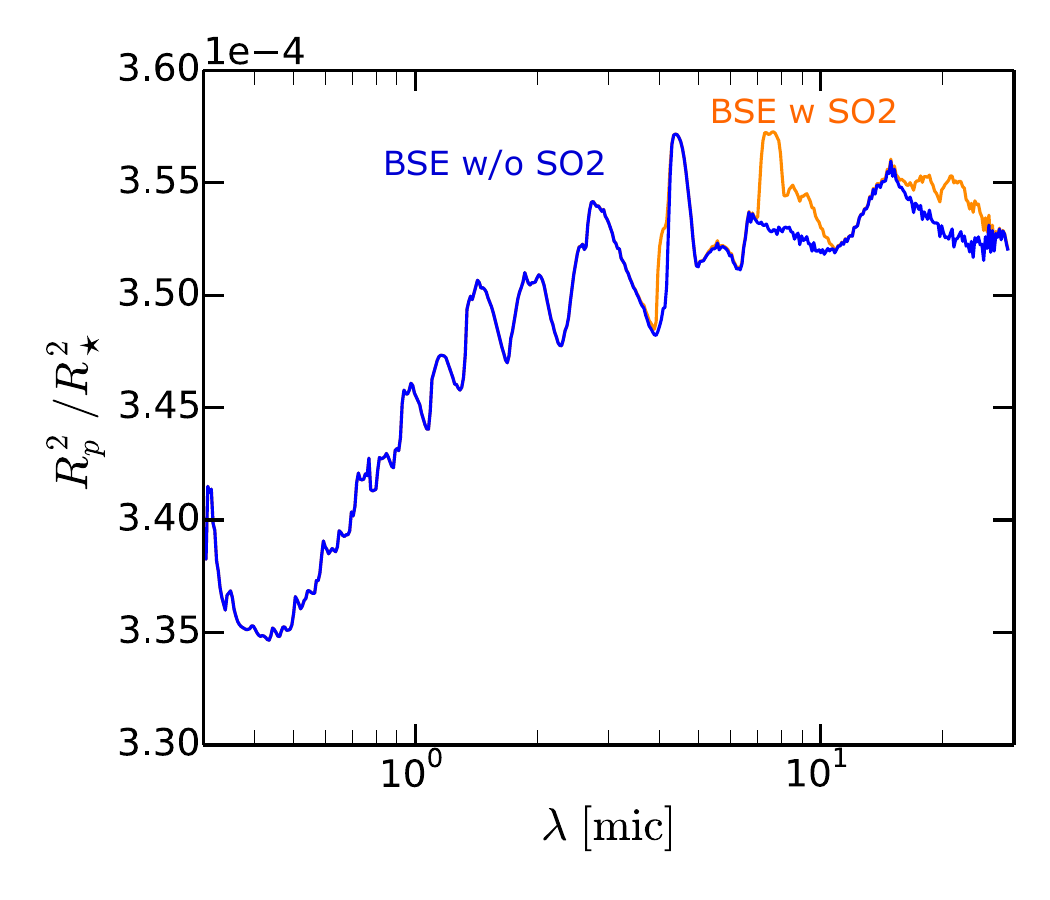} \\[-9mm]
\hspace*{-7mm}
\includegraphics[width=85mm,height=65mm,trim=4 0 0 18,clip]{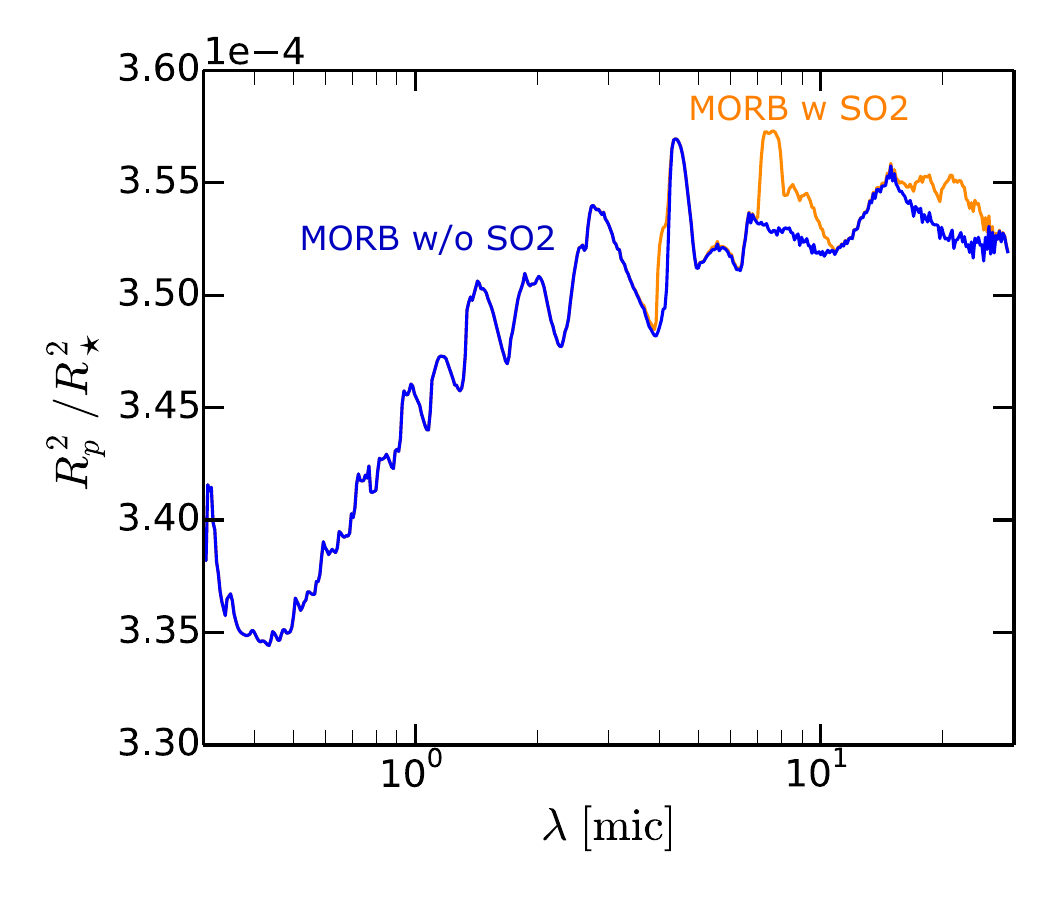} &
\hspace*{-7mm}
\includegraphics[width=85mm,height=65mm,trim=4 0 0 18,clip]{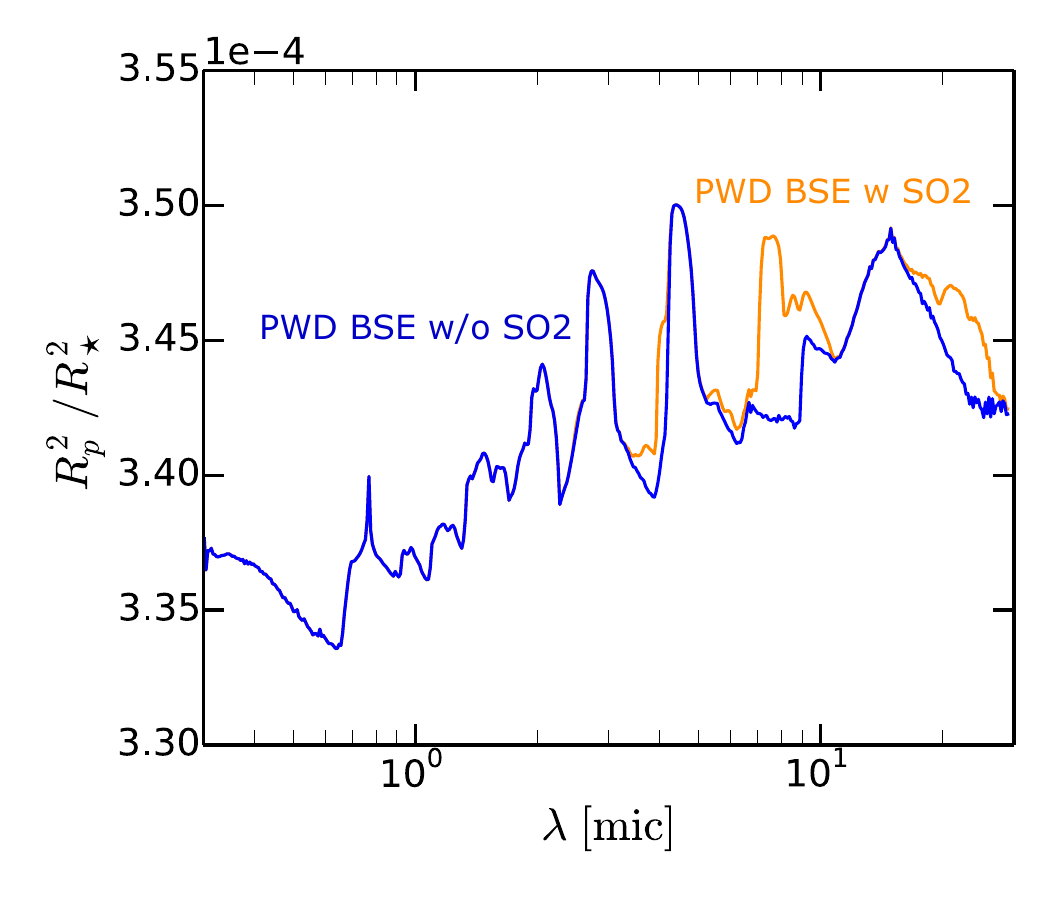} \\[-7mm]
\end{tabular}
\caption{Impact of \ce{SO2} on the transmission spectra generated with
  ARCiS for isothermal atmospheres with $T\!=\!2000\,$K and surface
  pressure $p_{\rm surf}\!=\!10\,$bar. The model input parameters are
  listed in Table~\ref{ARCIS_parms}.  The four compositions which show
  the strongest \ce{SO2} absorption features are selected: CC,
  BSE, MORB and PWD-BSE. The blue lines show the spectra when the
  \ce{SO2} line opacity is omitted.}
  %The mean molecular weight of the
  %respective molecules is still accounted for and the scale of the
  %y-axis is adapted in each row.}
\label{Fig_SO2}
\end{figure*}

\subsubsection{Mean molecular weight and scale height}
\label{mu}

The pressure scale height $H_p$ has a profound impact on the
observability of spectral features in exoplanet transmission spectra.
It is given by
\begin{equation}
  H_p = \frac{k_{\rm B}T}{g\,\mu\,m_{\rm H}}
\end{equation}
where $k_B$ is the Boltzmann constant, $T$ the temperature, $g\!=\!G
M_{\rm pl}/R_{\rm pl}^2$ the surface gravity of the planet, and $\mu$
the mean molecular weight in units of the mass of hydrogen $m_{\rm
  H}$. $G$, $M_{\rm pl}$, and $R_{\rm pl}$ are the gravitational
constant, the mass and radius of the planet, respectively, which were
adopted from 55\,Cnc\,e (Table~\ref{ARCIS_parms}). In our isothermal
models with constant molecular concentrations, we have $T\!=\!T_{\rm pl}$
(temperature of the planet's surface and atmosphere), and $H_p$ and
$\mu$ are constants.

Table~\ref{Tab_main_mols} lists the results of a series of ten models
with different element abundances at fixed $T\!=\!2000$\,K and fixed
surface pressure $p_{\rm surf}\!=\!10$\,bar. { Beside the resulting
molecular compositions, we also list the mean molecular
weights that varry between 9.2 and 61.6, where the abundant heavy
\ce{SO2} molecule can have a profound influence.}

\begin{figure*}
\centering
\begin{tabular}{cc}
\hspace*{-7mm}
\includegraphics[width=85mm,height=65mm,trim=4 0 0 18,clip]{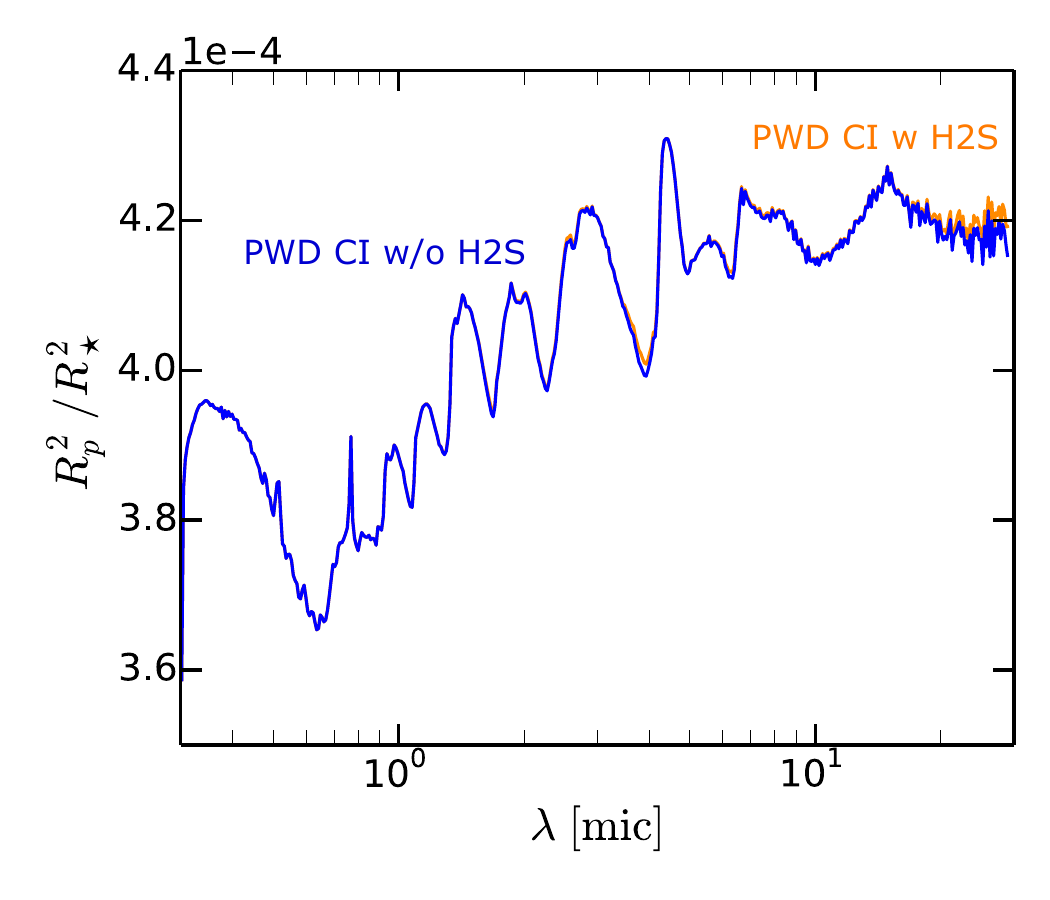} &
\hspace*{-7mm}
\includegraphics[width=85mm,height=65mm,trim=4 0 0 18,clip]{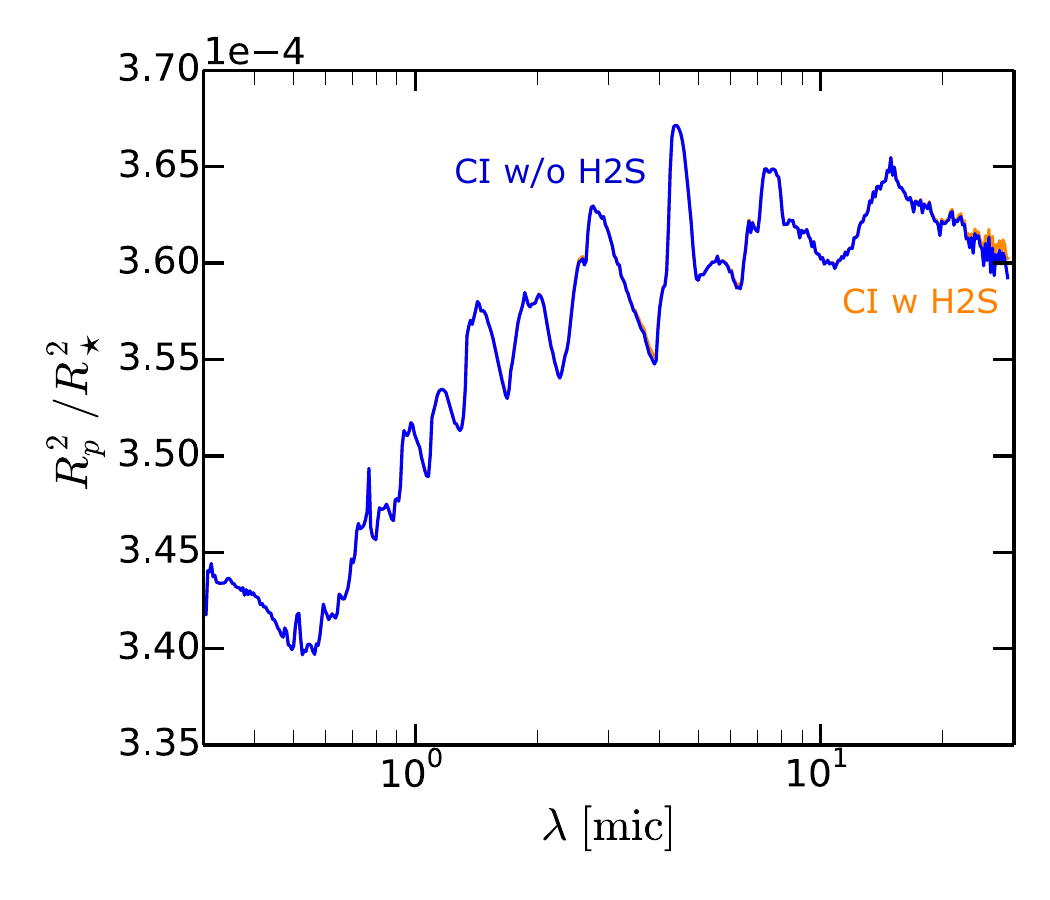} \\[-7mm]
\end{tabular}
\caption{Impact of \ce{H2S} on the transmission spectra generated with
  ARCiS for isothermal atmospheres with $T\!=\!2000\,$K and $p_{\rm
    surf}\!=\!10\,$bar. The model input parameters are listed in
  Table~\ref{ARCIS_parms}. The two compositions with visible \ce{H2S}
  features are selected here: CI and PWD-CI.  The blue line shows the
  spectrum when the \ce{H2S} line opacity is omitted.}
  %The mean
  %molecular weight of the respective molecules is still accounted for
  %and the scale of the y-axis is adapted in each figure.}
\label{Fig_H2S}
\end{figure*}

Figure ~\ref{composition_spectra} shows our predicted transmission
spectra for nine of the ten models (solar model excluded), where we
see a clear correlation between $\mu$ and the transit depths. For
example, the model with the largest transit depth is the PWD-CI
composition. The main species in its atmosphere is \ce{H2} with a
mixing ratio of about $67\%$, followed by \ce{H2O} (20\%) and \ce{CO}
(8.5\%), which results in a mean molecular weight of only
$\mu\!=\!9.2$ (Table \ref{Tab_main_mols}). Larger mean molecular
weights $\mu$ lead to smaller scale heights $H_p$, making it 
more difficult to detect any molecular features by transmission
spectroscopy. This effect limits our ability to find and
spectroscopically characterise sulphur-rich exoplanets, which
generally have large $\mu$.

\subsubsection{SO$_2$ and H$_2$S in transmission spectroscopy}

Table \ref{Tab_main_mols} shows that the most abundant gaseous sulphur
species in any of our models is either \ce{SO2} or \ce{H2S}. Both
molecules are non-linear and polar, with a bent structure similar to
\ce{H2O}. Whereas \ce{SO2} is more dominant in the BSE, MORB, CC,
present Earth, PWD-BSE and PWD-CC compositions, \ce{H2S} dominates in
the solar, Archean, CI and PWD-CI models.

%In Figs.~\ref{Fig_extrasolar}, \ref{Fig_ssystem}, \ref{FigPWDspectra} and \ref{Fig_solar} we show simulated transmission spectra for 2000\,K and 10\,bar for different total element abundances.
{ Figures~\ref{Fig_SO2} and \ref{Fig_H2S} show the simulated
  transmission spectra for 2000\,K and 10\,bar for different total
  element abundances. From our sample of 10 different rocky element
  compositions, we have selected the spectra with the most distinct
  absorption features of \ce{SO2} and \ce{H2S}. See \ref{A4} for all
  other spectra.}

Besides the full spectra based on all included opacities (orange), the
figures also show spectra for which the opacities of \ce{SO2} and
\ce{H2S} have been artificially set to zero (blue). { Thus, these
  molecules are still included in the calculation of the mean molecular
  weight in ARCiS}. The following spectral features of \ce{SO2} and
  \ce{H2S} occur:
\begin{itemize}
    \item a box-like absorption feature at $7\!-\!8\,\mu$m with a
      right shoulder extending to $\sim\!10\,\mu$m $\Rightarrow$
      \ce{SO2},
    \item a left shoulder on the main \ch{CO2} absorption band at
      $4.5\,\mu$m $\Rightarrow$ \ce{SO2}, { this is the feature
      that was recently detected by \citep{2023Natur.614..659R}
      in WASP\,39\,b},
    \item increased absorption longward of $15\,\mu$m $\Rightarrow$
      \ce{SO2}.
    %\item no clear observable features of \ce{H2S}.
\end{itemize}
{ None of our models predicts strong spectral features of \ce{H2S}.}
Whether \ce{H2S} or \ce{SO2} dominates in the gas depends on the O/H
ratio in the atmosphere. Our classification scheme for temperatures
below 600\,K (\ref{A1}) indicates no coexistence of \ce{H2S} and
\ce{SO2}. However, at larger temperatures, Table~\ref{Tab_main_mols}
shows that both molecules are important in the CI model at 2000\,K and
10\,bar, indicating a smooth transition from one molecule to the
other.  { At 100\,bar, Table~\ref{Tab_crusttypes_100bar} indicates
  that in the CI model, \ce{H2S} is abundant below 1700\,K and
  \ce{SO2} at higher temperatures.}

\subsubsection{Distinct SO$_2$ absorption features}

\ch{SO2} gives rise to some very distinct spectral features { as
  shown in Figs.~\ref{Fig_SO2}, ~\ref{Fig_H2S}, \ref{Fig_extrasolar}
  and \ref{FigPWDspectra}.} These absorption features are particularly
strong in the models for MORB, BSE and PWD-BSE compositions. These are
particularly rich in gaseous sulphur as illustrated by Figure
\ref{sulphur_gasform}. { The PWD-BSE model provides the sulphur richest
atmosphere, with a sulphur concentration of $\sim\!\!15\%$ in the gas
phase at 2000\,K and 1\,bar surface pressure}, see
Fig.~\ref{sulphur_gasform}), which also shows the strongest spectral
features due to \ch{SO2} absorption.  However, even though the overall
gaseous sulphur abundance is the highest there, the \ch{SO2} abundance
is not the highest in the PWD-BSE model. It's atmosphere reaches a
\ch{SO2} concentration of $\sim\!14\%$ at 2000\,K and 10\,bar surface
pressure, whereas it is by $4-5\%$ higher for the atmospheres of the
BSE and MORB compositions respectively, see
Table~\ref{Tab_main_mols}.

{ This puzzling behaviour of the \ch{SO2} absorption features
can be explained by looking at the dominant molecular
species in the three different atmospheres.} Whilst the dominating
molecule in the PWD-BSE atmosphere is \ch{CO2}, the BSE and MORB
atmospheres are mainly composed of \ce{H2O}.  Water produces
very strong absorption features all the way between $2$ and
$10\,\mu$m, { such that the \ch{SO2} features become undetectable.}

\subsubsection{\ch{H2S} absorption features}

{ Using ARCiS, we find only weak \ch{H2S} absorption
  features. These features are just about visible in the PWD-CI and
the CI models, see Fig.~\ref{Fig_H2S}.}
CI chondrites found on Earth date back to the time of the
 evolutionary formation stages of the solar system
\citep{Herbort2020} and are particularly rich in carbon.  Besides a
high C/O ratio, CI chondrites also have a high H/O ratio. This shows
to have a strong impact on the formation of \ce{H2S} rather than
\ce{SO2}.  { In the CI model,} \ce{H2S} is the major gaseous
sulphur species below 2400\,K at 100\,bar surface pressure. 
%As the temperature increases, the \ch{SO2} mixing ratio increases
%whereas the \ch{H2S} mixing ratio decreases, as shown in
%Table~\ref{Tab_crusttypes_100bar}.
Above 2400\,K, \ch{SO2} becomes the dominant sulphur species even in the CI
atmosphere.
%At 2000\,K and 10\,bar surface pressure \ch{H2S} and \ch{SO2} occur and
%are similarly abundant, see Table ~\ref{Tab_main_mols}. The features
%due to \ch{H2S} are very weak with the clearest absorption peaks in a
%wavelength range of $25-30\,\mu$m, see Fig.~\ref{Fig_H2S}.

The gaseous sulphur concentration in the CI and PWD-CI models is low
in comparison to other models such as the BSE composition (see
Fig.~\ref{sulphur_gasform}), even though the total sulphur abundance
for both hydrogen-rich models is high. Therefore, the mixing ratios of
all sulphur containing species is relatively low in these two
models. The \ch{H2S} content reaches $3.6\%$ in the CI model, compared
to $19\%$ for \ch{SO2} in the MORB model, which is the model with the
largest \ch{SO2} concentration. This is one of the reasons for the
shallowness of the \ch{H2S} features in Fig.~\ref{Fig_H2S}.

{ In addition, the \ce{H2S} features are located at wavelengths
  where the \ch{H2O} molecule is dominant.  The overlap with \ch{H2O}
  opacity is stronger for \ce{H2S} than it is for \ce{SO2}, such that
  the water, which dominates in all atmospheres with high \ch{H2S}
  concentrations, masks their features.}

\subsubsection{The strong absorption feature of PS}
\label{PS_features}

\begin{figure}
\centering
\includegraphics[width=86mm,height=67mm,trim=20 25 15 10,clip]{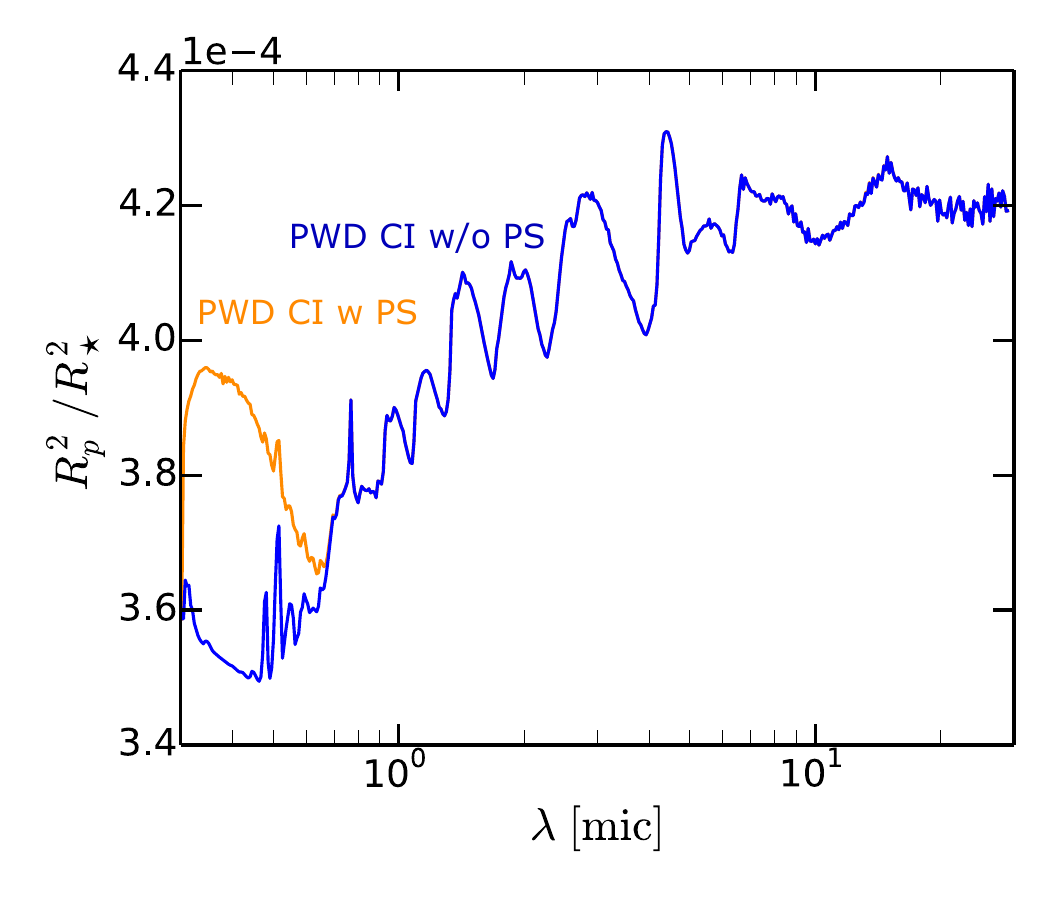}
\vspace*{-2mm}
\caption{Transmission spectrum obtained with ARCiS for PWD-CI abundances
  at 2000\,K and 10\,bar. The orange curve shows the complete results
  including PS, whilst the blue curve is the spectrum obtained with zero
  PS opacity.}
  %The ARCiS model parameter settings are listed in Table~\ref{ARCIS_parms}.}
\label{PS_CI}
\end{figure}

Table~\ref{Tab_PS} shows that all PWD models, the Archean model, and
the CI model contain $1-260\,$ppm of the PS molecule at 2000\,K
and 10\,bar surface pressure.  { We discovered that this molecule
  causes a surprisingly strong and broad absorption feature between
  0.3 and 0.8\,$\mu$m, see Fig.~\ref{PS_CI}. This figure shows the
  transmission spectrum of the PWD-CI model, which has the largest PS
  concentration among all models (260\,ppm).  The transmission
  spectra of all other models producing the PS absorption feature are
  shown in \ref{A4}.}
%Table~\ref{Tab_PS} indicates, that the solar model also contains
%some concentrations of PS which don't give rise
%to any features in Fig.~\ref{PS_effect}.

This unexpected result raises the question about the reliability of the thermo-chemical data used to calculate the PS concentrations.  A comparison of different data sources for the Gibbs free energies of PS \citep{Worters2017} shows deviations of about 5\,kJ/mol, which is a typical uncertainty, so this suspicion can be rejected. According to the NIST database, PS is energetically favored over \ce{P2} and \ce{S2} by about $15-20$\,kJ/mole at about 1500\,K.

\subsubsection{Dependence on total sulphur abundance}

\begin{figure}
\centering
\includegraphics[width=86mm,height=67mm,trim=20 25 15 10,clip]{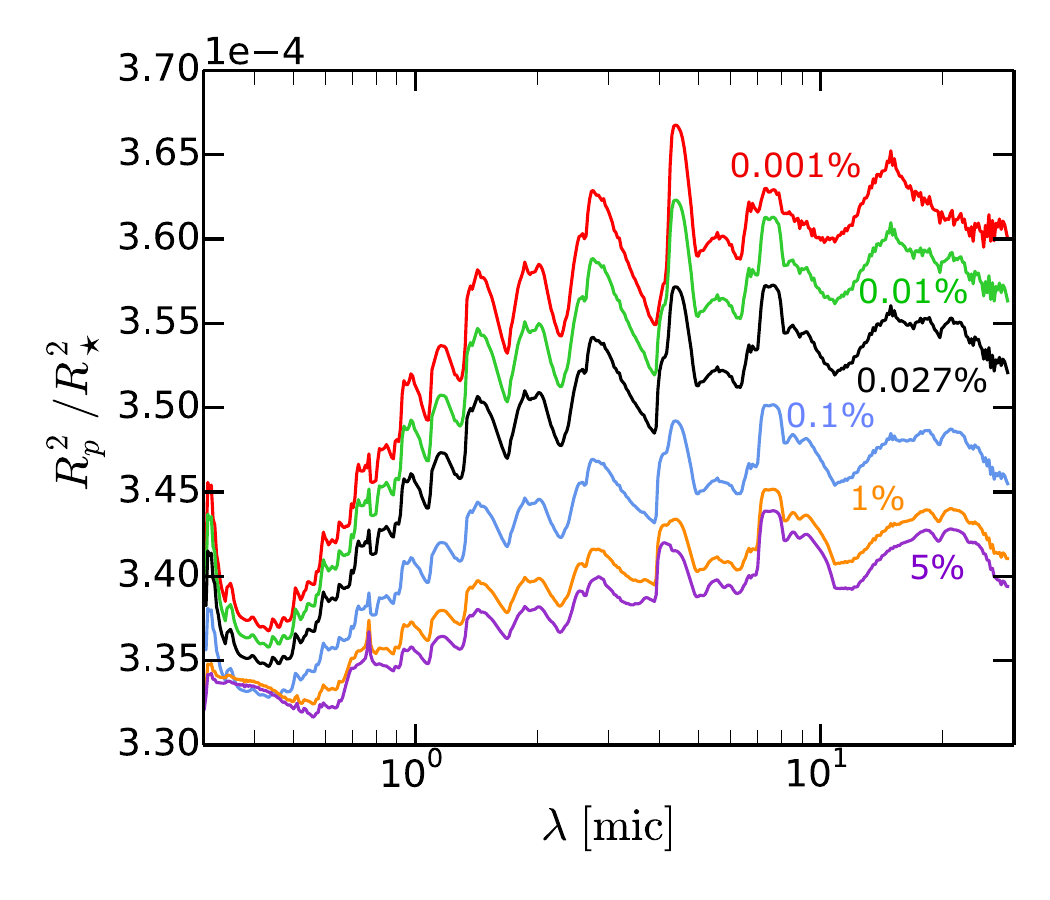}
\vspace*{-2mm}
\caption{Transmission spectra for a Bulk Silicate Earth composition
  with artificially varied total sulphur element mass fraction (gas
  and condensates). The spectra have been computed for a surface pressure of
  10\, bar and an isothermal atmosphere of 2000\, K. The numbers next
  to the graphs indicate the corresponding sulphur mass fraction for
  each modeled spectrum.}
\label{vary_sulph}
\end{figure}

{ Considering the BSE composition as an example, we have
  investigated how the transmission spectrum changes if we
  arbitrarily vary the total sulphur element abundance in the model.}
Figure~\ref{vary_sulph} shows spectra for sulphur mass fractions of
0.001, 0.01, 0.027 (unaltered BSE model), 0.1, 0.993
($\sim\!1.0$), and 4.77 ($\sim\!5.0$)\footnote{When we set the total
  element abundance of each individual element (crust and atmosphere)
  in percentage mass fraction, they do not sum up to exactly
  100$\%$. Before computing the equilibrium chemistry, {\sc GGchem}
  renormalises to $100\%$.}.
%This is why an input element abundance
%for 1\% leads to a final mass fraction percentage of 0.993\%.}

A larger total element abundance of sulphur allows for more
 \ch{SO2} and \ch{S2} to form in the gas phase, see
Table~\ref{Tab_main_mols}. As both are very heavy molecules, the mean
molecular weight of the atmosphere increases, leading to a drop in
scale height $H_p$ and therefore to a flattening of all features. This
is illustrated in Fig. ~\ref{vary_sulph}. The box-like feature around
$7-8\,\mu$m of \ch{SO2} is strongest for a total sulphur
mass fraction of $0.027\%$. For smaller values, the \ce{SO2} concentration in the atmosphere drops quickly. For larger values, the radial extent of the atmosphere shrinks substantially, 
see Table~\ref{Tab_main_mols}.

\begin{figure*}
\centering
\begin{tabular}{ccc}
\hspace*{-4mm}
\includegraphics[height=54mm,trim=20 20 15 0,clip]{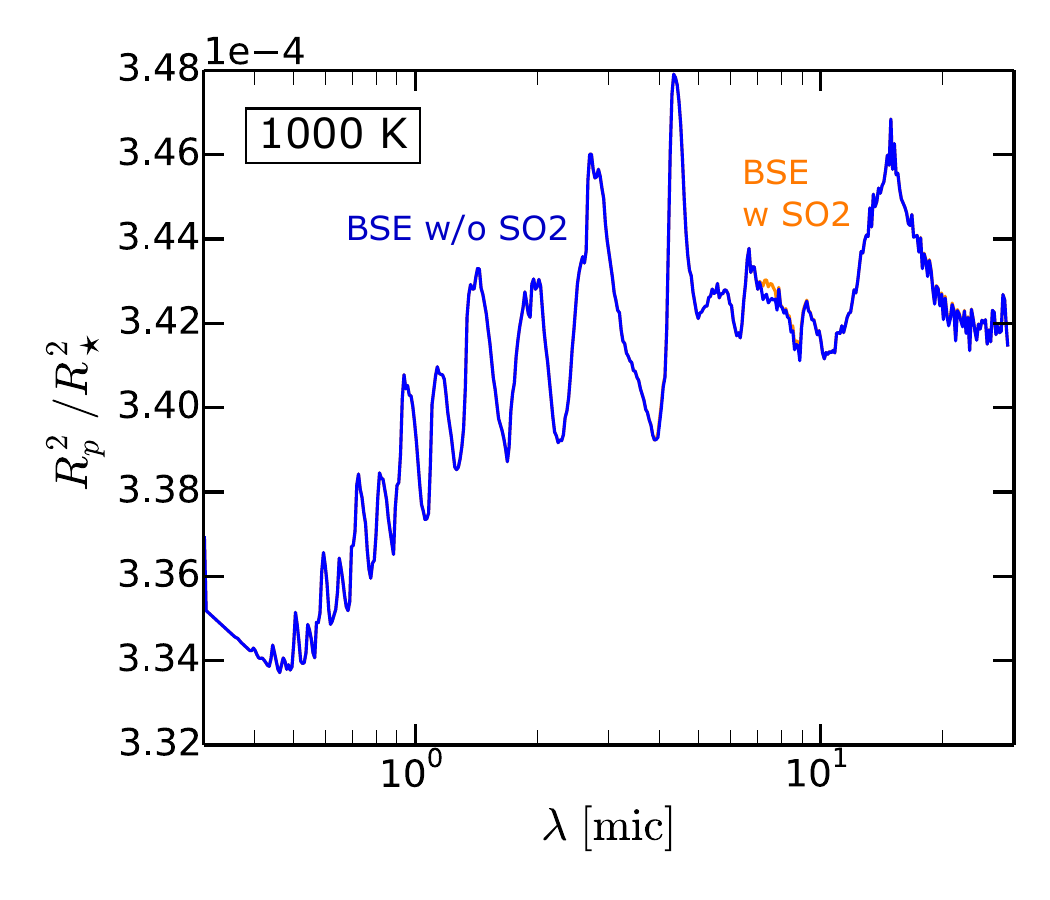} &
\hspace*{-4mm}
\includegraphics[height=54mm,trim=52 20 15 0,clip]{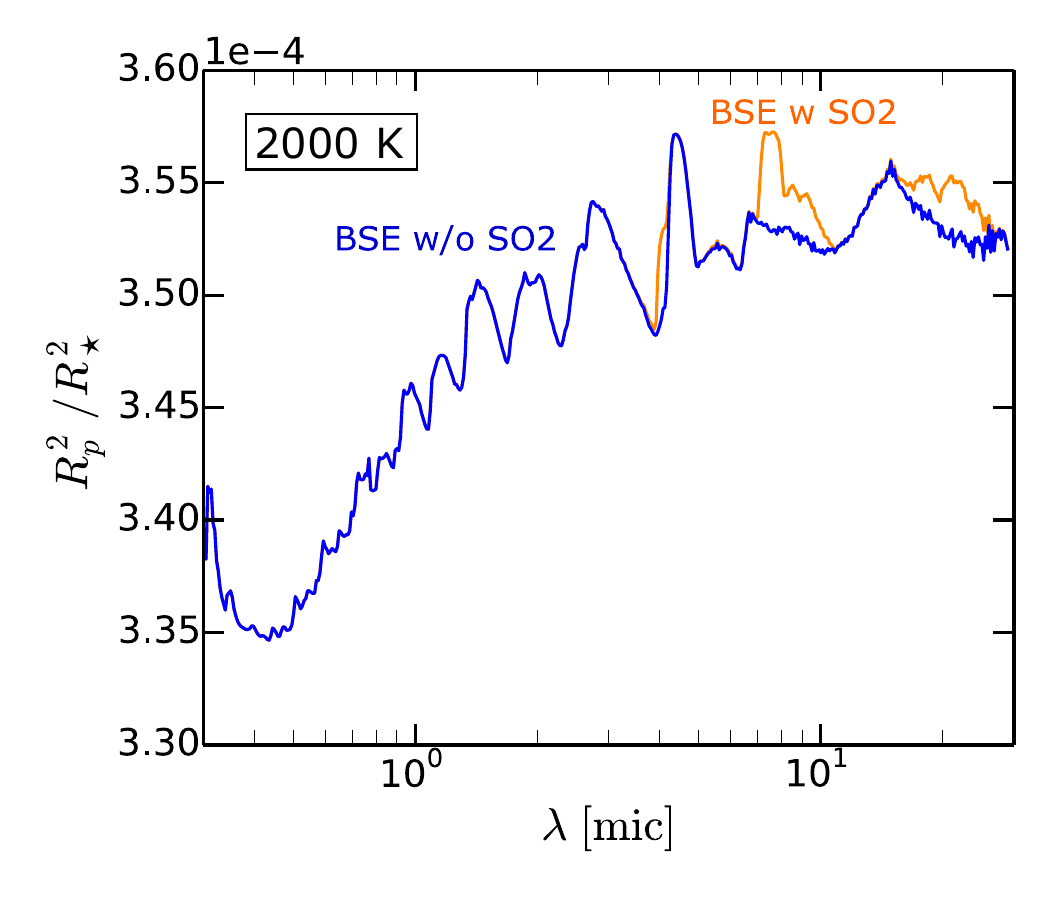}&
\hspace*{-4mm}
\includegraphics[height=54mm,trim=52 20 15 0,clip]{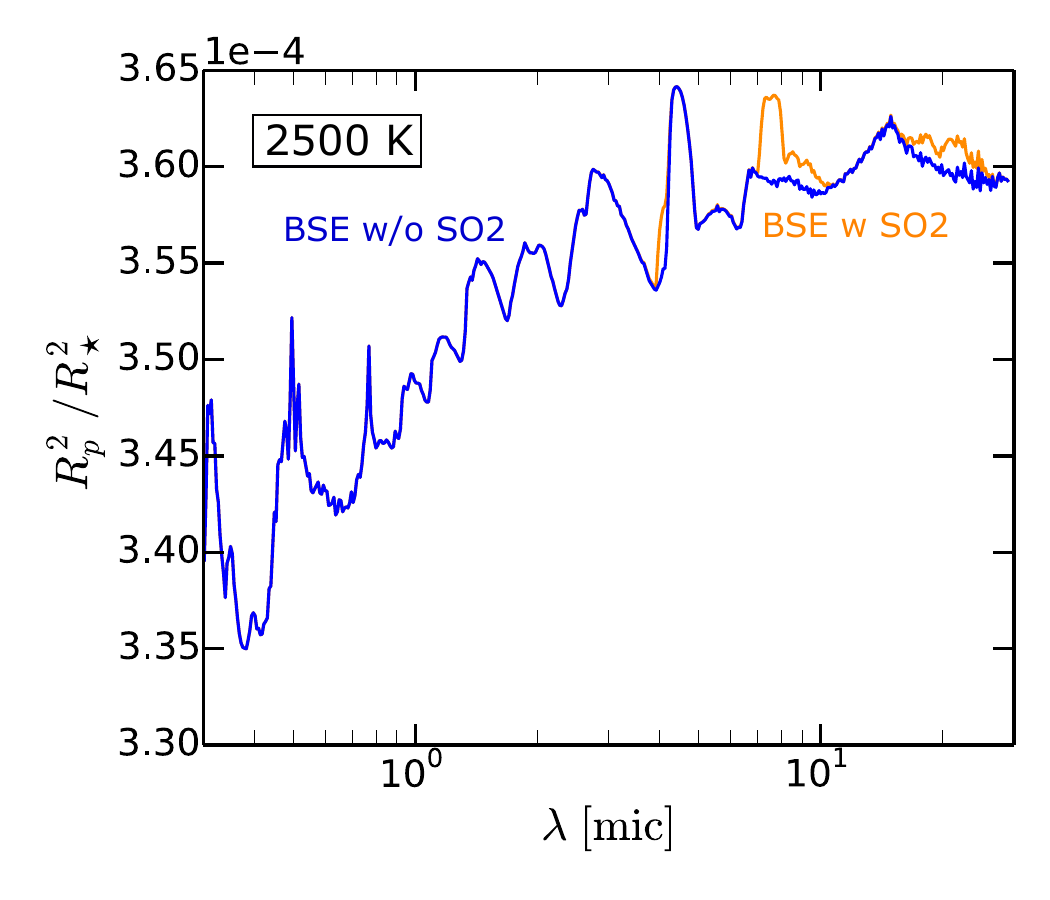}
\end{tabular}\\[-3mm]
\caption{Effect of temperature on the spectral appearance of \ch{SO2}
  in transmission spectra of isothermal atmospheres calculated for BSE
  total element abundances and a surface pressure of 10\,bar.}
\label{temp_effect}
\end{figure*}
\begin{figure*}
\centering
\vspace*{-5mm}  
\begin{tabular}{ccc}
\hspace*{-4mm}
\includegraphics[height=54mm,trim=20 20 15 0,clip]{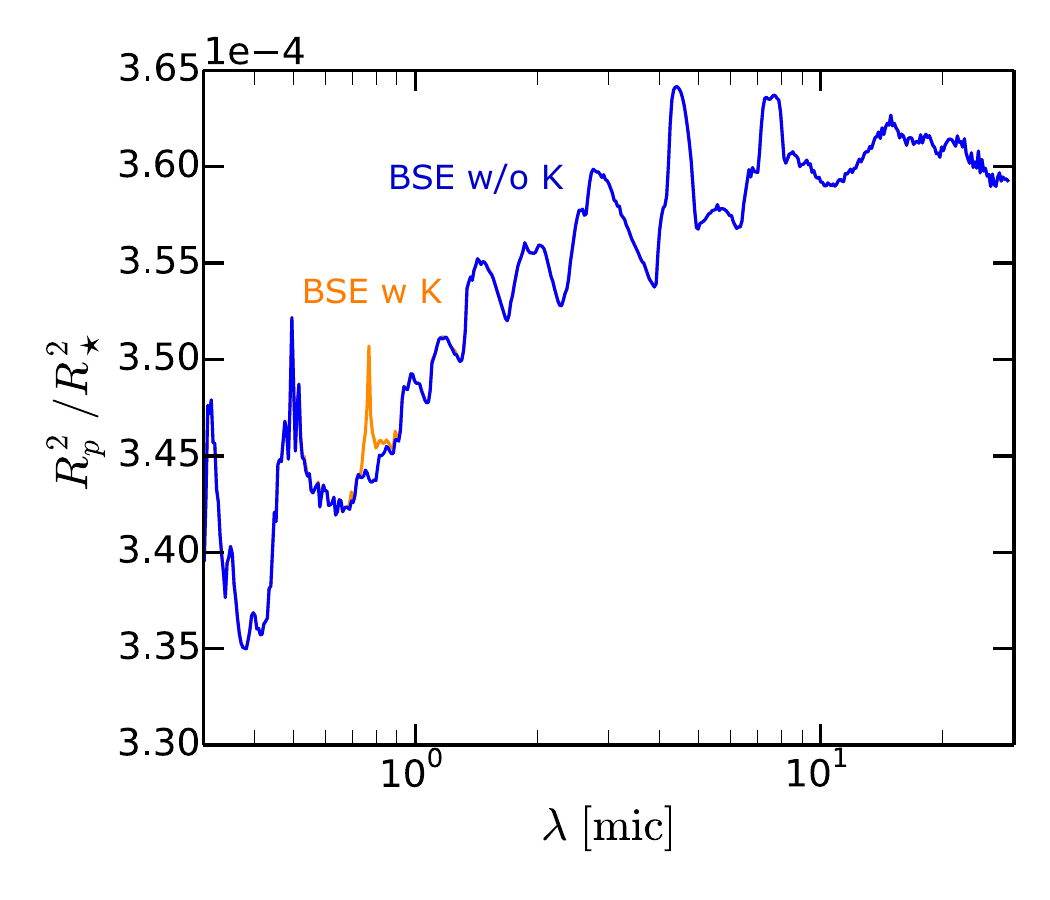} &
\hspace*{-4mm}
\includegraphics[height=54mm,trim=52 20 15 0,clip]{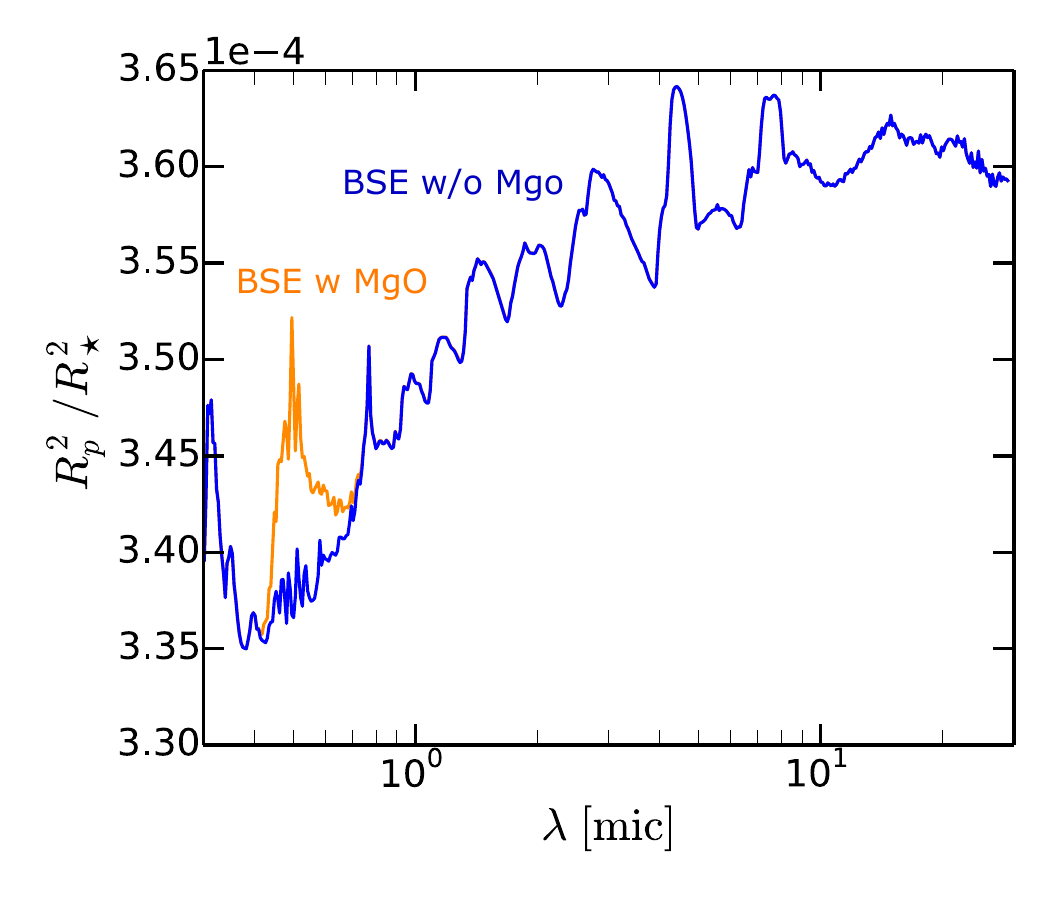}&
\hspace*{-4mm}
\includegraphics[height=54mm,trim=52 20 15 0,clip]{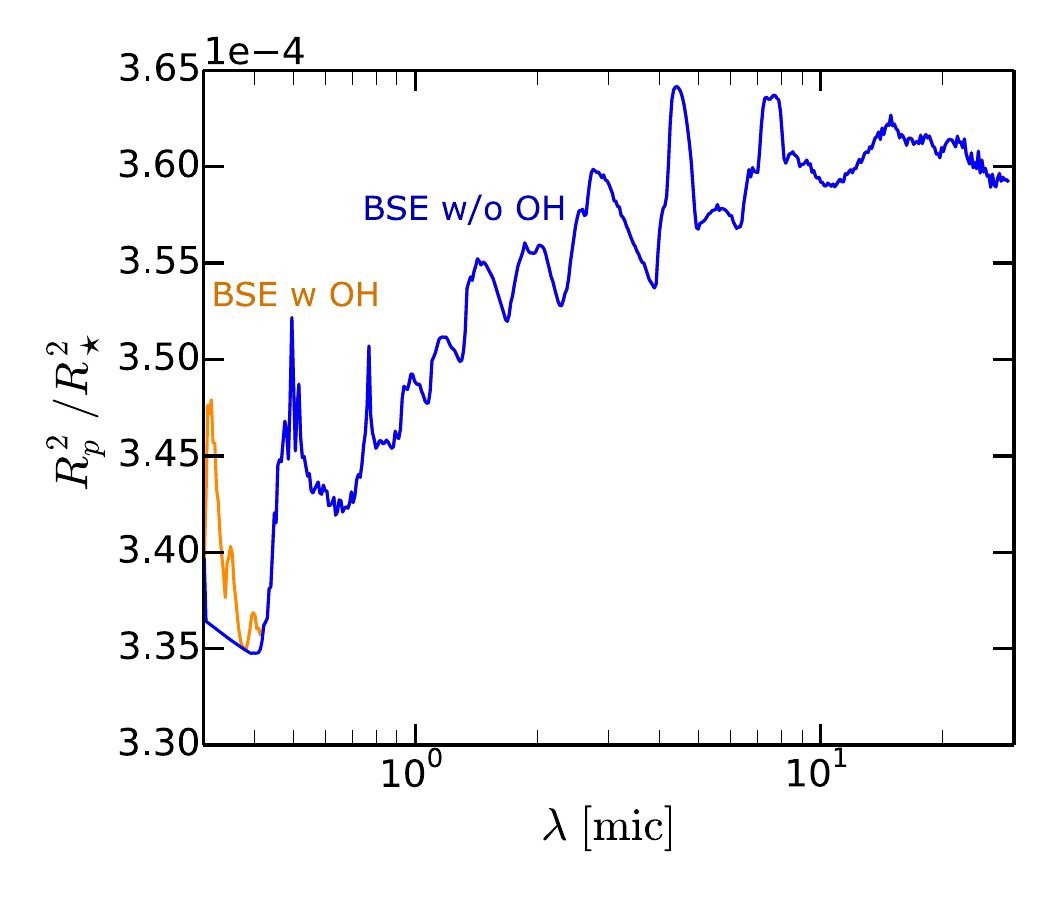}
\end{tabular}\\[-3mm]
\caption{Other spectral features in the BSE model with $T_{\rm
    surf}\!=\!2500\,$K and $p_{\rm surf}\!=\!10\,$bar.}
\label{Fig_2500K}
\end{figure*}

\subsubsection{The effect of temperature} \label{Temperature}

{ The temperature affects (i) the equilibrium chemistry in the gas
phase, (ii) the phase equilibrium at the planetary surface, and (iii)
the shape of the spectral features as more and more high-excitation
lines show up with increasing temperature.} As shown in \ref{A1}, the mixing ratios in a cold
atmosphere without condensation are constant up to about 600\,K
\citep{Woitke2020}. {At higher temperatures, the diversity of
gaseous species increases such that other sulphur bearing species
occur, for example \ch{COS} or \ch{SO}, in
addition to \ce{SO2} or \ce{H2S}, see Table~\ref{Tab_main_mols}.}
{ But most importantly, the amount of sulphur in the gas is
  controlled by condensation at the surface, which changes the sulphur
  content radically.}

{ Figure~\ref{temp_effect} shows spectra for the BSE model at 1000\,K,
2000\,K and 2500\,K.  As in Fig. ~\ref{Fig_extrasolar}, the influence
of \ce{SO2} is highlighted.  At a temperature of
1000\,K, the spectral features of \ce{SO2} become almost invisible,
which appear clearly in the high-$T$ spectra. This agrees with the
predictions in Fig.~\ref{sulphur_gasform}, which shows a drastic drop
in the gaseous sulphur concentration below $\sim\!1300$\,K for the BSE
composition with a surface pressure of 100\,bar.  Less \ch{SO2} can form
below this temperature as the condensate \ch{FeS} builds up.}

{ The overall transit depth, shape and distinctiveness of the absorption
features are affected as well.}  As the temperature is increased from
1000\,K to 2000\,K, the \ch{SO2} features appear in the IR wavelength
regime, as discussed. At shorter wavelengths, an optical feature of OH
at $\sim\!0.3-0.4\,\mu$m is present at 2000\,K but not at 1000\,K.  A
further increase in temperature from 2000\,K to 2500\,K mainly
affects these shorter wavelengths, see Fig.~\ref{temp_effect}. The
\ch{OH} peak becomes more distinct and two new peaks appear as shown
in Fig.~\ref{Fig_2500K}. These additional features are due to \ce{K}
(potassium) at $\sim\!0.8\,\mu$m and \ch{MgO} (Magnesium Oxide)
between $\sim\!0.45-0.8\,\mu$m.

The \ch{MgO} feature at 2500\,K is particularly strong. Magnesium
(\ch{Mg}) is a highly refractory element which predominantly
occurs in condensates.
%Figure~\ref{one_model} illustrates, how the relative
%element abundances in the atmosphere change as a function of
%temperature for the chosen constant pressure of 1\, bar.
%Below 2000\,K magnesium is mostly present in form of condensates.
However, as the temperature is increased above $\sim\!2000\,$K in the
10\,bar model, \ch{Mg} evaporates to form \ch{MgO} and \ch{MgH}.

\subsubsection{Dependence on surface pressure}

\begin{figure}
\centering
\vspace*{-2mm}  
\hspace*{-2mm}
\includegraphics[width=87mm,height=70mm,trim=15 20 15 5,clip]{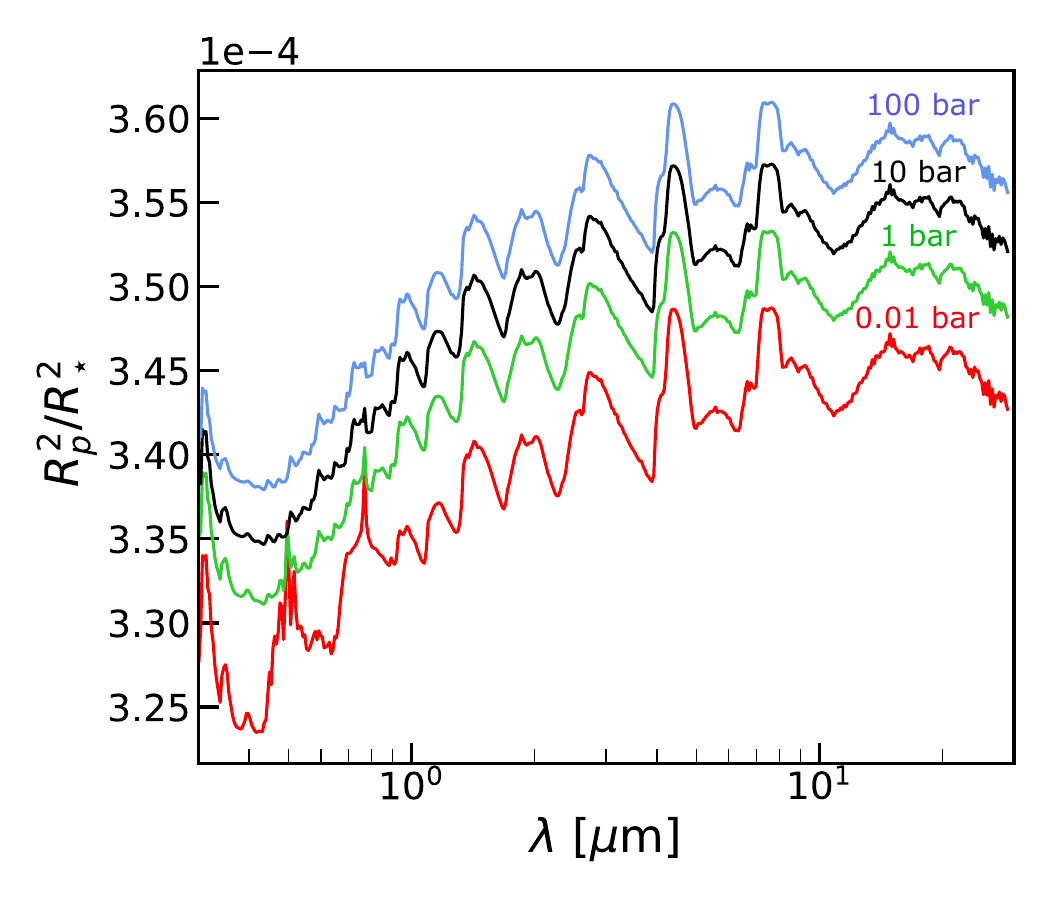}\\[-2mm]
\caption{Effect of surface pressure on transmission spectra of rocky
  planets for the BSE model with $T\!=\!2000\,$K. The surface
  pressures are annotated.}
\label{pressure_impact} 
\end{figure}

Figure~\ref{pressure_impact} shows spectra of the BSE model at 2000\,K
for four different surface pressures between 0.01\, and 100\,bar. The
main effect is a change in the total radial extent of the atmosphere,
which increases the overall transit depth.  The shape and magnitude of
the features is not much affected by the surface pressure between
$1-100$\,bar. For much lower pressures, however, other molecular
features appear, in particular at shorter wavelengths. The additional
peaks in the 0.01\,bar spectrum are due to higher concentrations of
\ch{MgO} and \ch{K} in the gas.  This effect is similar to a
temperature increase from 2000\,K to 2500\,K at 10\,bar surface
pressure as discussed in section \ref{Temperature}.
%This agrees with the $(T_{\rm surf},p_{\rm surf})$ degeneracy discussed by
%\citet{Herbort2020}.

\section{Observability with JWST}

In order to investigate the detectability of sulphurous molecules, we
use the modelling tool PandExo \citep{Batalha2017}. This allows us to
predict the size of the error bars of James Webb telescope (JWST)
measurements for atmospheres of rocky planets. We looked at the NIR
regime which will be covered by the MIRI LRS and NIRCam spectrographs
of JWST as well as the MIR regime covered by MIRI MRS.

\subsection{Observability of \ch{SO2} and \ch{H2S}}

Figures~\ref{Fig_extrasolar}, \ref{Fig_ssystem} and
\ref{FigPWDspectra} predict that the atmospheric compositions
originating from PWD-BSE crust models show the strongest \ch{SO2}
features of all our investigated compositions. Therefore we choose
this model to check the observability of \ch{SO2} with JWST.

\begin{figure}
\hspace*{-1mm}
\includegraphics[width=90mm,height=62mm,trim=8 0 22 0,clip]{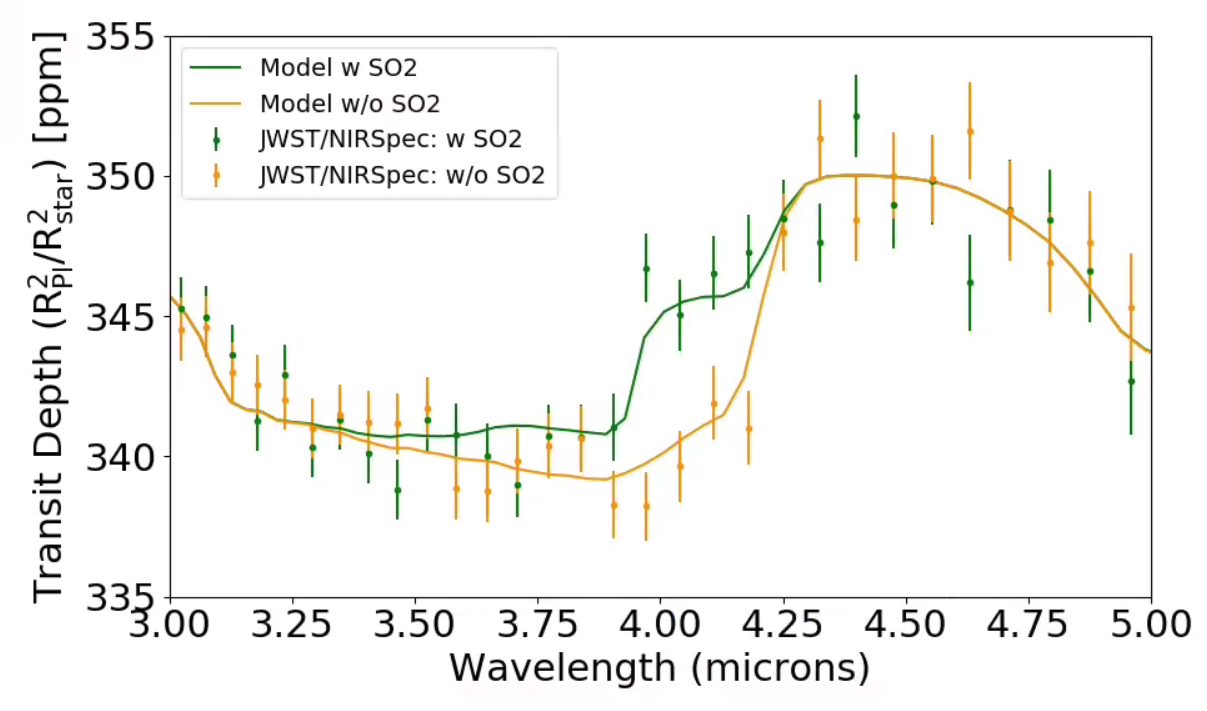}\\
\hspace*{-1mm}
\includegraphics[width=90mm,height=62mm,trim=0 0 5 0,clip]{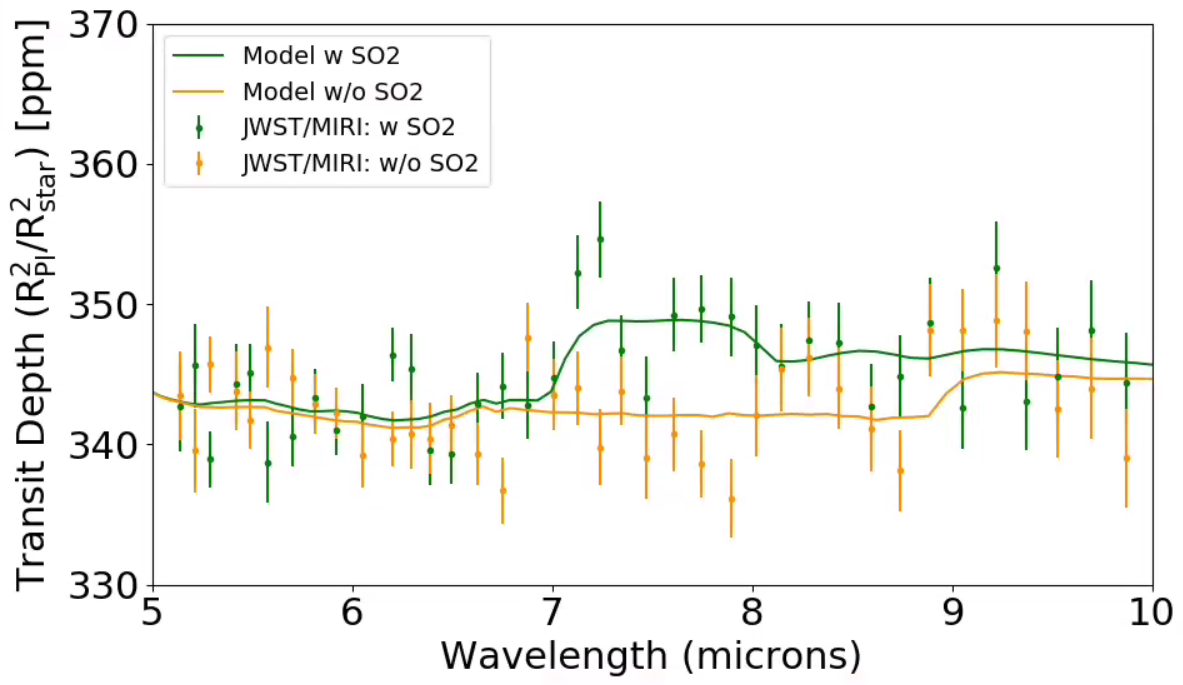}\\[-6mm]
\caption{Simulated transmission spectra for
  the PWD-BSE model with \ce{SO2} (green line) and without \ce{SO2}
  (orange line). Predicted JWST observations are based on combining
  30\, and 40\, transits, respectively, using PandExo
  \citep{Batalha2017} for NIRSpec/G395M in the top and for MIRI/LRS in
  the bottom panel. For clarity, the data was binned down to
  $R\!\sim\!30$ in their respective wavelength range for both
  simulated observations.}
\label{JWST_SO2}
\end{figure}

Figure~\ref{JWST_SO2} shows predicted \ch{SO2} observational
signatures of this composition assuming a hot equilibrium chemistry
atmosphere of 55\,Cnc\,e. The $\sim5$ ppm strong \ch{SO2} signal is in
principle observable with JWST using NIRSpec/G395M (left) and MIRI/LRS
(right) combining 30 and 40 transits, respectively. However, in
these calculations 'perfect' performance without systematic noise was
assumed.

In general, \ch{H2S} is found to show much weaker absorption lines
than \ch{SO2} such that it appears quite unlikely that present 
and future missions such as JWST, Ariel, or the Habitable World Observatory
(HWO) \citep{2020SPIE11443E..2AM,2020SPIE11443E..3NY,2019AGUFM.P51H3452D}
will be able to detect \ch{H2S} in an atmosphere like the one we modeled. 
The strongest \ch{H2S} feature we found is located at long wavelengths in
the MIR regime between $20-30\,\mu$m.

\subsection{Observability of PS}

\begin{figure}
\centering
\includegraphics[width=90mm,height=58.5mm,trim=5 0 16 0,clip]{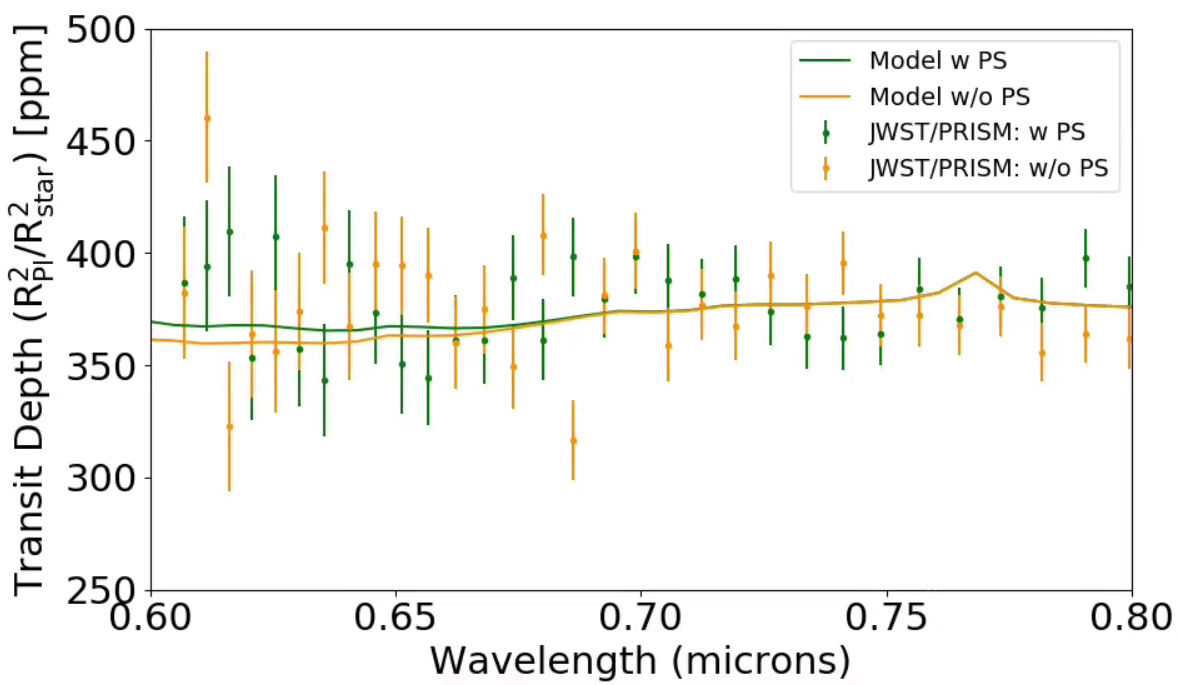}
\caption{Simulated transmission spectrum for the PWD-CI composition
  with \ce{PS} (green line) and without \ce{PS} (orange
  line). Predicted JWST observations are based on combining 40
  transits using PandExo \citep{Batalha2017}. The predictions are for
  NIRspec/Prism for a system similar to 55\,Cnc\,e with a star of
  maximal allowed brightness in the J-band of 10.5 mag. A detection of
  PS does not seem possible. Note the different scaling of the y-axis
  compared to Fig.~\ref{JWST_SO2}.}
\label{PS_pandexo}
\end{figure}

Our ARCiS spectral models show that the molecule PS may cause a broad
absorption feature between 0.3 and 0.65\,$\mu$m that is surprisingly
strong (up to $\sim 40$ ppm in the PWD-CI chondrite model), despite
being based on gaseous PS concentrations of only a few ppm to a few
100\,ppm, see Fig.~\ref{PS_effect}. The feature's wavelength interval
is at the edge of the observable range with JWST. Only NIRSPEC Prism
covers the longer wavelength part of this PS feature between
$0.6-0.65\,\mu$m, where it reaches 8\, ppm at maximum. Towards
shorter wavelengths the feature strength would increase as predicted
in Sect.~\ref{PS_features}, but this is completely out of the
detectable range of JWST. In addition, the instrument mode NIRSPEC
Prism is only usable for faint stars ($m_J \geq 10.5$). For such
targets, the PS feature won't be detectable in any of the compositions
according to PandExo predictions as demonstrated in
Fig.~\ref{PS_pandexo}.  However, future space telescopes operating at
optical wavelengths ($0.3-0.6\,\mu$m), such as LUVOIR-B, could
certainly observe a 40\,ppm strong PS feature. Finding PS in a rocky
planet atmosphere could also be an interesting science case for
ground-based instruments.

\subsection{Dependence on atmospheric structure}
\label{sec:thermal}
We ran a number of additional verification tests where we used
more realistic temperature structures for 55\,Cancri\,e. These
studies included atmospheric profiles where the temperature is
monotonously decreasing with height, similar to Fig.~2 of
\cite{20JiMoJa}, and profiles with a temperature inversion, similar
to Fig.~1 in \cite{Zilinskas2021}. In these models, we used {\sc
GGchem} to adjust the molecular concentrations to chemical
equilibrium at each atmospheric height to the local
temperature and pressure, without condensation. While the
transition depth and spectral shape of the various absorption
features can change substantially in these models, we found that our
general conclusions about the detectability of \ce{SO2} and \ce{PS}
in warm and hot rocky exoplanet atmospheres, and the non-detectability of
\ce{H2S}, remain the same.

\section{Summary and discussion}
\label{sec:Summary}
We have investigated how much sulphur can be expected in the
atmospheres of warm and hot rocky exoplanets, which sulphur molecules
are most abundant, and whether these molecules might be
detectable with JWST.  A large chemical diversity in planetary
environments is expected.  We considered various sets of total element
abundances, i.e.\ before condensation, as found in common types of
rocks \citep{Herbort2020, Herbort2022} and assumed chemical and phase
equilibrium at the planetary surface to calculate the chemical
composition of the atmospheric gas.

For surface temperatures between about 1000\,K and 3000\,K, we found
typical sulphur concentrations in the gas of a few percent, depending
on total element abundances assumed.  The most abundant sulphur
molecule in the atmosphere, reflecting its redox state, is normally
either \ce{SO2} or \ce{H2S}.  However, in a few cases, both molecules
can coexist beside \ce{S2}.  Other, less abundant sulphur molecules of
interest are \ce{SO}, \ce{SO3}, \ce{COS}, \ce{HS}, \ce{S2O},
\ce{Na2SO4} and \ce{PS}, all of which can occur with concentrations
$>\!100\,$ppm.

The abundance of sulphur in these atmospheres is controlled by
condensation at the surface-atmosphere interface in our models.  As
sulphur is among the more volatile elements, its gas abundance above
3000\,K first increases with falling temperature, as more of the
abundant refractory elements like Ca, Al, Ti, Si, Fe, Mn and Cr
condense at the surface. Below about 1000\,K to 2000\,K, dependent on
pressure, sulphur starts to form condensates as well, in particular
\ce{FeS[s]}, \ce{FeS2[s]}, \ce{CaSO4[s]}, \ce{MnS[s]} and
\ce{Na2S[s]}, causing the S abundance in the gas phase to drop
quickly.  However, at intermediate temperatures, the ability of
sulphur to condense, and hence the amount of sulphur left in the gas
phase is controlled by the availability of metals such as Fe, Mn, Ca
and Na, which depends on the assumed total element abundances in
complicated ways.

In order to determine whether or not the various sulphur
molecules might be detectable by transmission spectroscopy, we
utilise simple 1D isothermal hydrostatic model with constant
molecular concentrations and without clouds. We used the {\sc ARCiS}
modelling platform \citep{2020A&A...642A..28M} to predict such
transmission spectra with star-planet system parameters adopted from
55\,Cancri\,e. Based on these models, it seems most promising to
search for \ce{SO2} in the atmospheres of hot ($\approx\!2000\,$K)
rocky exoplanets.  \ce{SO2} produces a left shoulder on one of the
main \ce{CO2} absorption features at 4.5\,$\mu$m, a distinct
absorption feature at $7-8\,\mu$m, and a broad absorption feature at
$15-25\,\mu$m. These features are most prominent for rocky element
abundances derived from polluted white dwarf observations. Quick
simulations with PandExo \citep{Batalha2017} showed that \ce{SO2}
might be detectable with JWST in sources like 55\,Cnc\,e, when about
30-40 transits can be observed.  In comparison, the molecule \ce{H2S}
is more difficult to detect. { Interestingly, the atmospheres containing
lots of sulphur also have large mean molecular weights, and are hence
less extended, which limits our ability to detect these molecules in the sulphur-richest atmospheres.}

Among all other sulphur species included in our ARCiS models, only the
molecule PS is found to be possibly detectable via a strong, broad
spectral absorption feature around $0.3-0.65\,\mu$m. This could be a
case for optical ground-based instruments or LUVOIR, because of JWST's
limited possibilities to observe the UV and optical. The NIRspec
spectrograph, which we used for our detectability verifications, is
only sensitive down to 0.6 micron and designed for faint sources only.

Beside the sulphur molecules \ce{SO2} and \ce{PS}, we also find
detectable absorption features by \ce{TiO} and \ce{MgH} in the solar
abundance model, and by \ce{MgO}, \ce{OH} and the K resonance line in
our hot rocky exoplanet transmission spectra {between 0.3 and
0.8\,$\mu$m}, which are otherwise mostly dominated by \ce{H2O} and
\ce{CO2} absorption.  \ce{MgO}, and \ce{K} are visible only for
surface temperatures $\gtrsim\!2000\,$K and the OH features becomes
more distinct as the temperature rises from 2000\,K to 2500\,K. We
note that ground-based instruments, which can observe high-resolution
spectra of exoplanet atmospheres, might allow for the detection of some
of these species.

Since the sulphur content in rocky exoplanet atmospheres depends
critically on the properties of the planet surface (surface pressure, surface
temperature, surface bulk composition), an observational constraint on
the \ce{SO2} mixing ratio would be an important step towards a better
characterisation of the chemical conditions on the planet's surface
and habitability.

\bigskip
\noindent{\sl Acknowledgements:\ }
This project has received funding from the European Research Council (ERC) under the 	European Union’s Horizon 2020 research and innovation programme (grant agreement no. 101088557, N-GINE). P.\,W.\, Ch.\,H., M.M. and L.C. acknowledge funding from the
   European Union H2020-MSCA-ITN-2019 under Grant Agreement
   no.\,860470 (CHAMELEON). O.H. acknowledges PhD funding from the St
   Andrews Center for Exoplanet Science and financial support from the
   \"Osterreichische Akademie der Wissenschaften.

\bibliography{library}
\appendix

\section{Chemical classification of the CHNOS system\label{A1}}

\begin{table*}[!b]
\caption{Classification of atmospheric types including sulphur$^{(1)}$.}
\label{Tab_subtypes}
\vspace*{-5mm}
\begin{center}
\resizebox{130mm}{!}{
\begin{tabular}{c|cccccccccccc} 
\hline
&&&&&&&&&&\\*[-2.2ex] 
   type&\ce{H2}& \ce{NH3}&\ce{N2}&\ce{CH4}&\ce{CO2}&\ce{H2O}& \ce{H2S}&\ce{S8}&\ce{SO2}&\ce{SO3}&\ce{H2SO4}&\ce{O2}\\ 
\hline
&&&&&&&&&&\\*[-2ex]
   A1&\tick &\tick &\cross &\tick &\cross &\tick &\tick&\cross &\cross
   &\cross &\cross &\cross\\ 
   A2&\cross &\tick &\tick &\tick &\cross & \tick &\tick&\cross
   &\cross &\cross &\cross &\cross\\
\hline   
   B1&\cross&\cross &\tick &\cross&\tick &\cross&\cross&\cross &\cross
   &\tick &\tick &\tick \\ 
   B2&\cross &\cross &\tick &\cross &\tick & \tick &\cross&\cross
   &\cross &\cross &\tick &\tick\\ 
   BC1&\cross &\cross&\tick &\cross &\tick &\cross &\cross&\cross
   &\tick &\tick &\tick &\cross\\ 
   BC2&\cross&\cross &\tick &\cross &\tick & \tick
   &\cross&\cross&\tick &\cross &\tick &\cross \\ 
   BC3&\cross &\cross &\tick &\cross &\tick &\tick &\cross &\tick
   &\tick &\cross &\cross &\cross \\ 
   BC4&\cross &\cross &\tick &\cross &\tick &\tick &\tick&\cross
   &\cross &\cross &\cross &\cross\\
\hline   
   C&\cross &\cross &\tick &\tick &\tick & \tick &\tick &\cross
   &\cross &\cross &\cross &\cross\\ 
   CG&\cross &\cross &\tick &\tick &\tick &\cross &\tick &\tick
   &\cross &\cross&\cross &\cross\\ 
\hline   
\end{tabular}}
\end{center}
$^{(1)}$ {In the low-temperature limit, only five molecules coexist in
  chemical equilibrium with mixing ratios exceeding trace
  abundances. The three main types are A for hydrogen-rich, B for
  oxygen-rich, and C for atmospheres where \ce{H2O}, \ce{CH4},
  \ce{CO2} and \ce{N2} can coexist \citet{Woitke2020}. By including
  sulphur, four new sub-types occur between B and C, the BC sub-types,
  and a fifth sub-type between C type and the graphite condensation
  zone, named the CG sub-type.}
\end{table*}

\citet{Woitke2020} have classified the chemical composition of cold
($T<600\,$K) atmospheres which consist only of the four most abundant
elements (C, H, N, and O).  This resulted in three distinct
atmospheric types. Hydrogen-rich type A atmospheres are characterised
by the presence of \ce{CH4}, \ce{H2O}, \ce{NH3} and (\ce{H2} or
\ce{N2}).  Oxygen-rich type B atmospheres show the presence of
\ce{O2}, \ce{CO2}, \ce{H2O} and \ce{N2}, and type C atmospheres show
the coexistence of \ce{CH4}, \ce{CO2}, \ce{H2O}, and \ce{N2}.  All
other molecules have only trace concentrations in chemical equilibrium
at low temperatures.  In this appendix, we build upon this model and
extend this characterisation to additionally include sulphur.  In the
low temperature limit, the bulk of the atmosphere is consisting of
exactly the same number of molecules as the numbers of elements, so
for C, H, N, O, S we expect five abundant molecules in all cases.  The
element abundances are then sufficient to determine the atmospheric
type.  With increasing temperatures, other molecules gain importance
due to the entropy term in the Gibbs free energy.

\begin{figure*}
\begin{tabular}{ccc}
\hspace*{-7mm}
\includegraphics[width=65mm,height=50mm]{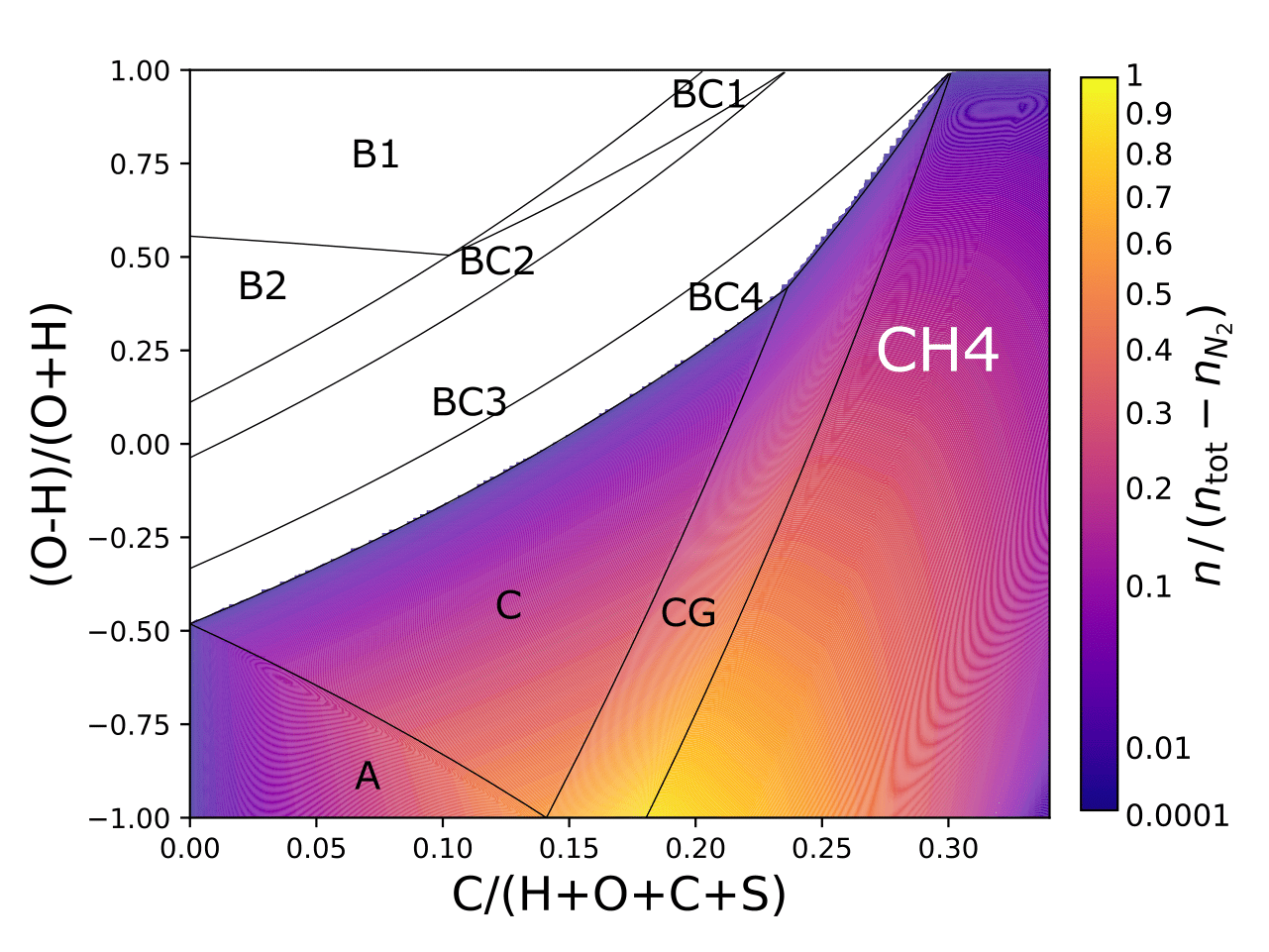} &
\hspace*{-6mm}
\includegraphics[width=65mm,height=50mm]{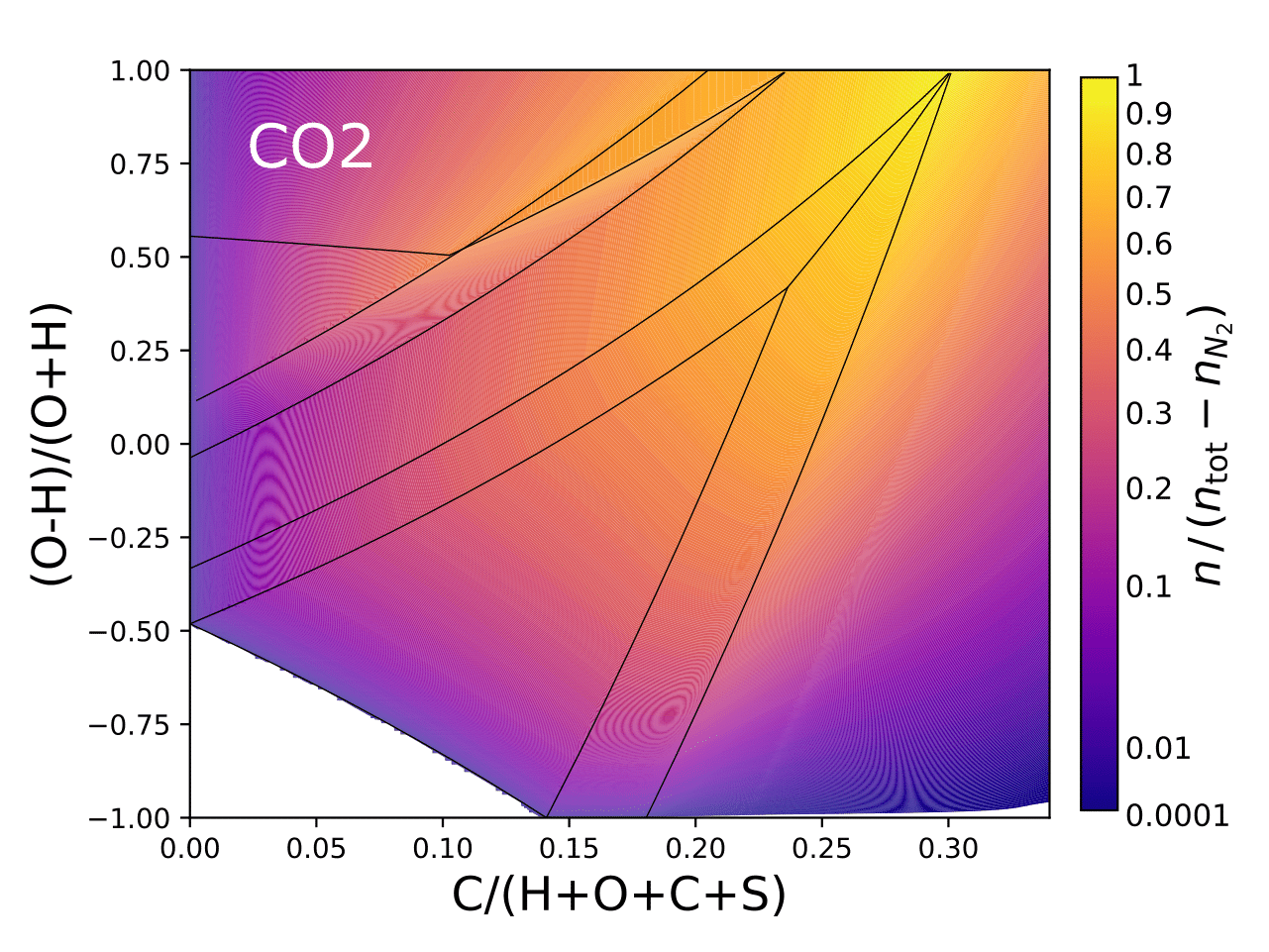} &
\hspace*{-6mm}
\includegraphics[width=65mm,height=50mm]{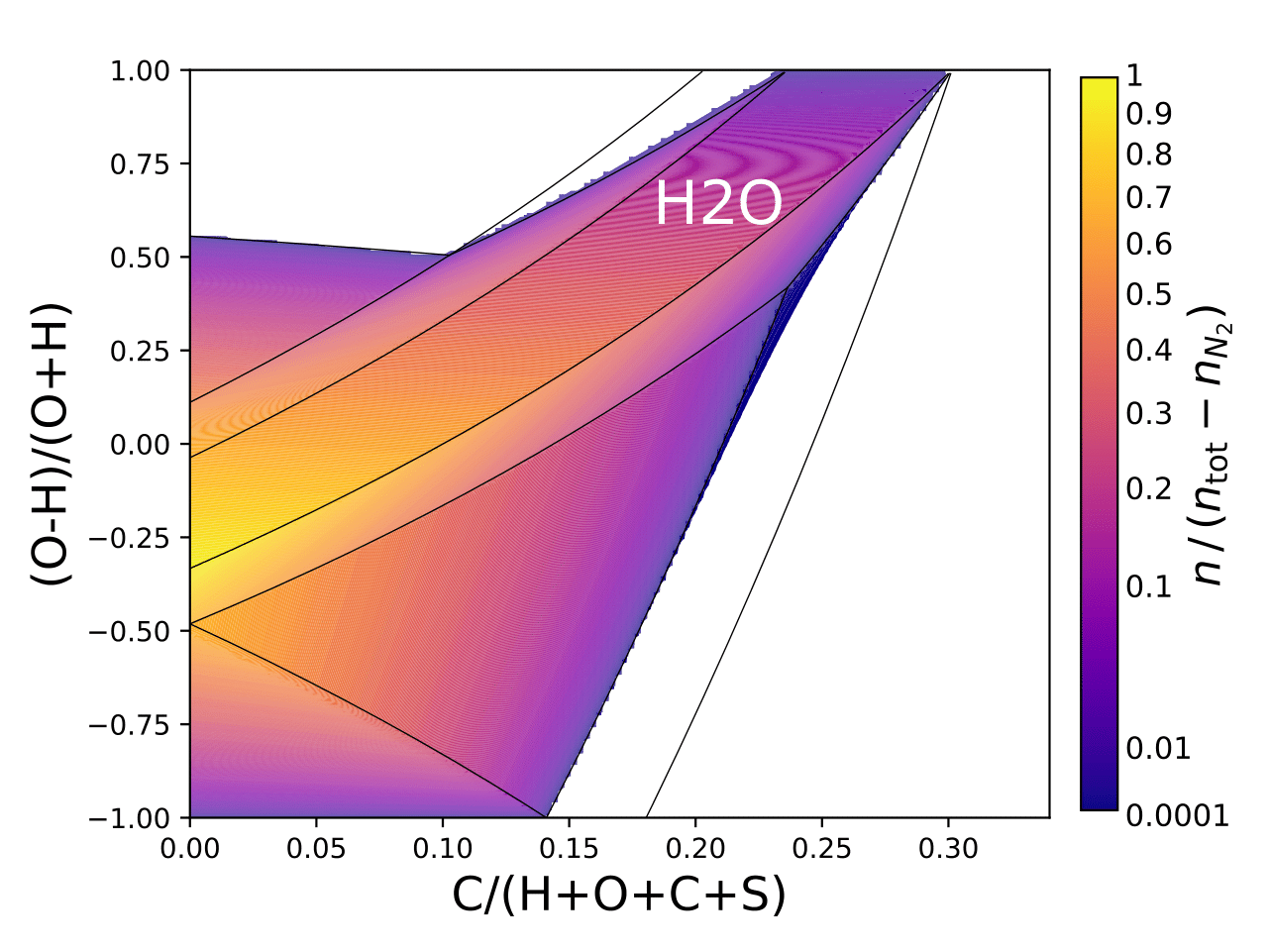} \\[-3mm]
\hspace*{-7mm}
\includegraphics[width=65mm,height=50mm]{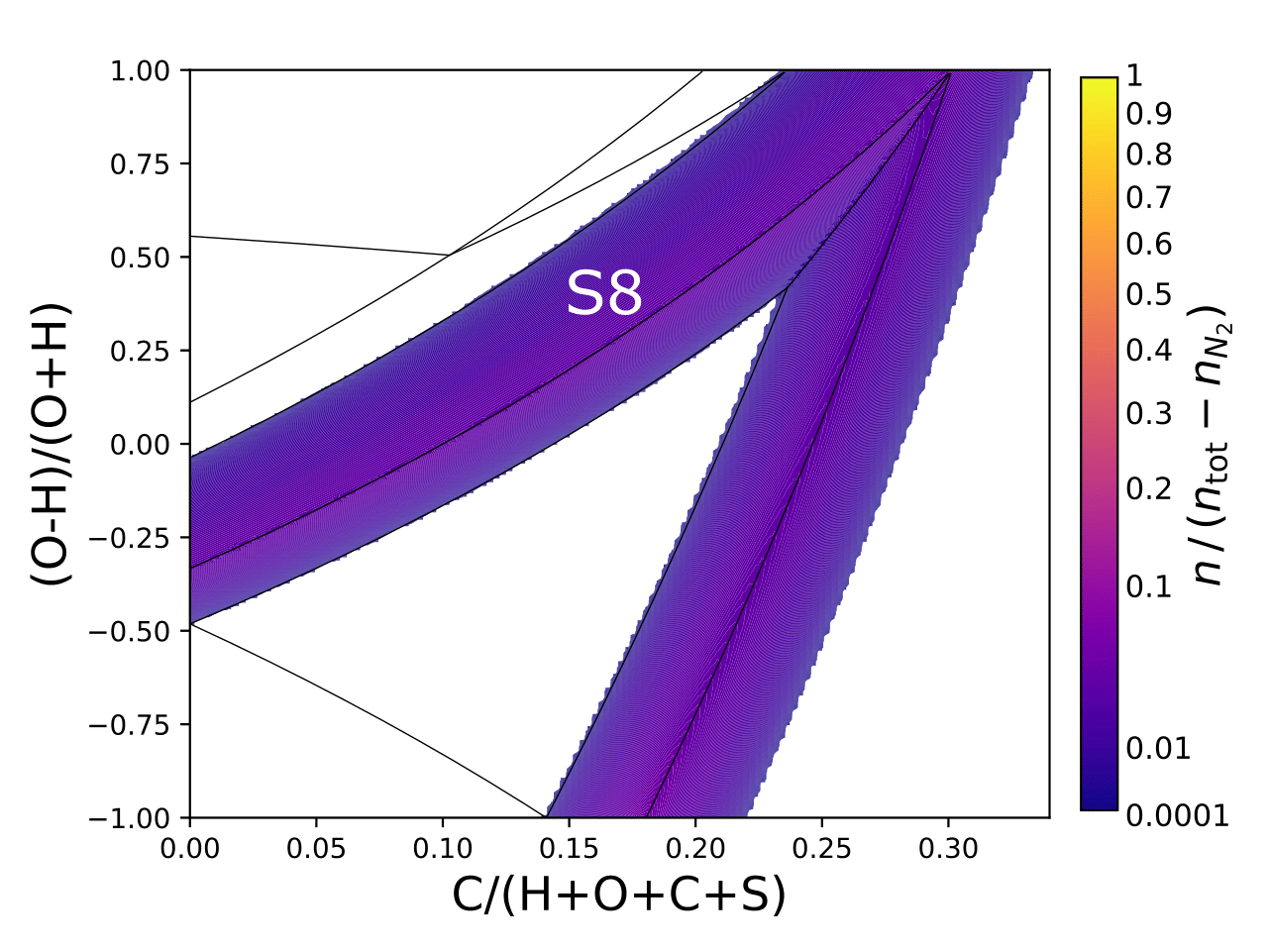} &
\hspace*{-6mm}
\includegraphics[width=65mm,height=50mm]{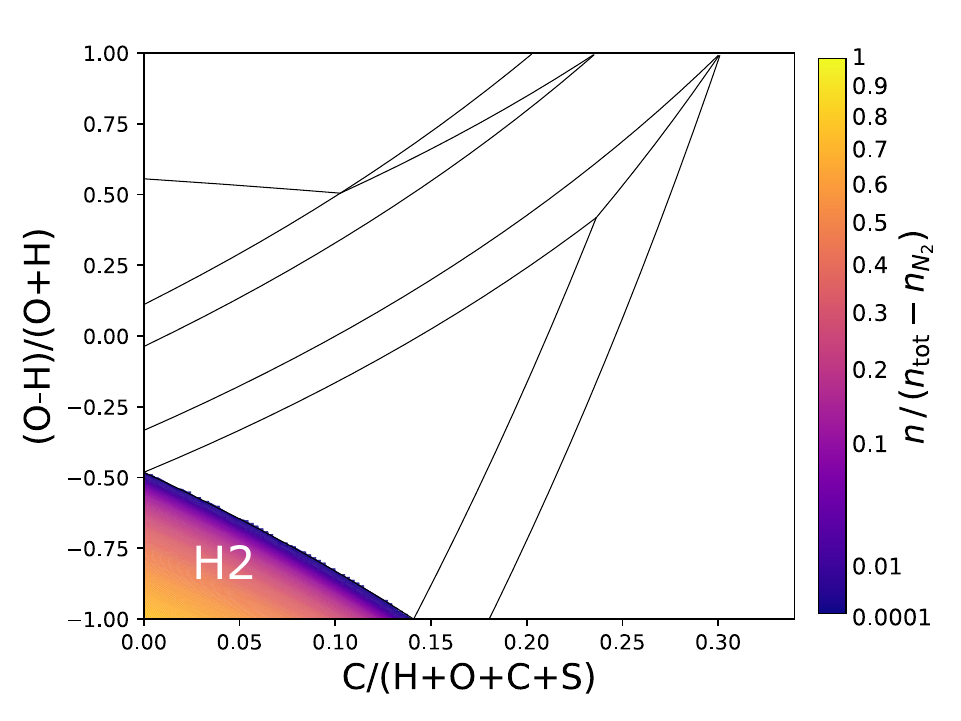} &
\hspace*{-6mm}
\includegraphics[width=65mm,height=50mm]{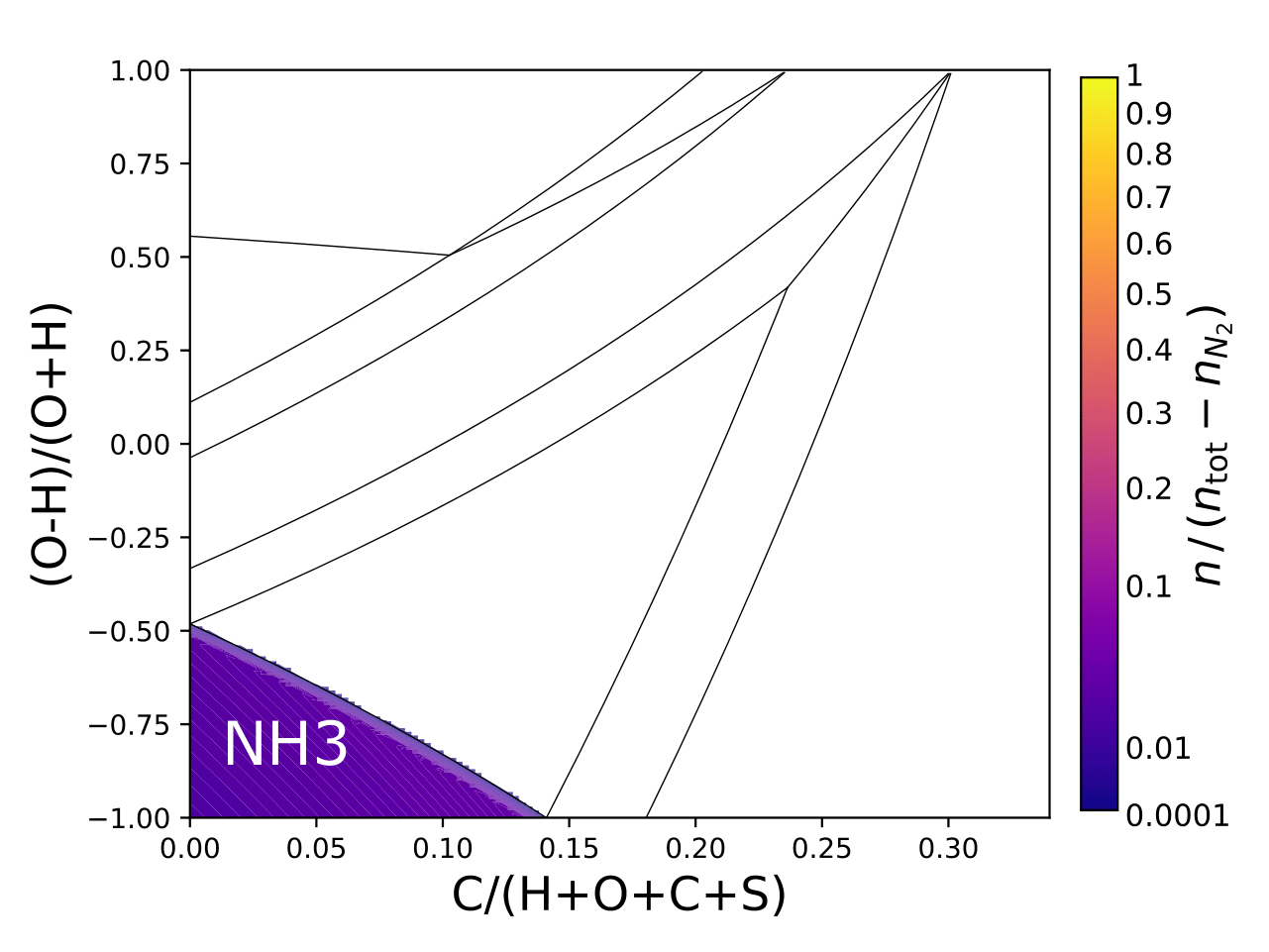} \\[-3mm]
\hspace*{-7mm}
\includegraphics[width=65mm,height=50mm]{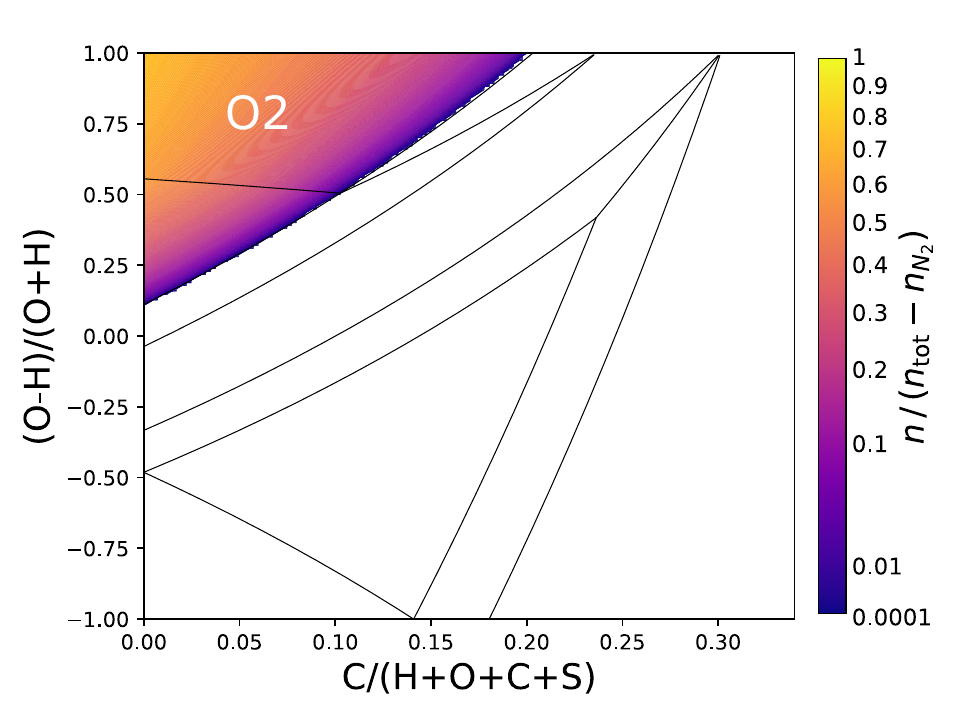} &
\hspace*{-6mm}
\includegraphics[width=65mm,height=50mm]{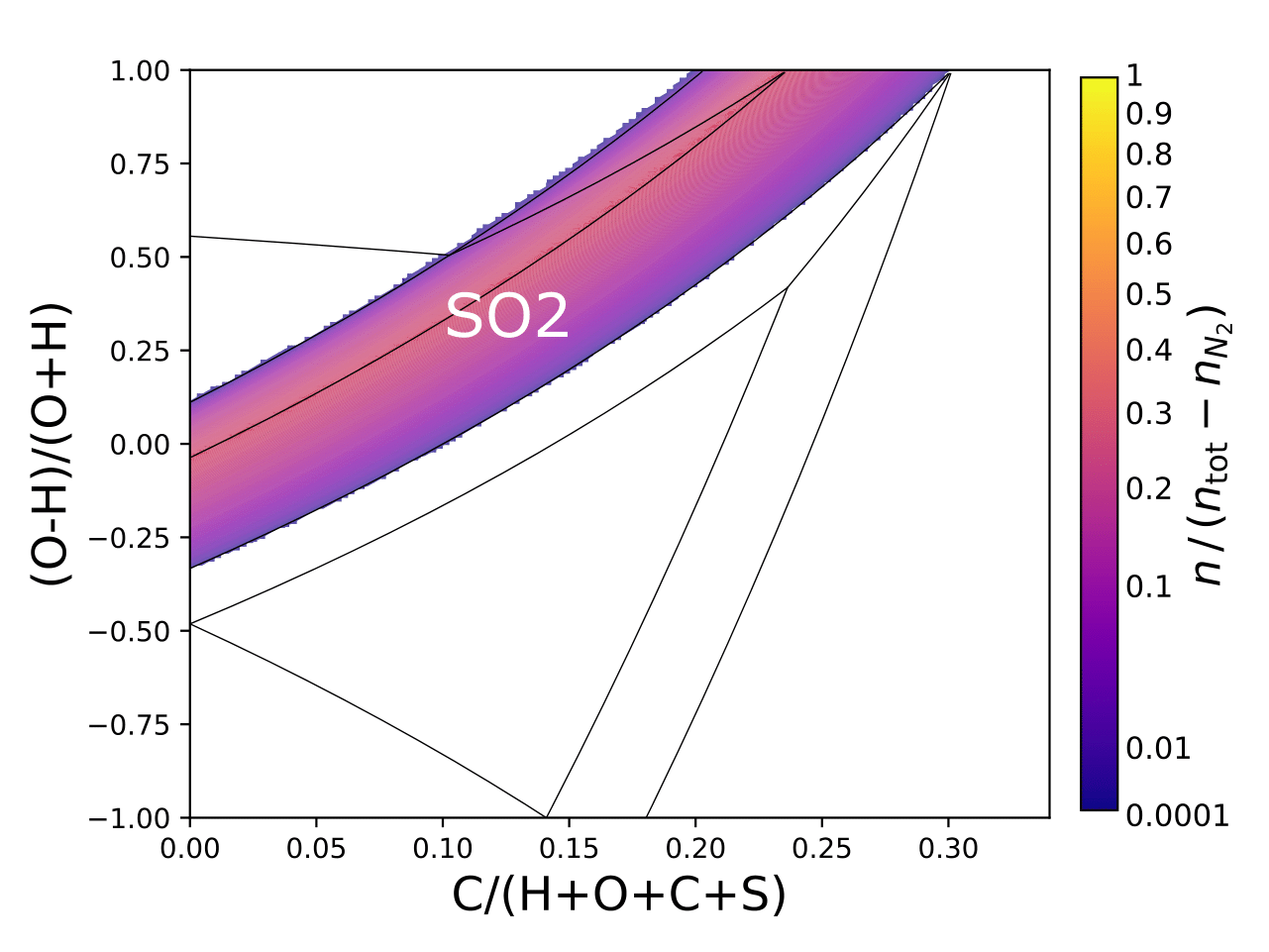} &
\hspace*{-6mm}
\includegraphics[width=65mm,height=50mm]{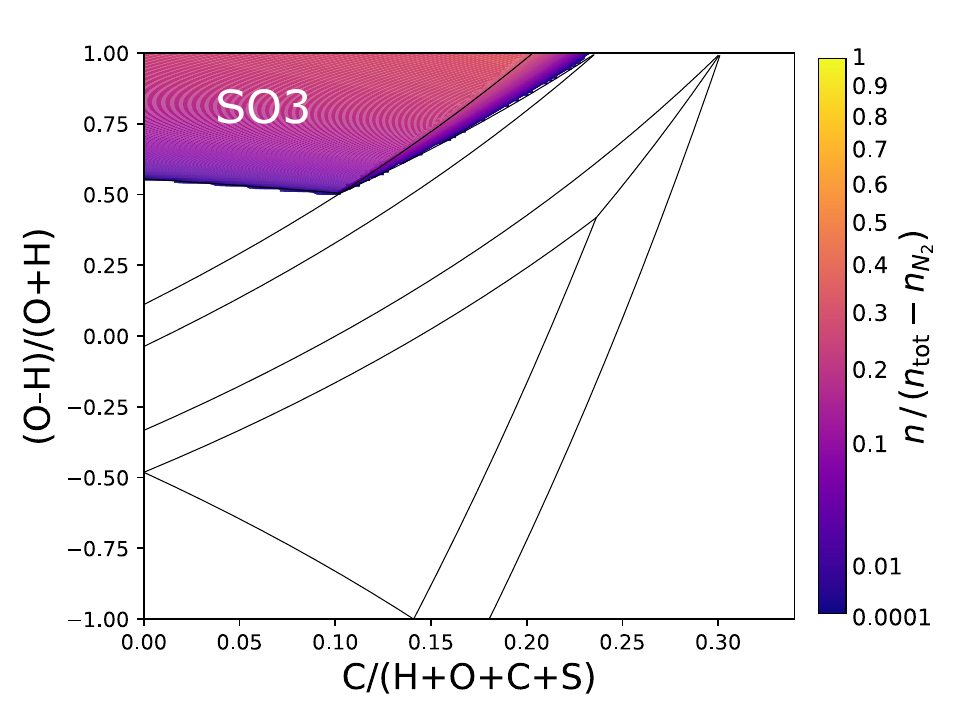} \\[-3mm]
\hspace*{-7mm}
\includegraphics[width=65mm,height=50mm]{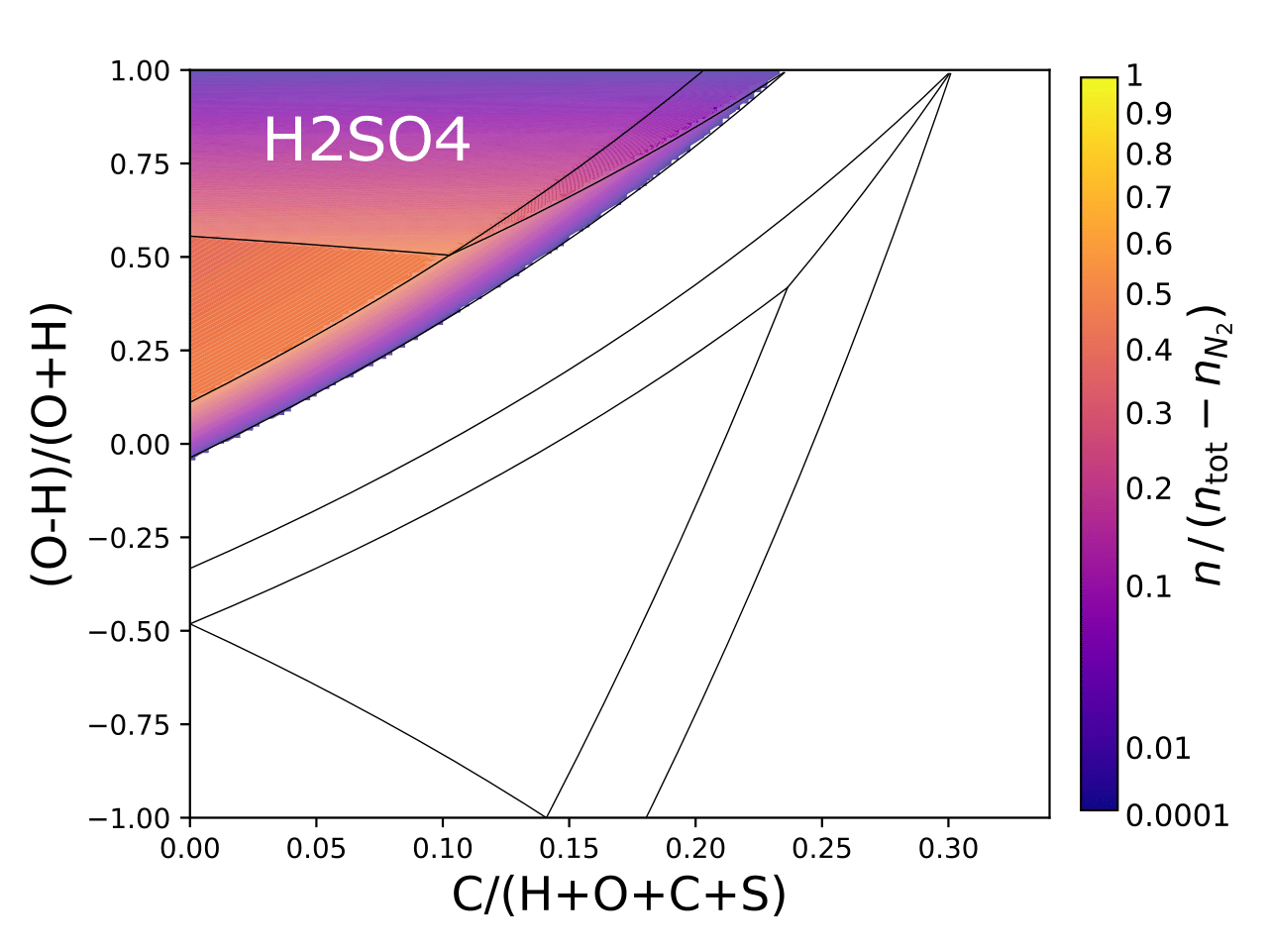} &
\hspace*{-6mm}
\includegraphics[width=65mm,height=50mm]{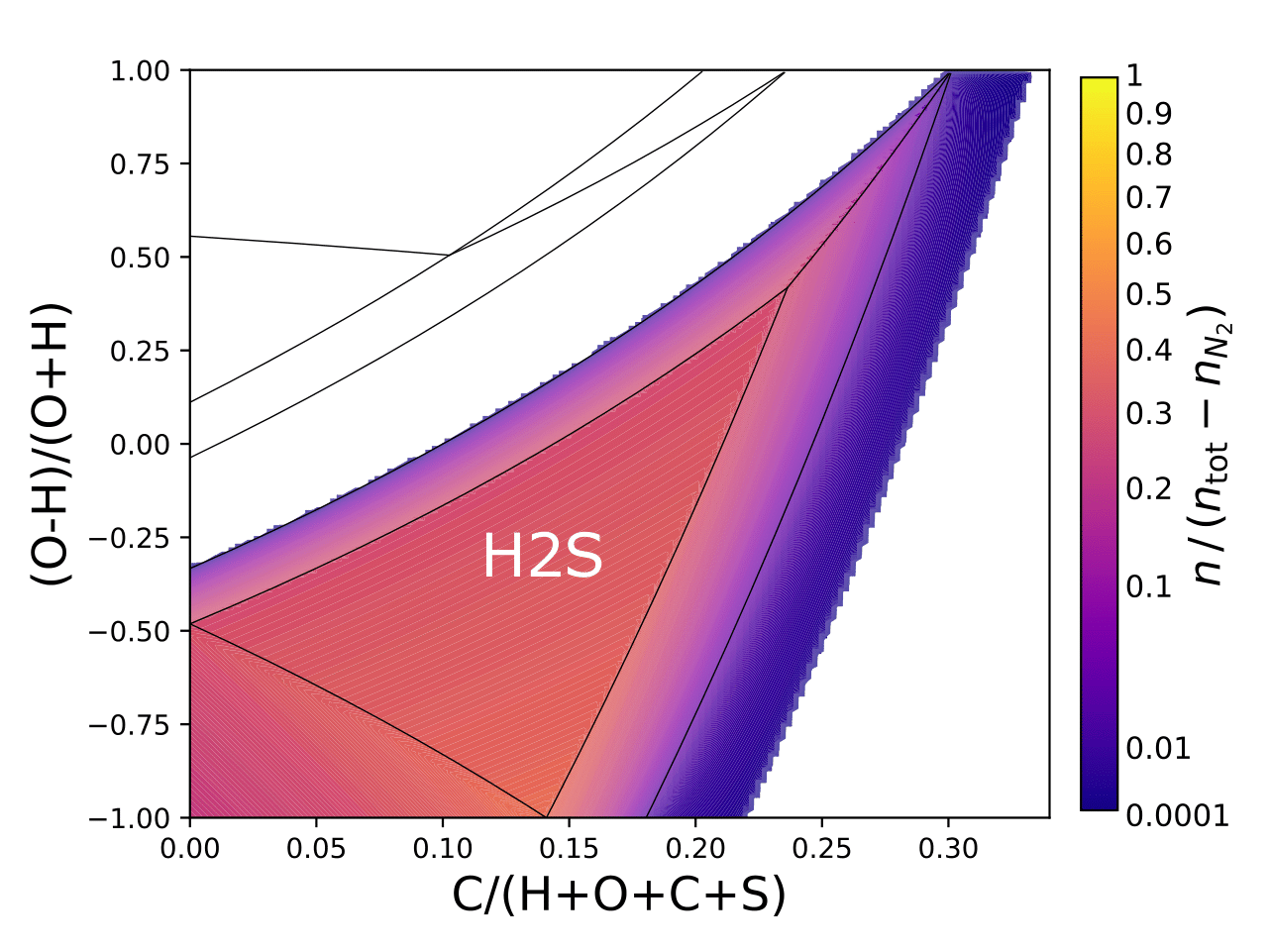} &
\hspace*{-6mm}
\includegraphics[width=65mm,height=50mm]{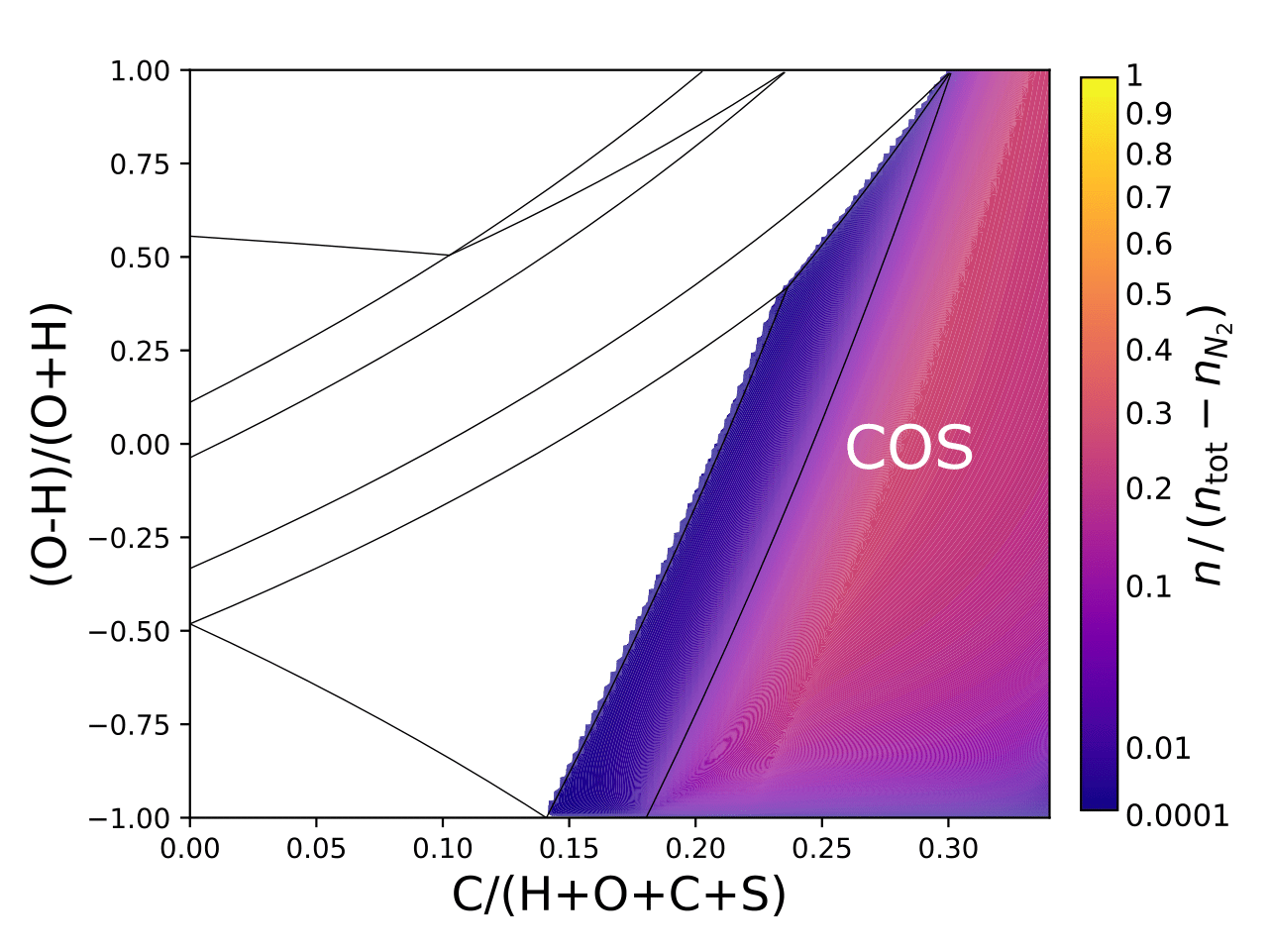} \\[-3mm]
\end{tabular}
\caption{Number densities of different molecules for various abundances in the CHNOS system. The N2 abundance is subtracted, as it does only interfere for the type A atmospheres. The solid lines are derived from theoretical calculations as discussed in Sect.~\ref{boundaries}. The coloured number densities are a result of equilibrium chemistry calculations with {\textsc{GGchem}} for 300\,K and 1\,bar. The element abundances of C, H, and O set according to the axis of the plots, while a fixed S abundance is investigated (S/(C+O+H+S)=0.1). Every subplot shows one distinct molecule.}
\label{atmos_types}
\end{figure*}

Similarly to \citet{Woitke2020}, we find different atmospheric types
for the gas phase compositions of the CHNOS system.  In
Fig.~\ref{atmos_types} the number densities of the most dominant gas
phase molecules for a sulphur abundance of $\rm {S/(C+H+O+S) = 0.1}$
are shown.  Hereby, the nitrogen abundance is not taken into account
as it does not interfere with the chemistry of the other elements but
mainly forms \ce{N2}.  The emerging atmospheric types are described in
the following as well as in Table \ref{Tab_subtypes}.

\smallskip
\noindent{\bf Type A:} The hydrogen dominated atmospheric, subtypes A1
(with \ce{NH3} and \ce{H2}) and A2 (with \ce{NH3} and \ce{N2}). These
types are the same as in \citet{Woitke2020}. Sulphur is present in
form of \ce{H2S}.

\smallskip
\noindent{\bf Type B:} The oxygen dominated atmospheric types are
characterised by the presence of \ce{O2}, \ce{CO2} and \ce{N2} in all
cases. Subtype B2 has \ce{H2SO4} and \ce{H2O} in addition, whereas the
new subtype B1 has no water but \ce{H2SO4} and \ce{SO3} in
addition. Subtype B1 occurs for high S/H ratios.

\smallskip
\noindent{\bf Type C:} The coexistence regime of \ce{CH4}, \ce{CO2},
\ce{H2O}, and \ce{N2} is completed by the presence of \ce{H2S} as the
major sulphur carrying species.

\smallskip
\noindent{\bf BC subtypes:} With increasing sulphur abundance, four
new subtypes emerge between type B and C atmospheres, called BC1 to
BC4.  All of these subtypes show the presence of \ce{CO2} and
\ce{N2}. None of them contains \ch{CH4}, \ce{NH3}, \ce{O2} or \ce{H2}.
Sulphur is present in form of certain combinations of \ce{H2S},
\ce{S8}, \ce{SO2}, \ce{SO3} and \ce{H2SO4}, see
Table~\ref{Tab_subtypes}. Thus, the subtypes BC1 to BC4 are forming a
sequence in redox potential.

\begin{figure*}
\begin{tabular}{ccc}
\hspace*{-7mm}
\includegraphics[width=65mm,height=50mm]{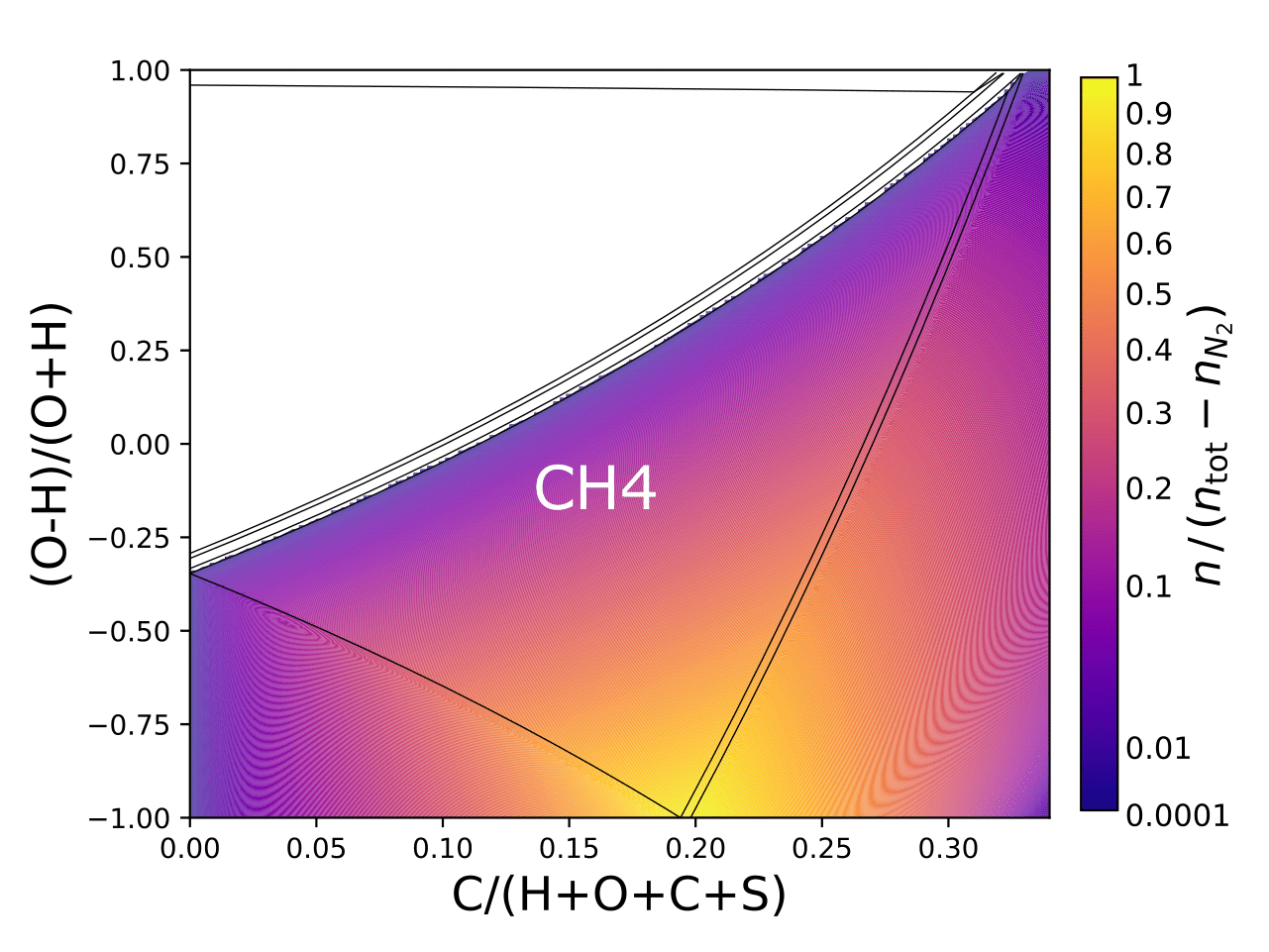} &
\hspace*{-6mm}
\includegraphics[width=65mm,height=50mm]{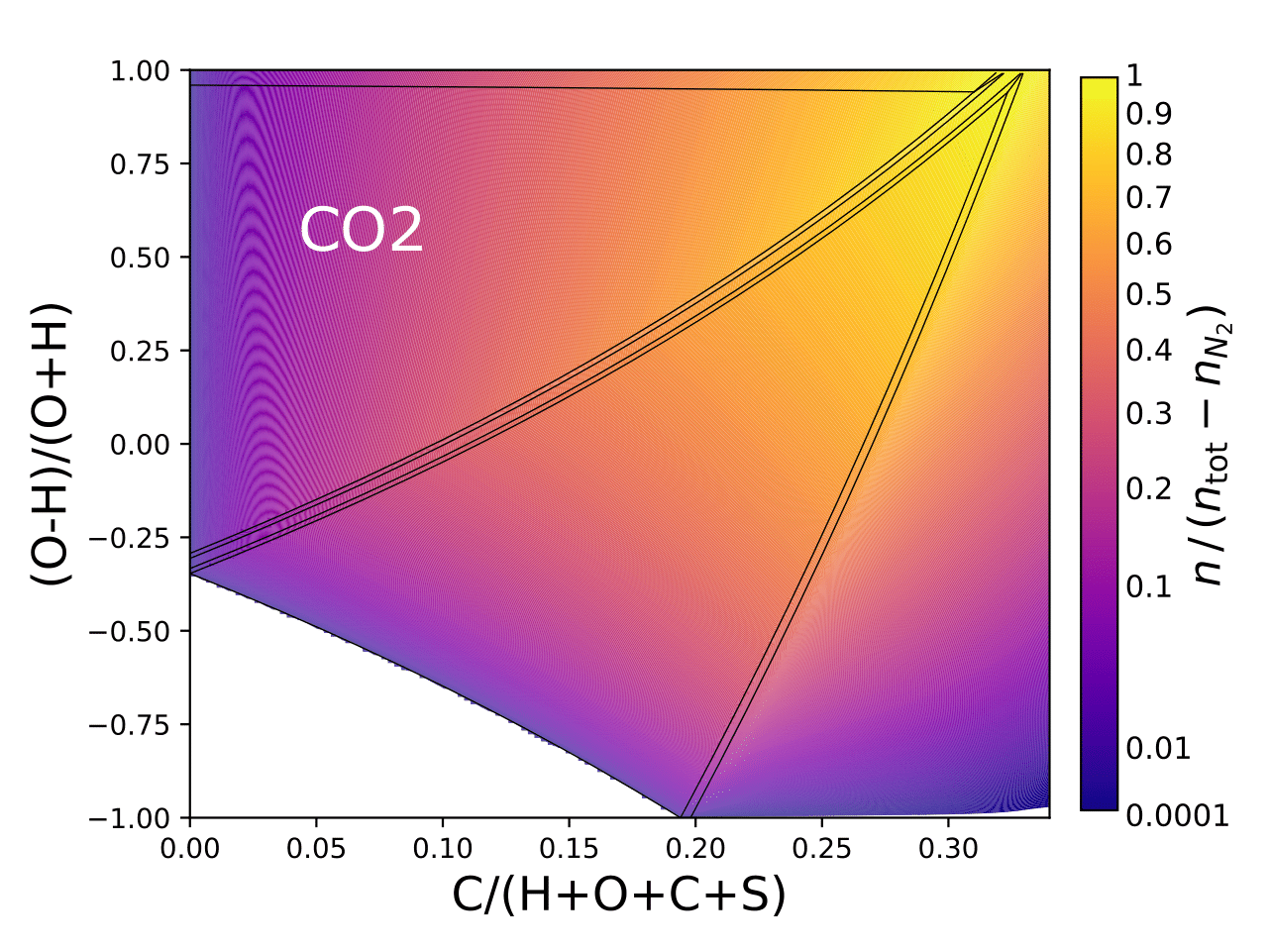} &
\hspace*{-6mm}
\includegraphics[width=65mm,height=50mm]{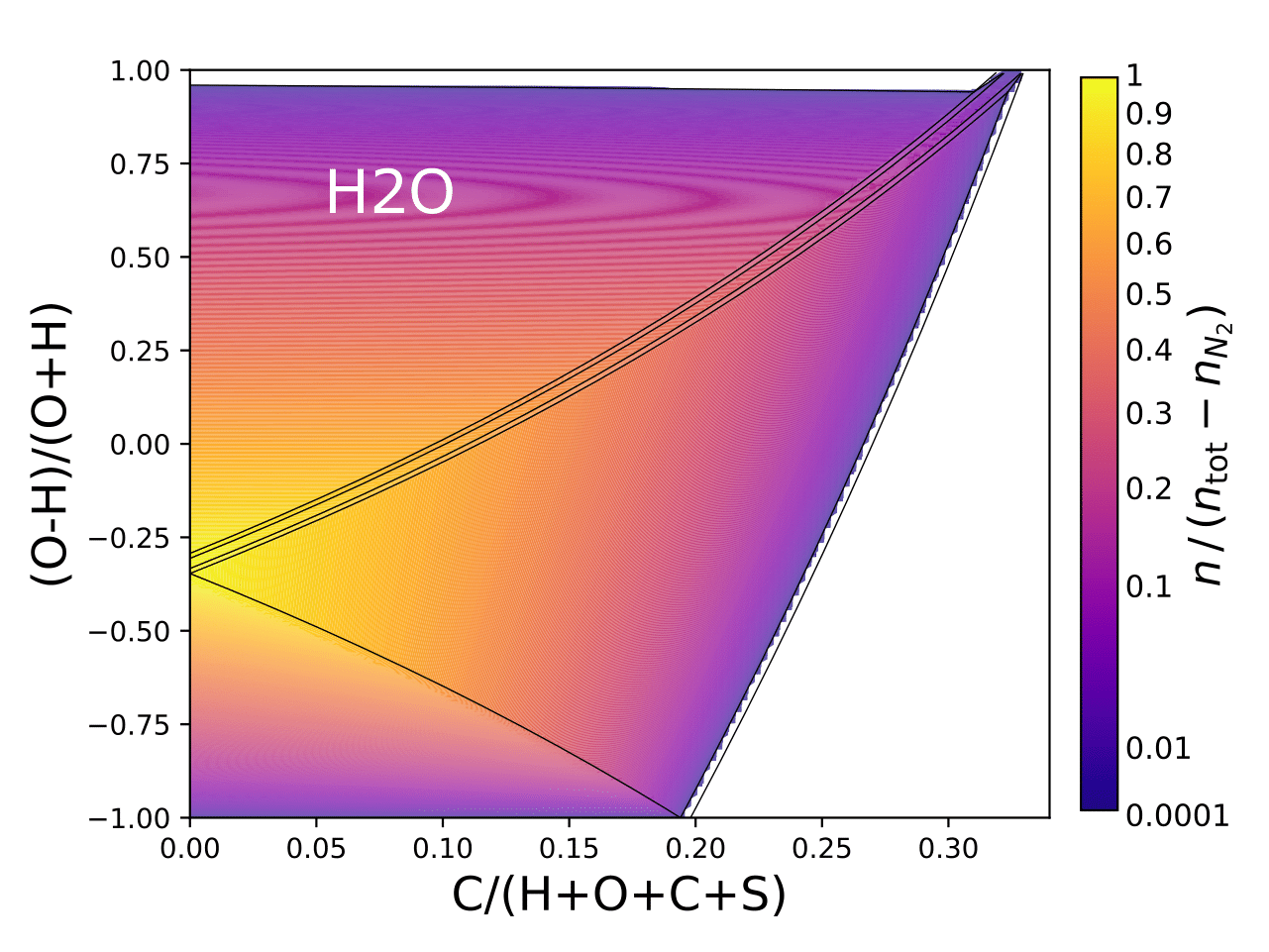} \\[-3mm]
\hspace*{-7mm}
\includegraphics[width=65mm,height=50mm]{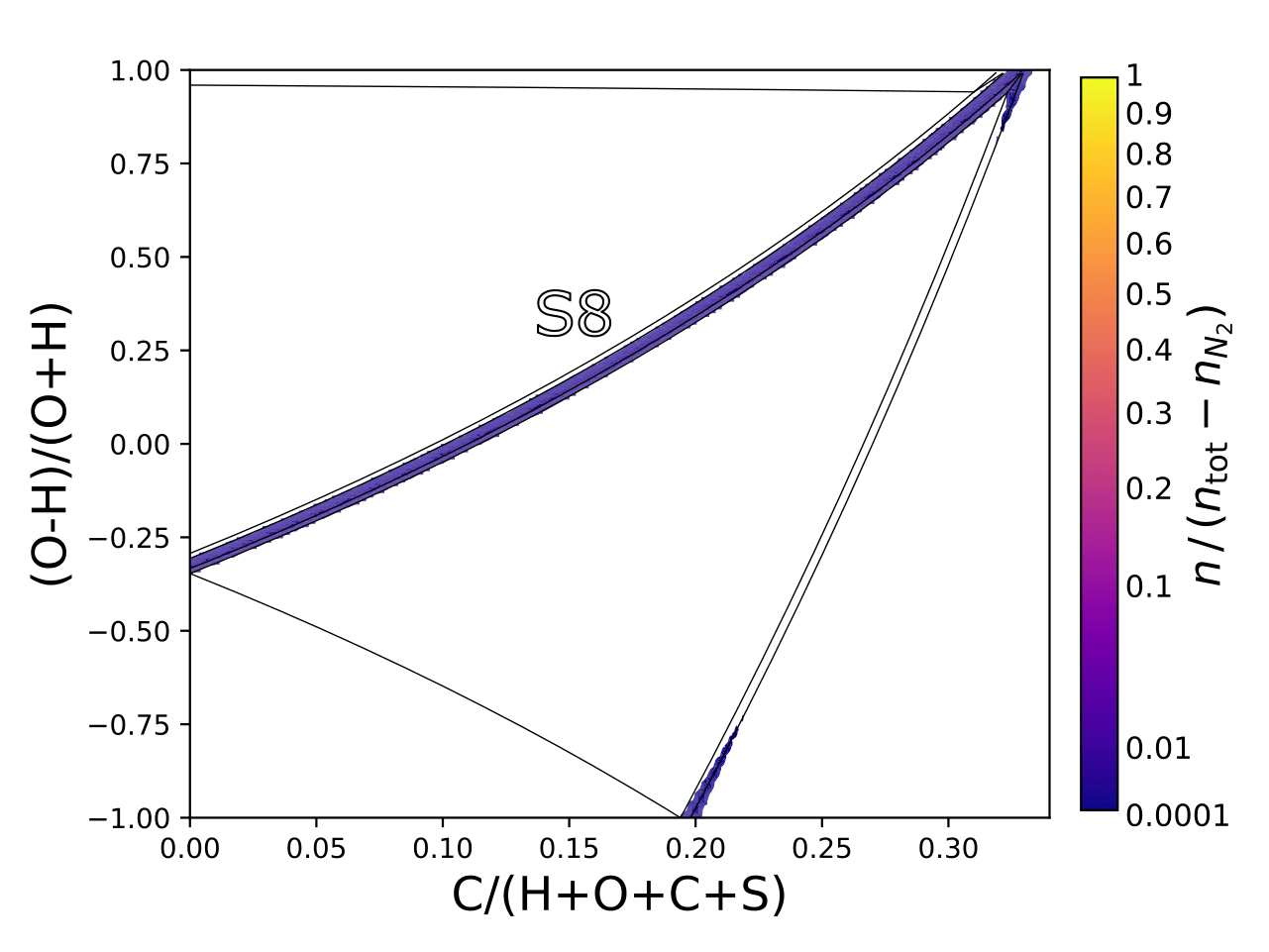} &
\hspace*{-6mm}
\includegraphics[width=65mm,height=50mm]{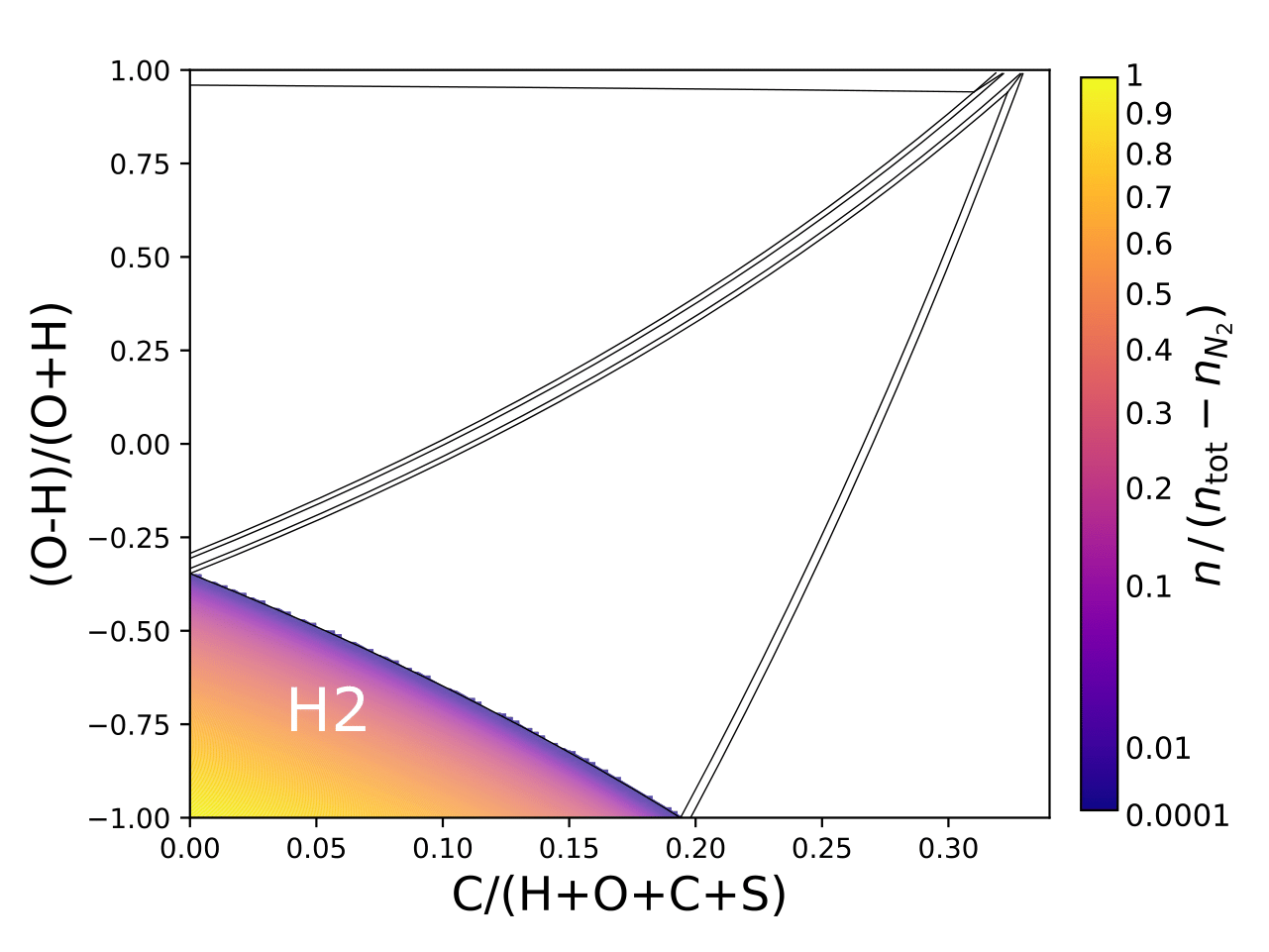} &
\hspace*{-6mm}
\includegraphics[width=65mm,height=50mm]{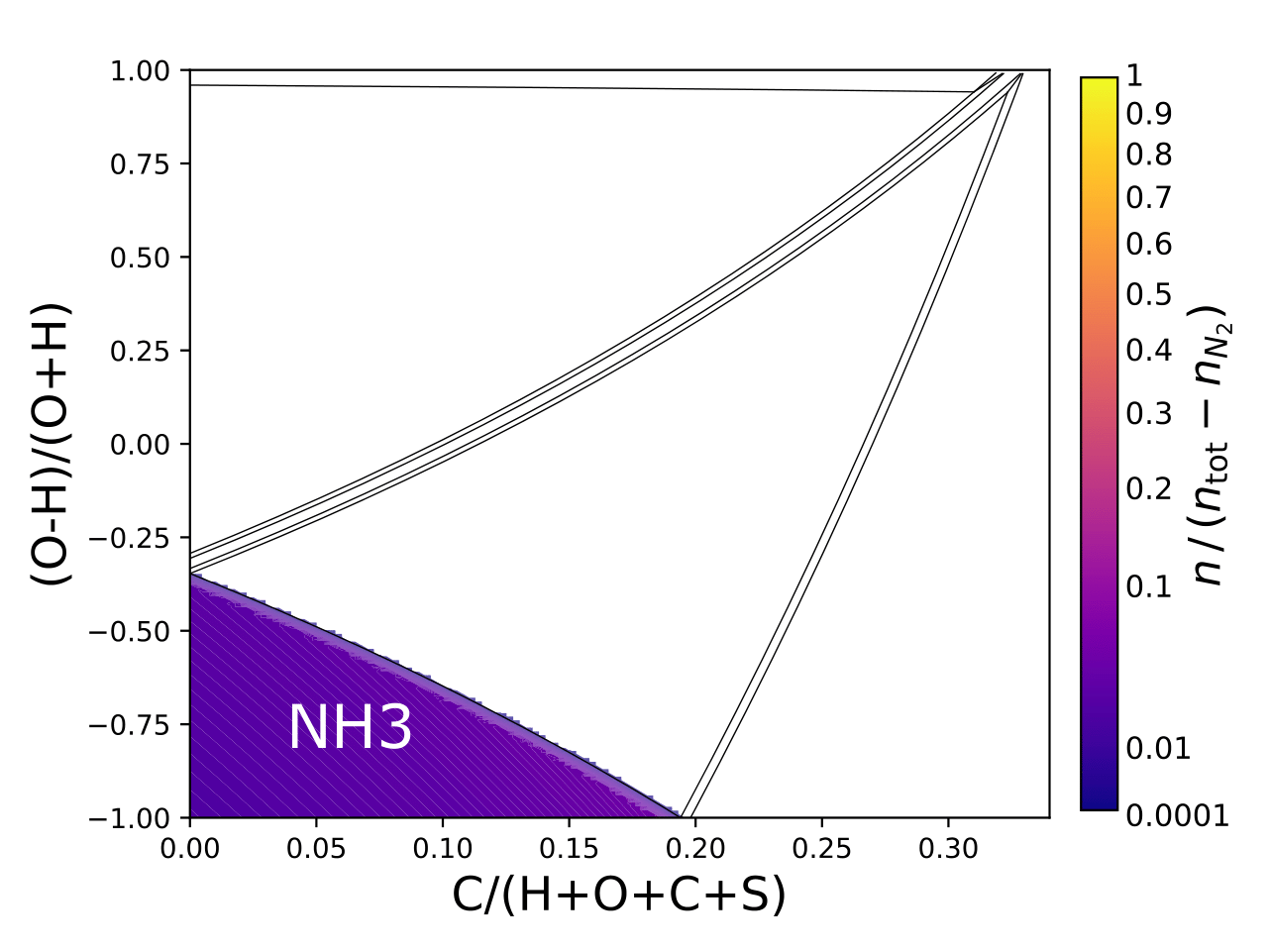} \\[-3mm]
\hspace*{-7mm}
\includegraphics[width=65mm,height=50mm]{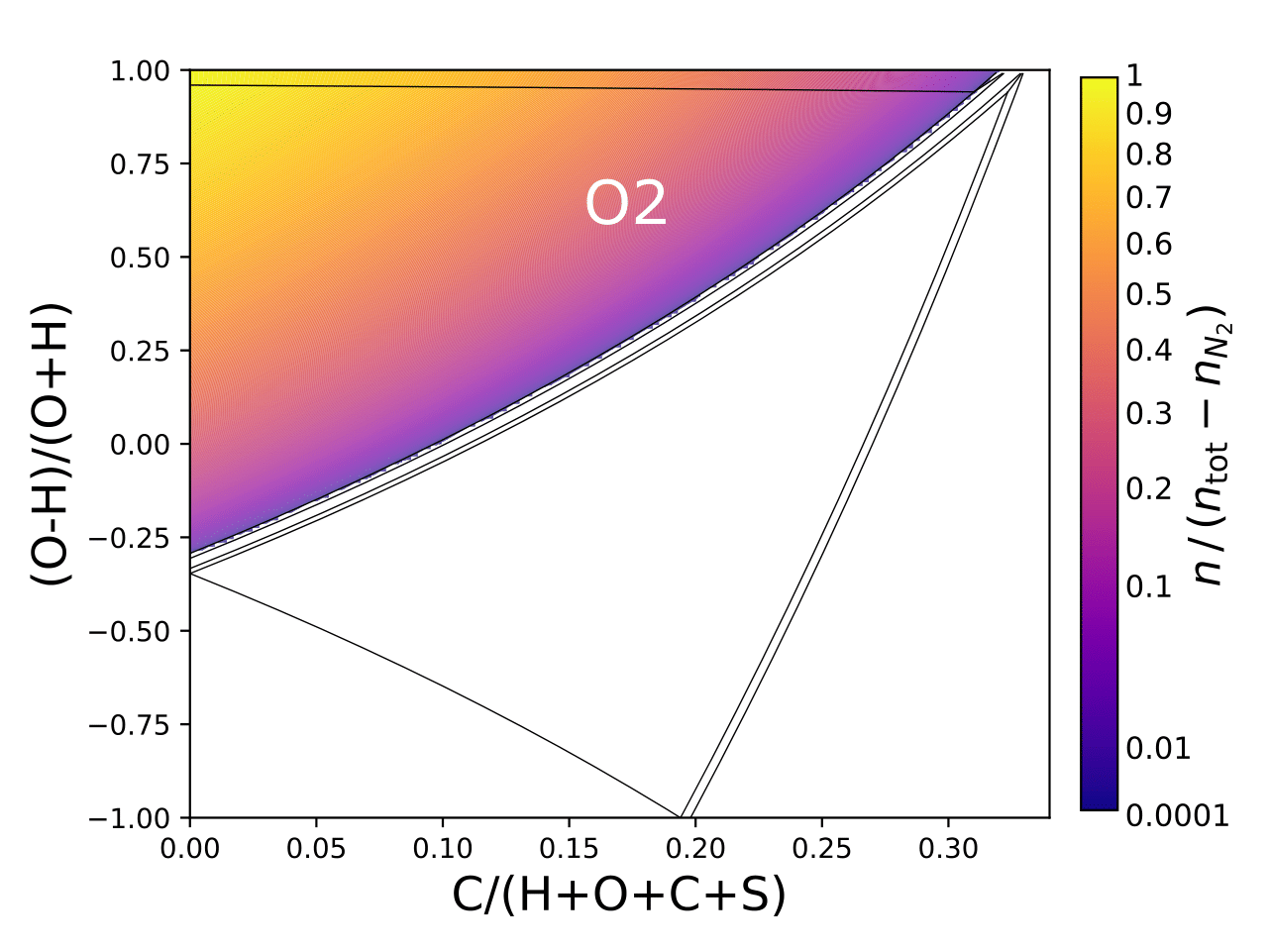} &
\hspace*{-6mm}
\includegraphics[width=65mm,height=50mm]{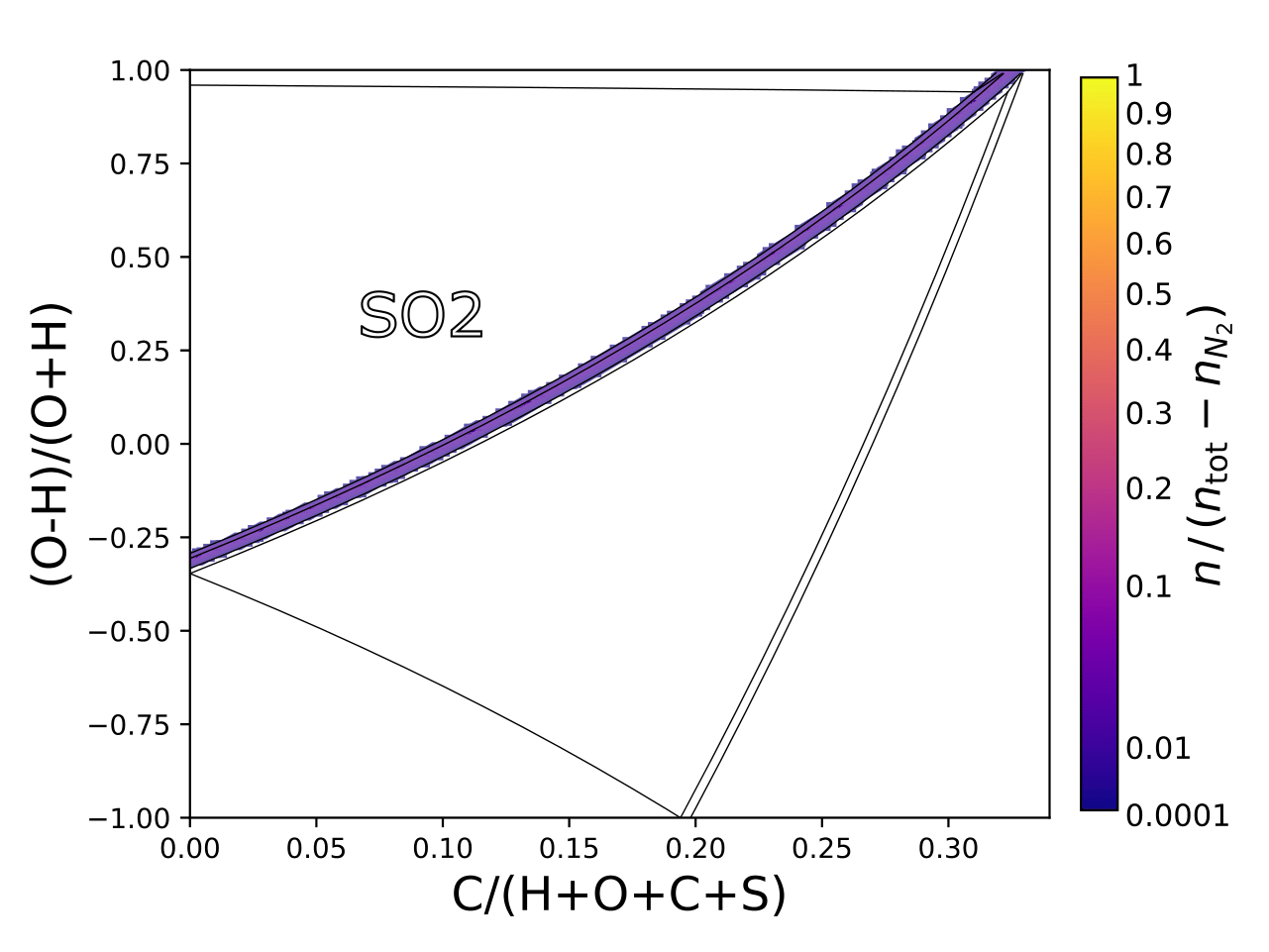} &
\hspace*{-6mm}
\includegraphics[width=65mm,height=50mm]{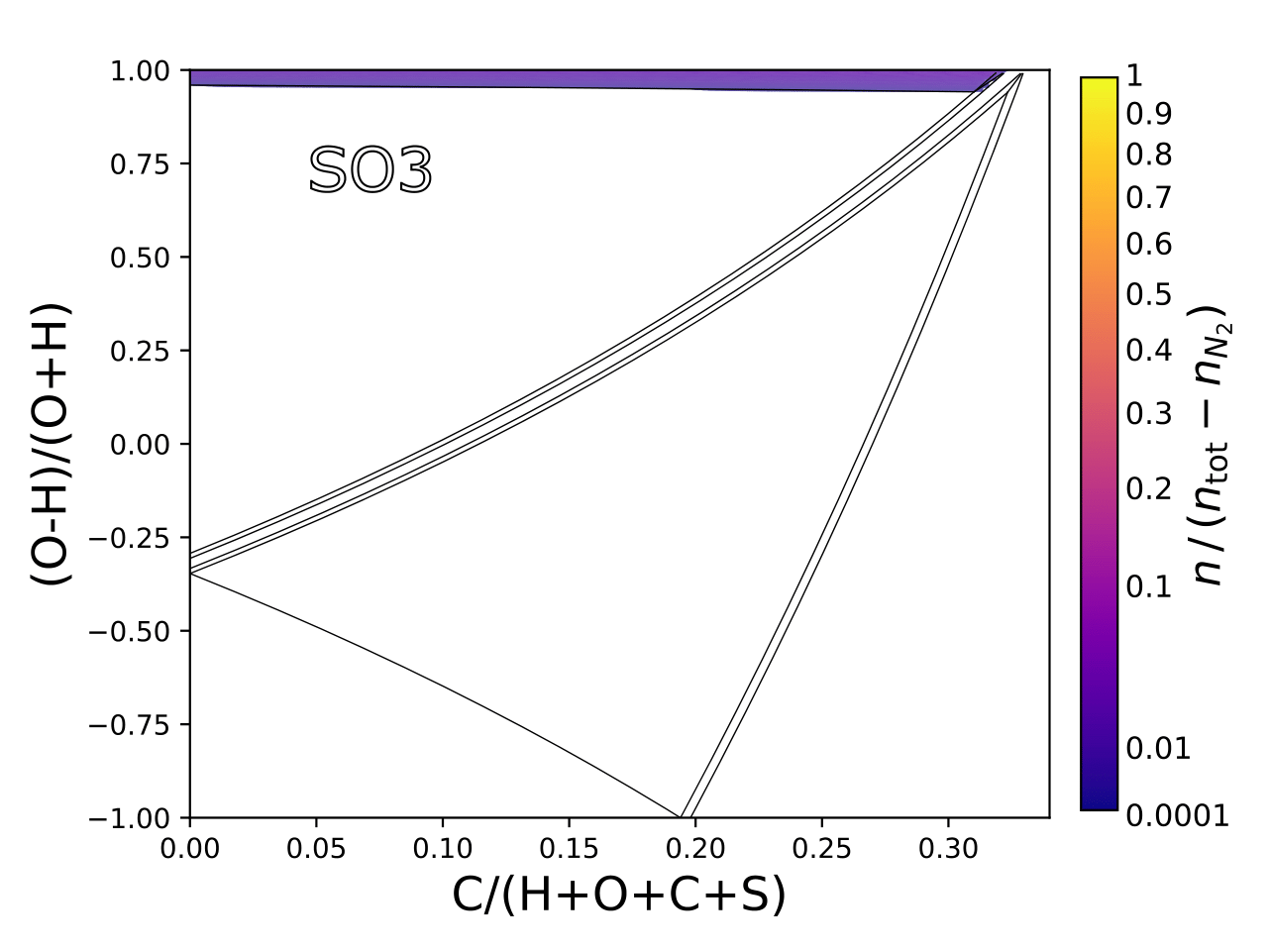} \\[-3mm]
\hspace*{-7mm}
\includegraphics[width=65mm,height=50mm]{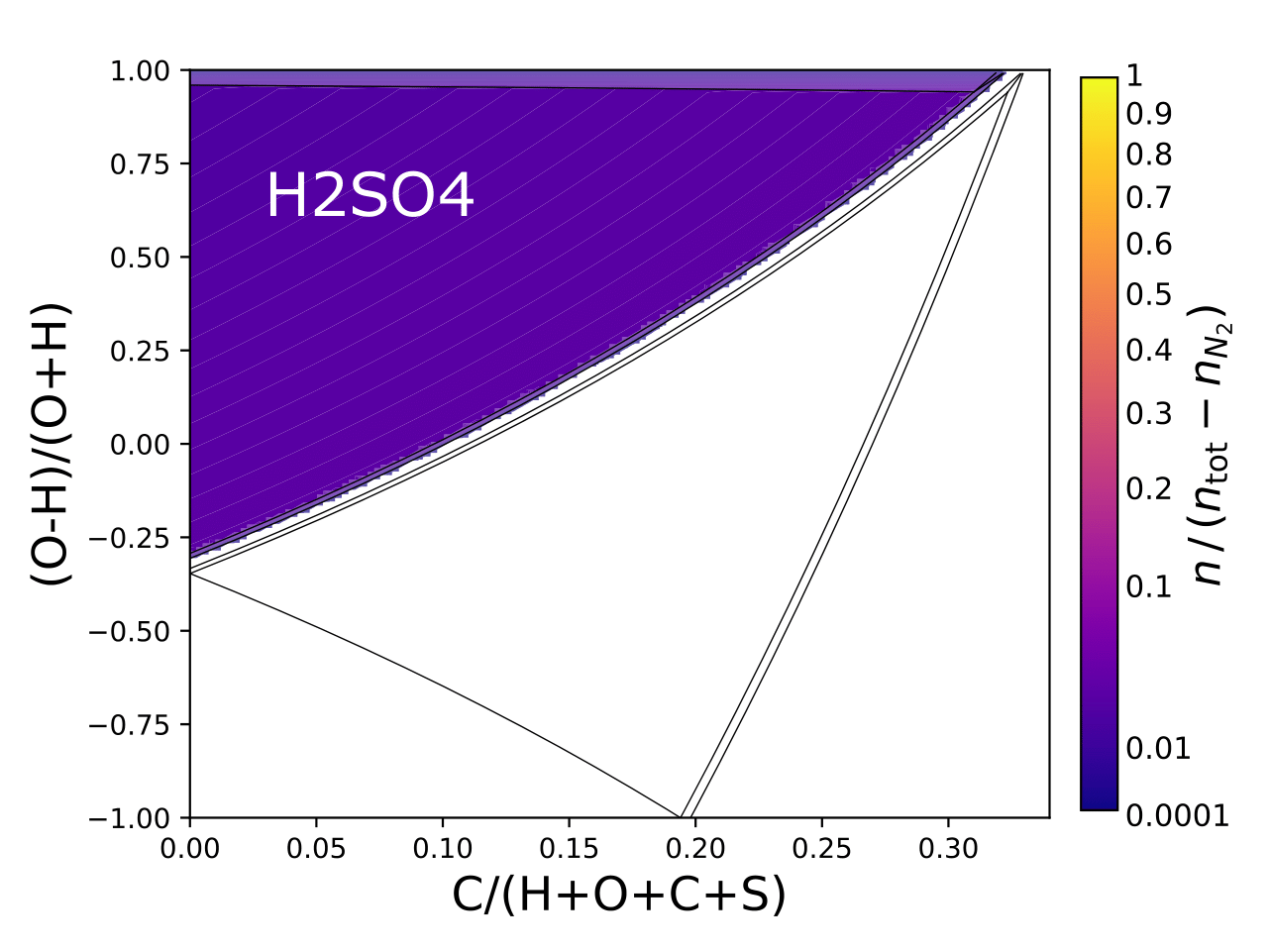} &
\hspace*{-6mm}
\includegraphics[width=65mm,height=50mm]{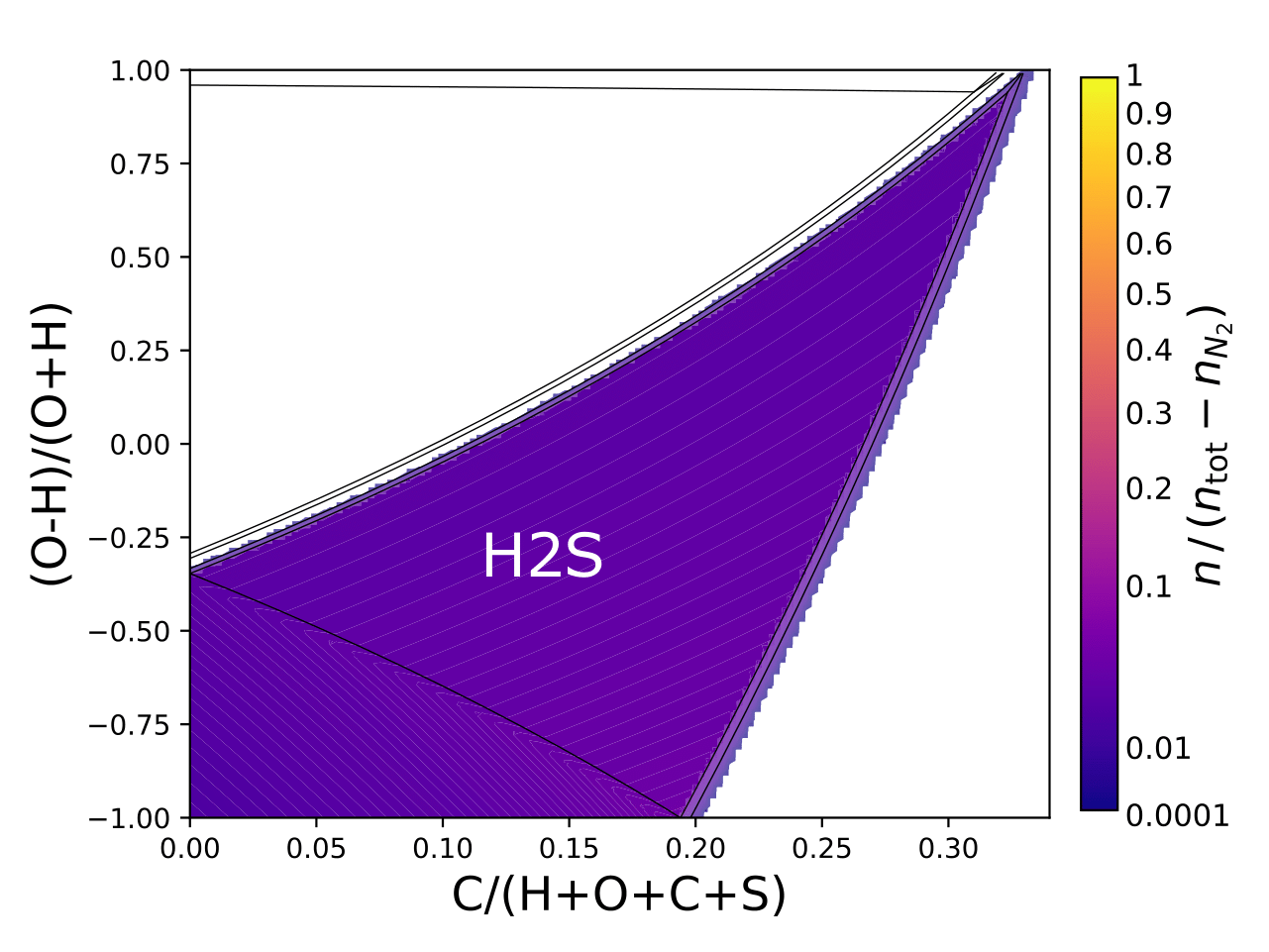} &
\hspace*{-6mm}
\includegraphics[width=65mm,height=50mm]{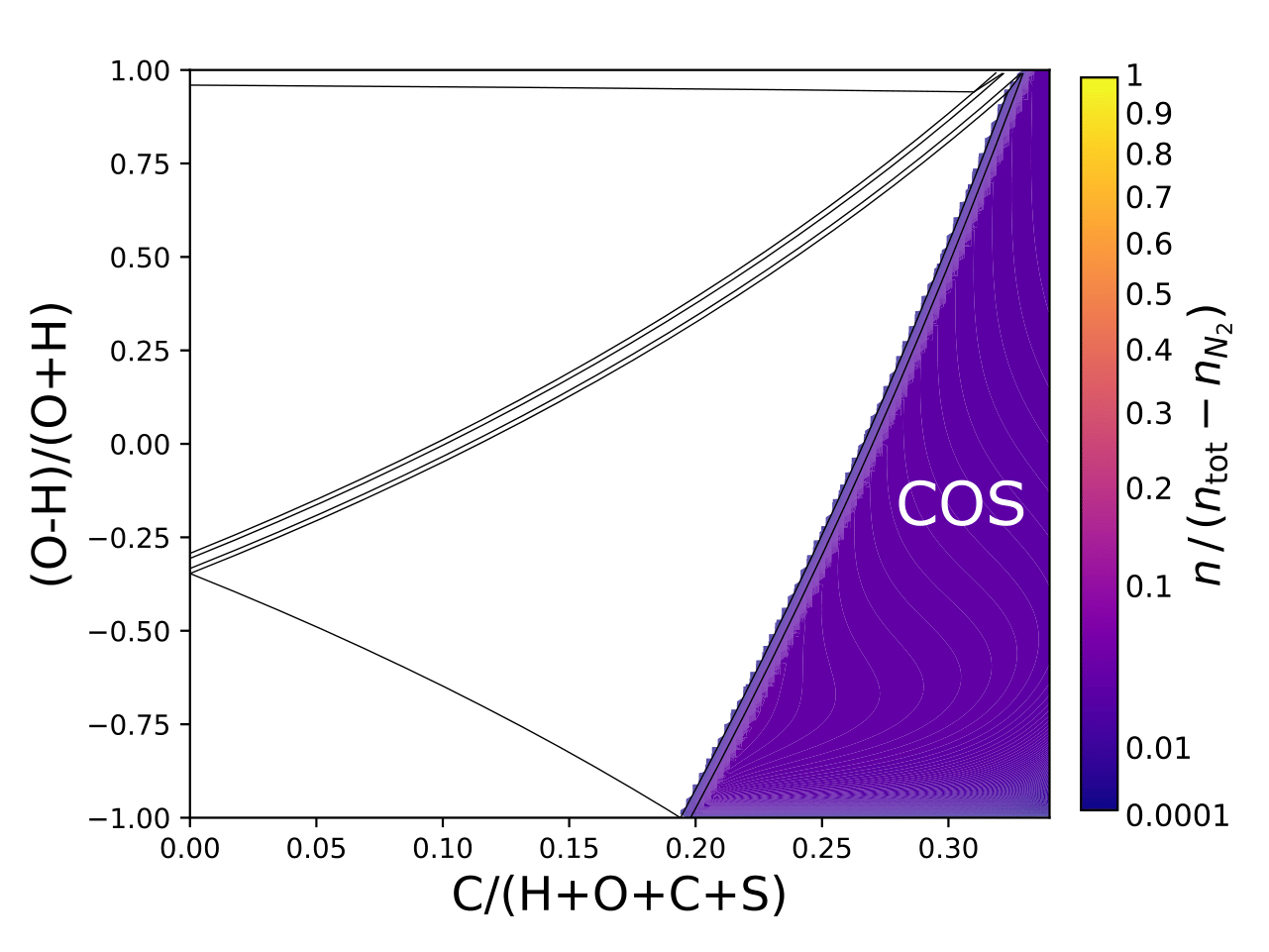} \\[-3mm]
\end{tabular}
\caption{As Fig.~\ref{atmos_types}, but for S/(C+H+O+S)=0.01.}
\label{atmos_types_lowS}
\end{figure*}

As in \citet{Woitke2020}, additional atmospheric types occur in
principle for very large carbon abundances, where graphite is
supersaturated. This includes the new peculiar subtype CG, which is
like type C except that water is replaced by the allotrope \ce{S8}.
For smaller sulphur abundances, see Fig.~\ref{atmos_types_lowS}, the
parameter space occupied by the new subtypes is shrinking, and in the
limiting case of very low sulphur element abundance, the original
types A, B and C are again revealed.

\subsection{Boundaries between two atmospheric types\label{boundaries}}
Here, we show how the boundaries of the different atmospheric
sub-types are derived.  As nitrogen does not interfere with the
chemistry of the other elements, we omit it for the calculation of the
boundaries of the atmospheric types.

At each of these boundaries the element abundances are such that two
molecules replace each other in the adjacent atmospheric types.  The
other present molecules remain unaltered.  The location of the
boundary is determined by the condition that
\begin{equation}
p_i > 0. \label{eqn:Eq12}
\end{equation}
Here, $p_i$ is the molecular partial pressure of a molecule $i$, which
is present in the investigated atmospheric types.

In the following, we show the derivation of the boundary conditions
for atmospheric types on the example of the type C atmosphere, which
consists only of the molecules \ce{CH4}, \ce{CO2}, \ce{H2O}, \ce{H2S},
and \ce{N2}.  The fictitious atomic pressure $p_{\rm atom}$ of the gas
mixture can be written as
\begin{equation}
 \label{Eq1}
 p_{\rm atom} = 3\,p_{\ce{H2O}} + 5\,p_{\ce{CH4}} + 3\,p_{\ce{CO2}} + 3\,p_{\ce{H2S}}.
\end{equation}
This atomic pressure can further be written in terms of the partial atomic pressures of the CHOS elements with
\begin{subequations}
 \begin{align}
  &\ce{H} \cdot p_{\rm atom}  =   p_{\ce{H_2O}} + 4p_{\ce{CH_4}} + 2p_{\ce{H_2S}},\label{Hpatom}\\
 &\ce{C} \cdot p_{\rm atom}  =   p_{\ce{CH_4}} +   p_{\ce{CO_2}}, \label{Cpatom}\\
 &\ce{O} \cdot p_{\rm atom}  = 2p_{\ce{H_2O}} + 2p_{\ce{CO_2}},\label{Opatom} \\
 &\ce{S} \cdot p_{\rm atom}  =  p_{\ce{H_2S}},\label{Spatom}
 \end{align}
\end{subequations}
where the element abundances H, C, O, and S are normalised by
\begin{equation}
    \ce{H + C + O + S} = 1.
\end{equation}
The total gas pressure of the system is given by the sum of all individual molecular partial pressures,
\begin{equation}
\label{Eq3}
p_{\rm gas}= p_{\ce{CO_2}} +  p_{\ce{H_2O}} +  p_{\ce{CH_4}} +  p_{\ce{H_2S}}.
\end{equation}
One can use Eqs.~(\ref{Hpatom})-(\ref{Spatom}) and Eq.~(\ref{Eq3}) to solve for the molecular partial pressures in terms of element abundances, which leads to
\begin{subequations}
 \label{Eq4}
 \begin{align}
 &\frac{p_{\ce{CO_2}}}{p_{\rm gas}} =\frac{4\ce{C} - \ce{H} + 2\ce{O} + 2\ce{S}}{2\ce{H} + 4\ce{O} + 4\ce{S}},\label{pCO2}\\
 &\frac{p_{\ce{H_2O}}}{p_{\rm gas}} =\frac{-4\ce{C} + \ce{H} + 2\ce{O} - 2\ce{S}}{\ce{H} + 2\ce{O} + 2\ce{S}},\label{pH2O}\\
 &\frac{p_{\ce{CH_4}}}{p_{\rm gas}} =\frac{4\ce{C} + \ce{H} - 2\ce{O} - 2\ce{S}}{2\ce{H}  + 4\ce{O} + 4\ce{S}},\label{pCH4}\\
 &\frac{p_{\ce{H_2S}}}{p_{\rm gas}} = \frac{4\ce{S}}{\ce{H} + 2\ce{O} + 2\ce{S}}.\label{pH2S}
 \end{align}
\end{subequations}
This shows clearly that sulphur  interferes with the chemistry of the C-H-N-O system. Nitrogen however, stays independent.

The boundary equations can be derived from Eqs.~(\ref{Eq4}), by requiring non-negative, non-zero partial pressures of the individual molecules.
This leads to
\begin{subequations}
\label{Eq5}
 \begin{align}
 &{\rm a)}\quad \ce{H} > 4\ce{C} + 2\ce{O} + 2\ce{S},\\
 &{\rm b)}\quad 2\ce{C} +\ce{S} > \ce{O} + 0.5\ce{H},\\
 &{\rm c)}\quad \ce{O} + \ce{S} > 2\ce{C} + 0.5\ce{H}.
 \end{align}
\end{subequations}
where:

\smallskip\noindent
 a) no \ce{CO_2} exists in the atmosphere\\
\noindent 
 b) no \ce{H_2O} exists in the atmosphere\\
\noindent 
 c) no \ce{CH_4} exists in the atmosphere

\medskip\noindent Beyond each of these boundaries, a new type of
atmosphere will occur.  Figure~\ref{atmos_types} illustrates these
boundaries in the C-H-O plane for a constant value of S. Each
sub-figure shows the occurrence of one particular molecule as a
function of (O-H)/(O+H) and carbon fraction (C/(C+H+O+S)) in the gas
phase.  The considered species are the molecules dominating the
atmosphere due to equilibrium at low temperatures.  At each black
line, which is one of the above mentioned boundaries, the element
ratios become such that two molecules replace each other.

\subsection{Condensation at low temperatures}

\begin{table}
\centering
  \caption{Condensates occurring in the various atmospheric types of
    the CHNOS system at low temperatures.}
  \label{Tab_condens}
  \vspace*{-2mm}\hspace*{-2mm}
  \resizebox{88mm}{!}{
  \begin{tabular}{c|ccccc} 
   Type & \ch{H2O}[l,s] & \ch{H2SO4}[l,s] &
   \ch{S}[l,s],\ch{S2}[s],\ch{S8}[s] & \ch{NH4SH}[s]&\ch{C}[s]\\ \hline
   A1 &\tick &\cross&\cross  &\tick&\cross \\
   B1 &\cross &\tick&\cross &\cross&\cross \\
   B2 &\tick&\tick&\cross &\cross& \cross\\
   BC1 &\cross&\tick&\cross&\cross& \cross\\
   BC2 &\tick &\tick&\cross &\cross&\cross\\
   BC3 &\tick&\tick&\tick&\cross&\cross\\
   BC4 &\tick& \tick&\tick&\cross& \cross \\
   C&\tick&\cross&\tick&\tick&\tick\\
   CG &\tick&\cross&\tick&\cross&\tick \\
  \end{tabular}}
\end{table}

In the low temperature limit, it is sub-type specific which
condensates can form. Table~\ref{Tab_condens} lists all of the seven
condensates and indicates which of them can form in which
sub-type. Highly oxidised CHNOS atmospheres such as Earth or Venus
will form \ch{H2O}[l,s] or \ch{H2SO4}[l,s] depending on sulphur
abundance, pressure and temperature. The atmospheric type of Earth is
the B-type \citep{Woitke2020}. However, not enough about the
atmospheric chemistry of Venus is known in order to determine its
exact type. Since gaseous water is certainly present in Venus'
atmosphere, it is not type BC1. If gaseous \ch{H2SO4} occurs beside
\ch{SO2}, Venus would be type BC2. If gaseous \ch{S2} or \ch{S8} occur
beside \ch{SO2} it would be type BC3.  The gas giants in our solar
system have more reduced atmospheres. Jupiter's atmospheric type is
known as an A-type with the corresponding sulphur condensate being
\ch{NH4SH}[s]. Predicting models for formation of the latter in
Jupiter's atmosphere exist \citep{Visscher2006}, however it has never
been confirmed by observations \citep{Sromovsky2018}.

\section{Tables of resulting gas phase abundances\label{A2}}

Table~\ref{Tab_crusttypes_100bar} lists the major molecules and
maximum sulphur fractions in the gas over different rock
materials at 100~bar, together with some specifics about the
temperature dependencies. 
Table~\ref{Tab_crusttypes_1bar} does the same for 1~bar.
Table~\ref{Tab_main_mols} shows the major molecules with
concentrations $>\!1$\% and trace species with concentrations
$>\!0.1$\%, $>\!100\,$ppm, and $>\!10\,$ppm, respectively, for different
rock materials, temperatures and pressures.
Table~\ref{condensates} explains the condensate families used in
Fig.\,\ref{one_model} as a reference.

\begin{table*}
\begin{center}
  \caption{Results of our chemical and phase equilibrium models for different element compositions at 100\,bar$^{(1)}$.}
  \label{Tab_crusttypes_100bar}
\vspace*{-1mm}  
\resizebox{16.5cm}{!}{
\begin{tabular}{c|c|c|cc|c|c|c} 
\hline  
&&&&&&&\\[-2ex]
   & major molecules
   & $\epsilon_{\rm S}^{\rm gas+cond}$ 
   & max.\ $\epsilon_{\rm S}^{\rm gas}/\sum_k \epsilon_k^{\rm gas}$
   & $T$-range\,[K] 
   & type$^{(2)}$ 
   & most abundant$^{(3)}$ 
   & condensed sulphur\\ 
   Model 
   & $n/n_{\rm tot}>10^{-1.5}$ 
   & [\%\,mf]
   & [mol.\,fractions]
   & with high $\epsilon^{\rm gas}_{\rm S}$$^{(1)}$
   & at $600$K
   & sulphur molecules
   & species \\
   & with $T$-range\,[K] &&
   & $T$\,[K] at $\epsilon_{\rm S}^{\rm gas}$-max.
   &
   & and $T$-range\,[K]
   & and $T$-range\,[K] \\
\hline
\hline
&&&&&&&\\[-2ex]
   CC & $220-630$:  \ce{N2} 
      & $0.07$ & $0.0176$ & $1110-3770$ &C& \ce{H_2S}: $200-800$ & $\leq 640$: \ce{FeS2}[s] \\
   & $620-3930$: \ce{H_2O} &&&1150&&\ce{SO_2}: $800-4600$ & $650-1150$: \ce{CaSO_4}[s]\\
   & $510-3950$: \ce{CO_2} &&&&&\ce{SO}: $\geq 4600$&\\
   &$1120-3220$: \ce{SO_2} &&&&&&\\
   &$2070-5000$: \ce{O_2}  &&&&&&\\
\hline
&&&&&&&\\[-2ex]
   CI & $260-550$: \ce{N2}
   & $5.411$ & $0.0448$ & $1920-4220$&C&\ce{H_2S}: $200-2400$ & $\leq520$: \ce{FeS2}[s] \\
   &$1780-2470$: \ce{H2S} &&&2310&&\ce{SO_2}: $2400-4300$& $480-1480$: \ce{FeS}[s]\\
   &$2310-4110$: \ce{SO2} &&&&&\ce{SO}: $\geq 4300$& $1500-2220$: \ce{FeS}[l]\\
   & $410-5000$: \ce{H2O} &&&&&& $2160-2280$: \ce{MnS}[s]\\
   & $490-4060$: \ce{CO2} &&&&&&\\
   & $390-720$:  \ce{CH4} &&&&&&\\
   &$1350-5000$: \ce{CO}  &&&&&&\\
   & $830-5000$: \ce{H2}  &&&&&&\\
\hline
&&&&&&&\\[-2ex]
   BSE & $\leq780$: \ce{N2}
    & $0.027$ & $0.0662$ & $1290-2250$ &C&\ce{H_2S}: $\leq 1200$& $\leq1310$: \ce{FeS}[s] \\
   & $750-3690$: \ce{H2O} &&&1320&&\ce{SO_2}: $1200-4600$&\\
   &$1270-3670$: \ce{SO2} &&&&&\ce{SO}: $\geq 4600$&\\
   &$2840-4140$: \ce{NaOH}&&&&&&\\
   & $690-3360$: \ce{CO2} &&&&&&\\
   &$2190-5000$: \ce{O2}  &&&&&&\\
\hline
&&&&&&&\\[-2ex]   
   MORB & $\leq 660$: \ce{N2}
        & $0.110$ & $0.0707$ & $1280-3750$ &C&\ce{H_2S}: $\leq 1200$& $\leq 1300$: \ce{FeS}[s] \\
   & $620-3950$: \ce{H2O} &&&1310&&\ce{SO_2}: $1200-4600$&\\
   & $580-3580$: \ce{CO2} &&&&&\ce{SO}: $\geq 4600$&\\
   &$1230-3930$: \ce{SO2} &&&&&&\\
   &$2780-5000$: \ce{O2}  &&&&&&\\
\hline
&&&&&&&\\[-2ex]   
   solar & $100-5000$: \ce{H2} 
         & $0.041$ & $1.31\cdot10^{-5}$ & $\geq 650$& A1 &\ce{H_2S}: $\leq 3600$& \ce{FeS}[s] $<10^{-4}$\\ 
   &&&&1140&&\ce{HS}: $3600-4800$& \ce{MnS}[s] $<10^{-4}$\\
   &&&&&&\ce{S}: $\geq4800$&\\
\hline
&&&&&&&\\[-2ex]      
   Archean & $411-5000$: \ce{H2O}
           & 0.027 & $0.00024$ & $650-4160$ & A1 &\ce{H_2S}: $\leq 3150$& $\leq 660$: \ce{FeS}[s] \\ 
   Earth & $100-5000$: \ce{H2}& &&$700$&&\ce{SO_2}: $3150-4400$&\\
    & \ce{CH_4} reaches peak abund. &&&&&\ce{SO}: $\geq 4400$&\\
    & of $n/n_{\rm tot}=10^{-2.96}$ only &&&&&&\\
\hline
&&&&&&&\\[-2ex]      
   Earth & $700-770$: \ce{NH3}
         & $4.841 ?$ & $0.0165$ & $\geq 2810$ &B1&\ce{H_2SO_4}: $\leq 900$& $\leq1540$: \ce{MgS}[s]\\ 
   &  $130-1760$: \ce{CH4} &&&3160&&\ce{SO_3}: $900-950$& $\leq2340$: \ce{CaS}[s]\\
   & $2440-5000$: \ce{CO}  &&&&&\ce{SO_2}: $950-4600$& $1540-1800$: \ce{MnS}[s]\\
   &  $100-5000$: \ce{H2}  &&&&&\ce{SO}: $\geq 4600$&\\
\hline
&&&&&&&\\[-2ex]      
   PWD & $270-460$: \ce{NH3}
       & $3.3$ & $0.0392$ & $2570-4060$ &A1&\ce{H_2S}: $\leq 3400$& $\leq2530$: \ce{MnS}[s]\\
   CI & $2470-3160$: \ce{H2S} &&&3000&&\ce{HS}: $3400-3800$& $\leq1480$: \ce{FeS}[s]\\
   &  $130-1330$: \ce{CH4} &&&&&\ce{SO}: $\geq 3800$& $1500-2960$: \ce{FeS}[l]\\
   &  $490-5000$: \ce{H2O} &&&&&&\\
   & $1270-4060$: \ce{CO}  &&&&&&\\
   &  $100-5000$: \ce{H2}  &&&&&&\\
\hline
&&&&&&&\\[-2ex]  
   PWD & $510-3120$: \ce{H2O}
       & $3.3$& $0.1611$ & $2890-4110$&C&\ce{H_2S}: $\leq 2300$& $970-2700$: \ce{MnS}[s]\\
  CC & $630-3330$: \ce{CO2} &&&3420&&\ce{SO_2}: $2300-2500$& $210-1480$: \ce{FeS}[s]\\
   &  $470-750$:  \ce{CH4} &&&&&\ce{S_2}: $2500-3500$& $1500-3330$: \ce{FeS}[l] \\
   & $2850-4500$: \ce{Na}  &&&&&\ce{SO}: $\geq 3500$& $\leq 280$: \ce{FeS2}[s]\\
   &  $900-4500$: \ce{SO}  &&&&&& $3420-3250$: \ce{Na_2S}[s]\\
   & $2700-3650$: \ce{S2}  &&&&&& \\
   &  $100-560$:  \ce{N2}  &&&&&& \\
   &  $930-3080$: \ce{H2}  &&&&&& \\
 \hline
&&&&&&&\\[-2ex]  
   PWD & $100-690$: \ce{N2}
       & $3.3$ & $0.2425$& $2370-4110$ &C&\ce{H_2S}: $\leq 800$& $810-1030$: \ce{CaSO4}[s]\\
   BSE & $1540-4750$: \ce{CO} &&&3290&&\ce{SO_2}: $800-2900$& $350-1480$: \ce{FeS}[s]\\
   &  $570-4270$: \ce{CO2} &&&&&\ce{S_2}: $2900-3500$& $\leq 800$: \ce{FeS2}[s]\\
   & $2600-4330$: \ce{SO}  &&&&&\ce{SO}: $\geq 3500$& $1500-3290$: \ce{FeS}[l]\\
   & $1270-4330$: \ce{SO2} &&&&&& $2220-2700$: \ce{MnS}[s]\\
   & $2080-3850$: \ce{S2}  &&&&&& $2220-2700$: \ce{MnS}[s]\\
\hline   
\end{tabular}}
\end{center}
\vspace*{-2mm}
{\footnotesize $^{(1)}$: \ The total (gas and condensate) sulphur abundances $\epsilon_{\rm S}^{\rm gas+cond}$ (input) are listed in mass fraction percentages (\%\,mf). The sulphur element abundances in the gas phase $\epsilon_{\rm S}^{\rm gas}/\sum_k \epsilon_k^{\rm gas}$ (output) are given in molar fractions, i.e.\ number of S-nuclei divided by the sum of the numbers of nuclei of all elements. 
\\  
$^{(1)}$: where $\epsilon_{\rm S}^{\rm gas}$ is at least $\frac{1}{2}$ of its maximum\\
$^{(2)}$: the atmospheric type of the model, see Appendix~\ref{atmos_types}\\
$^{(3)}$: considering $T\!<\!3000$K}
\end{table*}

\begin{table*}
\begin{center}
  \caption{Results of our chemical and phase equilibrium models for different element compositions at 1\,bar$^{(1)}$.}
  \label{Tab_crusttypes_1bar}
\vspace*{-1mm} 
\resizebox{16.5cm}{!}{
\begin{tabular}{c|c|c|cc|c|c|c} 
\hline  
&&&&&&&\\[-2ex]
   & major molecules
   & $\epsilon_{\rm S}^{\rm gas+cond}$ 
   & max.\ $\epsilon_{\rm S}^{\rm gas}/\sum_k \epsilon_k^{\rm gas}$
   & $T$-range \,[K] with high $\epsilon_{\rm S}$;
   & Type
   & most abundant sulphur
   & condensed sulphur species\\ 
Model 
   & $n/n_{\rm tot}>10^{-1.5}$ 
   & [\%\,mf]
   & [mol.\,fractions]
   & $T$\,[K] at $\epsilon_{\rm S}^{\rm gas}$-maximum
   &  at $600$K
   & molecule: $T$-range\,[K]
   & in crust with $T$-range\,[K] \\
   & with $T$-range\,[K] &&&&&&\\
\hline
\hline
&&&&&&&\\[-2ex]
   CC & $100-500$:  \ce{N2} 
      & $0.07$ & $0.0175$ & $1110-3770$ &C& \ce{H2S}: $\leq 1000$ &$650-880$: \ce{CaSO4} [s] \\
   & $500-2900$: \ce{H2O} &&&$890$&&\ce{SO2}: $1000-3600$ &$\leq640$: \ce{FeS_2}[s]\\
   & $400-2900$: \ce{CO_2} &&&&&\ce{SO}: $3600-4000$&\\
   &$900-2400$: \ce{SO_2} &&&&&\ce{S}: $\geq 4000$&\\
   &$1750-4300$: \ce{O_2}  &&&&&&\\
   &$2650-2800$: \ce{OH}  &&&&&&\\
   &$1050-1700$: \ce{HF}  &&&&&&\\
   &$2800-3100$: \ce{K}  &&&&&&\\
\hline
&&&&&&&\\[-2ex]
   CI & $100-370$: \ce{N_2}
   & $5.411$ & $0.0450$ & $1920-4220$&C&\ce{H_2S}: $\leq 1700$ &$\leq460 $\ce{FeS_2}[s] \\
   &$2850-4200$: \ce{OH} &&&$1800$&&\ce{SO_2}: $17000-3400$&$460-1480$: \ce{FeS}[s]\\
   &$1800-3200$: \ce{SO2} &&&&&\ce{SO}: $3400-4000$&$1500-1780$: \ce{FeS}[l]\\
   & $300-3700$: \ce{H2O} &&&&&\ce{S}: $\geq 4000$&\\
   & $400-3100$: \ce{CO2} &&&&&&\\
   & $280-550$:  \ce{CH4} &&&&&&\\
   &$1300-5000$: \ce{CO}  &&&&&&\\
   & $600-3900$: \ce{H2}  &&&&&&\\
   %&$2800-5000$: \ce{H}  &&&&&&\\
\hline
&&&&&&&\\[-2ex]
   BSE & $\leq650$: \ce{N2}
    & $0.027$ & $0.0658$ & $1290-2250$ &C&\ce{H_2S}: $\leq 1100$&$\leq1170$: \ce{FeS}[s] \\
   & $600-3050$: \ce{H2O} &&&$1200$&&\ce{SO_2}: $1000-3500$&\\
   &$1150- 2700$: \ce{SO2} &&&&&\ce{SO}: $3500-4000$&\\
   & $600-2550$: \ce{CO2} &&&&&\ce{S}: $\geq 4000$&\\
   &$1900-4300$: \ce{O2}  &&&&&&\\
   & $700-800$: \ce{H2}  &&&&&&\\
\hline
&&&&&&&\\[-2ex]   
   MORB & $\leq 550$: \ce{N2}
        & $0.110$ & $0.0703$ & $1280-3750$ &C&\ce{H_2S}: $\leq 1000$&$\leq1100$: \ce{FeS}[s]\\
   & $500-2850$: \ce{H2O} &&&$1100$&&\ce{SO_2}: $1000-3600$&\\
   & $500-2700$: \ce{CO2} &&&&&\ce{SO}: $3600-4000$&\\
   &$1050-2800$: \ce{SO2} &&&&&\ce{S}: $\geq 4000$&\\
   &$1900-4300$: \ce{O2}  &&&&&&\\
   &$2700-2800$: \ce{OH}  &&&&&&\\
   &$2700-2800$: \ce{K}  &&&&&&\\
\hline
&&&&&&&\\[-2ex]   
   solar & $100-4800$: \ce{H2} 
         & $0.041$ & $1.31\cdot10^{-5}$ & $\geq 650$& A1 &\ce{H_2S}: $100-2600$&\\ 
   &$2600-5000$: \ce{H} &&&$1300$&&\ce{HS}: $2600-3100$&\\
   &&&&&&\ce{S}: $\geq 3100$&\\
\hline
&&&&&&&\\[-2ex]      
   Archean & $300-3750$: \ce{H2O}
           & 0.027 & $0.0024$ & $650-4160$ & A1 &\ce{H_2S}: $\leq 2100$&\ce{Fes}[s]: $100-650$\\ 
   Earth & $100-3950$: \ce{H2}& &&&&\ce{SO_2}: $2100-3400$&\\
    & $2850-4200$: \ce{OH}&&&&&\ce{SO}: $3400-4400$&\\
   & \ce{CH_4} reaches peak abund. &&&&&\ce{S}: $\geq 4400$&\\
& of $n/n_{\rm tot}=10^{-2.98}$ only &&&&&&\\
\hline
&&&&&&&\\[-2ex]      
   Earth & $100-370$: \ce{N2}
         & $4.841 ? $ & $0.0165$ & $\geq 2810$ &B1&\ce{H_2SO_4}: $\leq 650$& $\leq1170$: \ce{CaSO_4}[s]\\ 
   &$300-3150$: \ce{H2O} &&&&&\ce{SO_3}: $650-8000$&\\
   & $350-3150$: \ce{CO2}  &&&&&\ce{SO_2}: $800-3600$&\\
   &  $100-4200$: \ce{O2} &&&&&\ce{SO}: $3600-4000$&\\
   &  $2450-5000$: \ce{CO}  &&&&&\ce{S}: $\geq4000$&\\
   &  $2650-3650$: \ce{OH}  &&&&&&\\
   &   $2800-3000$: \ce{K}  &&&&&&\\
\hline
&&&&&&&\\[-2ex]      
   PWD & $200-350$: \ce{NH3}
       & $3.3$ & $0.0399$ & $2570-4060$ &A1&\ce{H_2S}: $\leq 2400$&$\leq1500$: \ce{FeS}[s]\\
   CI & $2650-5000$: \ce{H}  &&&$2300$&&\ce{S2}: $2400-2500$&$\leq2080$: \ce{MnS}[s]\\
   &  $100-950$: \ce{CH4} &&&&&\ce{HS}: $2500-3700$&$1500-2250$: \ce{FeS}[l]\\
   &  $400-3650$: \ce{H2O} &&&&&\ce{SO}: $3700-3900$&\\
   & $900-3100$: \ce{CO}  &&&&&\ce{S}: $\geq3900$&\\
   &  $100-3900$: \ce{H2}  &&&&&&\\
\hline
&&&&&&&\\[-2ex]  
   PWD & $400-2250$: \ce{H2O}
       & $3.3$& $0.1746$ & $2890-4110$&C&\ce{H_2S}: $\leq 1700$&$210-1480$: \ce{FeS}[s]\\
  CC & $500-3150$: \ce{CO2} &&&$2400$&&\ce{S_2}: $1700-2600$&$1500-2370$: \ce{FeS}[l]\\
   &  $400-700$:  \ce{CH4} &&&&&\ce{S_2}: $2600-2800$&$\leq280$: \ce{FeS_2}[s]\\
   & $2100-3250$: \ce{K}  &&&&&\ce{SO_2}: $2800-3500$&$900-2250$: \ce{MnS}[s]\\
   &  $2450-3100$: \ce{SO}  &&&&&\ce{SO}: $3500-4000$&\\
   & $1950-2750$: \ce{S2}  &&&&&\ce{S}: $\geq4000$& \\
   &  $100-450$:  \ce{N2}  &&&&&& \\
   &  $650-2250$: \ce{H2}  &&&&&& \\
   &  $2250-3300$: \ce{SO_2}  &&&&&& \\
   &  $750-3300$: \ce{CO}  &&&&&& \\
 \hline
&&&&&&&\\[-2ex]  
   PWD & $100-550$: \ce{N2}
       & $3.3$ & $0.2586$& $2370-4110$ &C&\ce{H_2S}: $\leq 700$&$\leq800$: \ce{FeS_2}[s] \\
   BSE & $1300-3300$: \ce{CO} &&&$2400$&&\ce{SO_2}: $700-2000$&$350-1480$: \ce{FeS}[s] \\
   &  $450-3200$: \ce{CO2} &&&&&and $2900-3500$&$1500-2370$: \ce{FeS}[l]\\
   & $2150-3200$: \ce{SO}  &&&&&\ce{S_2}: $2000-2700$&$2220-2340$: \ce{MnS}[s]\\
   & $1150-3300$: \ce{SO2} &&&&&\ce{SO}: $3600-4000$&$810-830$: \ce{CaSO_4}[s]\\
   & $1550-2850$: \ce{S2}  &&&&&\ce{S}: $\geq 4000$&\\
\hline   
\end{tabular}}
\end{center}
\vspace*{-2mm}
{Like Tab. \ref{Tab_crusttypes_100bar} but for a surface pressure of 1\, bar.\\
\footnotesize$^{(1)}$: where the sulphur abundance is at $\frac{1}{2}$ of its maximum.}
\end{table*}

\begin{table*}
\begin{center}
  \caption{Molecular composition of the atmospheric gas in our rocky exoplanet models with varying element composition, surface temperature and pressure. Sulphur molecules are highlighted in bold face.}
  \label{Tab_main_mols}
\vspace*{-2mm}  
\resizebox{9.0cm}{!}{
\begin{tabular}{c|c|c|c|c|c} 
\hline  
&&&&&\\[-2ex]
   Model
   & $n_{\rm mol}/n_{\rm tot}$ $\ge 1\%$ [$\%$]
   &$\ge 0.1\%$ [$\%$]
   &$\ge 100$\,ppm [ppm];
   &$\ge 10$\,ppm [ppm]
   & $\mu$\,[amu] \\ 
\hline
\hline
&&&&&\\[-2ex]
   CC & \ce{H2O}: $44$
      & \ce{AlF2O}: $0.90$ &\ce{NaF}: $590$ &\ce{(NaCl)2}: $84$&  \\
  2000 K & \ce{CO2}: $34$ &\ce{HCl}: $0.74$ &\ce{NaOH}: $490$ &\ce{O}: $69$& $32.155$\\
  10 bar & \ce{O2}: $11$&\ce{KCl}: $0.62$ &\ce{CO}: $430$ &\ce{FeCl2}: $64$&\\
   &\textbf{\ce{SO_2}: 4.5 }&\ce{N2}: $0.42$ &\ce{NO}: $420$ &\ce{(KCl)2}: $23$&\\
   &\ce{HF}: $2.2$&\ce{OH}: $0.21$&\ce{KF}: $330$ &\ce{HO2}: $12$&\\
   &\ce{NaCl}: $1.3$ &&\textbf{\ce{SO3}: 270}&&\\
   &&&\ce{Cl}: $250$&&\\
   &&&\ce{KOH}: $210$&&\\
   &&&\ce{Fe(OH)2}: $150$&&\\
   &&&\ce{H2}: $120$&&\\
\hline
&&&&&\\[-2ex]
     CI & \ce{H2O}: $55$
      &\ce{N2}: $0.77$ &\ce{NaCl}: $940$ &\ce{K}: $86$&  \\
  2000K & \ce{H2}: $13$& \textbf{\ce{HS}: 0.39}&\ce{Na}: $880$ &\ce{OH}: $81$& $23.300$\\
   10 bar& \ce{CO}: $11$ &\textbf{\ce{SO}: 0.38}  &\textbf{\ce{S2O}: 850}& \textbf{\ce{S3}: 41}&\\
   &\ce{CO2}: $10$&\ce{NaOH}: $0.17$ &\ce{KOH}: $540$&\ce{PO2}: $24$&\\
   &\textbf{\ce{H2S}: 3.6}&\textbf{\ce{COS}: $0.11$} &\ce{KCl}: $320$&\ce{NaF}: $16$&\\
   &\textbf{\ce{SO2}: 3.2}&&\ce{Fe(OH)2}: $220$&&\\
   &\textbf{\ce{S2}: 2.2}&&\ce{HF}: $200$&&\\
   &&&\ce{HCl}: $190$&&\\
   &&&\ce{H}: $180$&&\\
   &&&\textbf{\ce{S}: 150}&&\\
\hline
&&&&&\\[-2ex]
      BSE & \ce{H2O}: $64$
       &\ce{KCl}: $0.53$ &\ce{N2}: $640$&\ce{NO}: $85$&\\
  2000\, K & \textbf{\ce{SO2}: 18 }&\ce{HF}: $0.35$ &\ce{KOH}: $580$&\textbf{\ce{Na2SO4}: 68}& $30.853$\\
  10\, bar & \ce{CO2}: $11$&\ce{HCl}: $0.33$&\textbf{\ce{SO3}: 560}&\ce{Cl}: $68$&\\
   & \ce{O2}: $2.8$&\ce{N2}: $0.42$ &\ce{H2}: $350$ &\textbf{\ce{SO}: 53}&\\
   &\ce{NaCl}: $1.6$&\ce{NaOH}: $0.19$&\ce{CO}: $270$&\ce{Na}: $43$&\\
   &&\ce{OH}: $0.18$&\ce{Fe(OH)2}: $260$&\ce{O}: $35$&\\
   &&&\ce{NaF}: $250$&\ce{(KCl)2}: $17$&\\
   &&&\ce{AlF2O}: $160$&\ce{FeCl2}: $10$&\\
   &&&\ce{(NaCl)2}: $120$&&\\
   &&&\ce{KF}: $100$&&\\
\hline
&&&&&\\[-2ex]   
      MORB& \ce{H2O}: $63$
       &\ce{HCl}: $0.68$&\textbf{\ce{SO3}: 780}&\ce{N2}: $89$&\\
  2000 K & \textbf{\ce{SO2}: 19}&\ce{KCl}: $0.48$&\ce{NaOH}: $580$&\ce{(NaCl)2}: $51$& $30.895$\\
  10\, bar &\ce{CO2}: $8.9$ &\ce{OH}: $0.21$&\ce{NaF}: $280$&\ce{O}: $47$&\\
   &\ce{O2}: $4.9$&\ce{AlF2O}: $0.16$&\ce{H2}: $260$ &\ce{FeCl2}: $44$&\\
   &\ce{HF}: $1.2$&&\ce{Fe(OH)2}: $250$&\ce{NO}: $42$&\\
   &\ce{NaCl}: $1.0$&&\ce{KOH}: $250$&\textbf{\ce{SO}: 42}&\\
   &&&\ce{CO}: $170$&\ce{(KCl)2}: $14$&\\
   &&&\ce{Cl}: $160$&\ce{Na}: $12$&\\
   &&&\ce{KF}: $150$&&\\
\hline
&&&&&\\[-2ex]      
       Archean & \ce{H2O}: $64$&\ce{NaOH}: $0.19$ &\textbf{\ce{H2S}: 560}&\ce{HF}: $73$&\\
  2000\, K & \ce{H2}: $36$ &\ce{Na}: $0.14$&\ce{H}: $310$&\ce{K}: $70$& $12.421$\\
  10\, bar &&&\ce{KOH}: $300$&\ce{NaCl}: $61$&\\
   &&&\ce{CO}: $270$&\ce{OH}: $56$&\\
   &&&\ce{Fe(OH)2}: $250$&\ce{PO2}: $37$&\\
   &&&\ce{CO2}: $110$&\textbf{\ce{HS}: 36}&\\
   &&&&\textbf{\ce{SO2}: 30}&\\
   &&&&\ce{NaH}: $19$&\\
   &&&&\ce{HCl}: $13$&\\
   &&&&\ce{KCl}: $11$&\\
   &&&&\ce{PO}: $10$&\\
\hline
&&&&&\\[-2ex]      
  & \ce{CO2}: $47$&\textbf{\ce{SO2}: 0.57} &\ce{HCl}: $970$&\ce{KF}: $65$&\\
  Earth & \ce{H2O}: $46$ &\ce{HF}: $0.44$&\ce{CO}: $890$&\ce{O}: $46$& $31.386$\\
  2000\, K &\ce{O2}: $4.8$&\ce{N2}: $0.25$&\ce{KCl}: $800$&\ce{Cl}: $27$&\\
   10\, bar&&\ce{OH}: $0.18$&\ce{NaOH}: $500$&\textbf{\ce{SO3}: 23}&\\
   &&\ce{NaCl}: $0.17$&\ce{AlF2O}: $290$&\ce{Na}: $12$&\\
   &&&\ce{NO}: $220$&&\\
   &&&\ce{KOH}: $210$&&\\
   &&&\ce{H2}: $190$&&\\
   &&&\ce{Fe(OH)2}: $180$&&\\
   &&&\ce{NaF}: $120$&&\\
\hline
&&&&&\\[-2ex]      
      PWD & \ce{H2}: $67$&\ce{N2}: $0.95$ &\textbf{\ce{HS}: 950}&\textbf{\ce{COS}: 92}&\\
  CI & \ce{H2O}: $20$&\ce{CO2}: $0.55$&\ce{K}: $420$ & \ce{NaH}: $87$ & $9.2369$\\
  2000\, K &\ce{CO}: $8.5$&\ce{Na}: $0.45$&\ce{H}: $420$&\ce{PO2}: $86$&\\
   10\, bar&\textbf{\ce{H2S}: 2.0}&\ce{P4O6}: $0.15$&\ce{KCl}: $410$&\ce{Mn}: $63$&\\
   &&\ce{NaOH}: $0.14$&\ce{KOH}: $410$&\ce{Fe(OH)2}: $51$&\\
   &&\ch{NaCl}: $0.13$&\textbf{\ce{PS}: $260$}&\ce{MnH}: $48$&\\
   &&&\textbf{\ce{S2}: 250}&\ce{NaF}: $39$&\\
   &&&\ce{HF}: $220$&\ce{Fe}: $37$&\\
   &&&\ce{PO}: $140$&\textbf{\ce{SO}: 28}&\\
   &&&\ce{PN}: $120$&\ce{P2}: $18$&\\
   &&&\ce{HCl}: $110$&\textbf{\ce{S}: 16}&\\
   &&&&\textbf{\ce{SO2}: $15$}&\\
   &&&&\ce{KF}: $15$&\\
   &&&&\ce{SiO}: $14$&\\
   &&&&\ce{OH}: $13$&\\
   &&&&\ce{NH3}: $11$&\\
\hline   
\end{tabular}}
\end{center}
\vspace*{-2mm}
%{\footnotesize ffz}
\end{table*}

\addtocounter{table}{-1}
\begin{table*}
\begin{center}
  \caption{continued. Entries with listed $\epsilon_{\rm S}^{\rm tot}$ [\%\,mf] indicate models where the total sulphur abundance has been modified artificially. For comparison, the base value for BSE is $\epsilon_{\rm S}^{\rm tot}\!=\!0.027$.}
  \label{Tab_main_mols2}
\vspace*{-2mm}  
\resizebox{11.0cm}{!}{
\begin{tabular}{c|c|c|c|c|c} 
\hline  
&&&&&\\[-2ex]
   & $n_{\rm mol}/n_{\rm tot}$ $\ge 1\%$ [$\%$]
   &$\ge 0.1\%$ [\%]
   &$\ge 100$\,ppm [ppm];
   &$\ge 10$\,ppm [ppm];
   & mean molecular \\ 
   Model 
   & 
   & 
   &
   &
   & weight [amu]  \\
\hline
\hline  
   &&&&&\\
      PWD & \ce{CO}: $47$&\ce{H2S}: $0.8$ &\ce{NaF}: $390$&\ce{Fe(OH)2}: $49$&\\
  CC & \ce{CO2}: $31$ &\ce{KCl}: $0.72$&\textbf{\ce{S2O}: 280}&\ce{OH}: $32$& $32.187$\\
  2000\, K &\ce{H2O}: $12$&\textbf{\ce{COS}: 0.33}&\ce{HCl}: $200$&\ce{(KCl)2}: $31$&\\
   10\, bar&\ce{H2}: $4.1$&\ce{NaCl}: $0.29$&\textbf{\ce{S}: 100}&\ce{PO2}: $22$&\\
   &\textbf{\ce{S2}: 1.1}&\ce{KOH}: $0.25$&\ce{H}: $100$&\textbf{\ce{S3}: 14}&\\
   &\textbf{\ce{SO2}: 1.0}&\textbf{\ce{SO}: 0.18}&&&\\
   &&\ce{HF}: $0.18$&&&\\
   &&\ce{N2}: $0.17$&&&\\
   &&\textbf{\ce{HS}: 0.15}&&&\\
   &&\ce{Na}: $0.14$&&&\\
   &&\ce{KF}: $0.12$&&&\\
   &&\ce{NaOH} $0.11$&&&\\
   &&\ce{K}: $0.10$&&&\\
 \hline
&&&&&\\[-2ex]  
      PWD & \ce{CO2}: $47$&\textbf{\ce{COS}: 0.50 }&\ce{Na}: $920$&\ce{K}: $90$&\\
  BSE & \ce{CO}: $30$ &\textbf{\ce{S2O}: 0.37}&\textbf{\ce{H2S}: 730}&\ce{PO2}: $36$& $43.085$\\
  2000\, K &\textbf{\ce{SO2}: 14}&\ce{H2}: $0.16$&\textbf{\ce{HS}: 710}&\ce{HF}: $34$&\\
   10\, bar&\textbf{\ce{S2}: 6.0}&&\ce{NaCl}: $640$&\ce{NaF}: $25$&\\
   &\ce{H2O}: $1.1$&&\ce{NaOH}: $330$&\ce{N2}: $24$&\\
   &\textbf{\ce{SO}: 1.0}&&\ce{S}: $250$&\ce{H}: $20$&\\
   &&&\ce{KCl}: $220$&\ce{OH}: $15$&\\
   &&&\textbf{\ce{S3}: 180}&\ce{HCl}: $13$&\\
   &&&\ce{KOH}: $100$&\textbf{\ce{CS2}: 13}&\\
   &&&&\ce{KF}: $10$&\\
\hline
&&&&&\\[-2ex]
      BSE & \ce{H2O}: $79$
       &\textbf{\ce{SO2}: 0.83} &\ce{N2}: $790$&\ce{NO}: $94$&\\
  2000\, K & \ce{CO2}: $13$&\ce{KCl}: $0.64$ &\ce{KOH}: $640$&\ce{Cl}: $82$& $23.615$\\
  10\, bar & \ce{O2}: $2.8$&\ce{HCl}: $0.45$&\ce{H2}: $430$&\ce{Na}: $43$&\\
   $\epsilon_s^{tot} = 0.001$& \ce{NaCl}: $1.9$&\ce{HF}: $0.43$ &\ce{CO}: $330$ &\ce{O}: $35$&\\
   mass fraction [$\%$]&&\ce{NaOH}: $0.21$&\ce{Fe(OH)2}: $310$&\textbf{\ce{SO3}: 26}&\\
   &&\ce{OH}: $0.20$&\ce{NaF}: $280$&\ce{(KCl)2}: $25$&\\
   &&&\ce{AlF2O}: $200$&\ce{FeCl2}: $15$&\\
   &&&\ce{(NaCl)2}: $170$&\ce{H}: $11$&\\
   &&&\ce{KF}: $110$&&\\
\hline
&&&&&\\[-2ex]
      BSE & \ce{H2O}: $73$
       &\ce{KCl}: $0.60$ &\ce{N2}: $730$&\ce{NO}: $91$&\\
  2000\, K & \ce{CO2}: $12$&\ce{HF}: $0.40$ &\ce{KOH}: $620$&\ce{Cl}: $77$& $26.461$\\
  10\, bar & \textbf{\ce{SO2}: 7.7}&\ce{HCl}: $0.40$&\ce{H2}: $400$&\ce{Na}: $43$&\\
   $\epsilon_s^{tot} = 0.01$&\ce{O2}: $2.8$&\ce{NaOH}: $0.20$&\ce{CO}: $310$ &\ce{O}: $35$&\\
   mass fraction [$\%$]&\ce{NaCl}: $1.8$&\ce{OH}: $0.19$&\ce{Fe(OH)2}: $290$&\textbf{\ce{Na2SO4}: 29}&\\
   &&&\ce{NaF}: $270$&\textbf{\ce{SO}: 22}&\\
   &&&\textbf{\ce{SO3}: 240}&\ce{(KCl)2}: $21$&\\
   &&&\ce{AlF2O}: $180$&\ce{FeCl2}: $13$&\\
   &&&\ce{(NaCl)2}: $150$&\ce{H}: $10$&\\
   &&&\ce{KF}: $110$&&\\
\hline
&&&&&\\[-2ex]
      BSE & \textbf{\ce{SO2}: 45}
       &\ce{KCl}: $0.36$ &\ce{KOH}: $470$&\ce{KF}: $81$&\\
  2000\, K & \ce{H2O}: $43$&\ce{HF}: $0.23$&\ce{N2}: $420$&\ce{NO}: $69$& $41.824$ \\
  10\, bar & \ce{CO2}: $7.2$&\ce{HCl}: $0.19$&\ce{H2}: $230$&\ce{(NaCl)2}: $56$&\\
   $\epsilon_s^{tot} = 0.1$&\ce{O2}: $2.8$&\ce{NaOH}: $0.15$&\ce{NaF}: $200$ &\ce{Cl}: $46$&\\
   mass fraction [$\%$]&\ce{NaCl}: $1.1$&\ce{OH}: $0.15$&\ce{CO}: $180$&\ce{Na}: $43$&\\
   &&\textbf{\ce{SO3}: 0.14}&\ce{Fe(OH)2}: $170$&\ce{O}: $35$&\\
   &&&\textbf{\ce{Na2SO4}: 170}&\textbf{\ce{K2SO4}: 16}&\\
   &&&\textbf{\ce{SO}: 130}&&\\
   &&&\ce{(AlF2O)}: $100$ &&\\
\hline
&&&&&\\[-2ex]
      BSE & \textbf{\ce{SO2}: 61}
       &\textbf{\ce{H2S}: 0.82} &\ce{NaOH}: $820$&\ce{OH}: $64$&\\
  2000\, K & \textbf{\ce{S2}: 16}&\ce{CO}: $0.69$ &\textbf{\ce{S3}: 800}&\ce{H}: $53$& $56.361$\\
  10\, bar & \ce{H2O}: $13$&\ce{NaCl}: $0.39$&\ce{HF}: $780$&\ce{K}: $52$&\\
   $\epsilon_s^{tot} = 0.993$&\textbf{\ce{SO}: 2.8}&\textbf{\ce{HS}: 0.30} &\ce{Na}: $540$ &\ce{KF}: $51$&\\
   mass fraction [$\%$]&\ce{CO2}: $1.8$&\ce{KCl}: $0.13$&\textbf{\ce{S}: 410}&\ce{Fe(OH)2}: $50$&\\
   &\textbf{\ce{S2O}: 1.7}&&\ce{HCl}: $370$&\ce{PO2}: $15$&\\
   &\ce{H2}: $1.1$&&\ce{KOH}: $260$&\textbf{\ce{SO3}: 12}&\\
   &&&\textbf{\ce{COS}: 190} &\textbf{\ce{S4}: 12}&\\
   &&&\ce{N2}: $160$ &&\\
    &&&\ce{NaF}: $130$ &&\\
\hline
&&&&&\\[-2ex]
      BSE & \textbf{\ce{SO2}: 71}
       &\ce{CO2}: $0.56$ &\textbf{\ce{S3}: 930}&\ce{HCl}: $65$&\\
  2000\, K & \textbf{\ce{S2}: 18}&\ce{H2}: $0.32$ &\ce{Na}: $520$&\ce{NaF}: $62$& $61.557$\\
  10\, bar & \ce{H2O}: $3.9$&\textbf{\ce{H2S}: 0.25}&\ce{NaOH}: $460$&\textbf{\ce{COS}: 60}&\\
   $\epsilon_s^{tot} = 4.77$&\textbf{\ce{SO}: 3.1}&\ce{CO}: $0.21$ &\textbf{\ce{S}: 430}&\ce{K}: $51$&\\
   mass fraction [$\%$]&\textbf{\ce{S2O}: 1.9}&\textbf{\ce{HS}: 0.17}&\ce{KCl}: $430$&\ce{N2}: $48$&\\
   &&\ce{NaCl}: $0.12$&\ce{HF}: $210$&\ce{OH}: $36$&\\
   &&&\ce{KOH}: $140$&\ce{H}: $29$&\\
   &&&&\ce{KF}: $25$ &\\
   &&&&\ce{PO2}: $18$ &\\
   &&&&\ce{Fe(OH)2}: $15$ &\\
   &&&&\textbf{\ce{SO3}: 15} &\\
   &&&&\textbf{\ce{S4}: 14} &\\
\hline
\end{tabular}}
\end{center}
\vspace*{-2mm}
%{\footnotesize ffz}
\end{table*}

\addtocounter{table}{-1}
\begin{table*}
\begin{center}
  \caption{continued.}
\vspace*{-2mm}  
\resizebox{12.0cm}{!}{
\begin{tabular}{c|c|c|c|c|c} 
\hline  
&&&&&\\[-2ex]
   & $n_{\rm mol}/n_{\rm tot}$ $\ge 1\%$ [$\%$]
   &$\ge 0.1\%$ [\%]
   &$\ge 100$\,ppm [ppm]
   &$\ge 10$\,ppm [ppm]
   & mean molecular \\ 
   Model 
   & 
   & 
   &
   &
   & weight [amu]  \\
\hline
\hline  
   &&&&&\\
      BSE & \ce{O2}: $49$&\ce{NaCl}: $0.62$ &\ce{H2}: $800$&\ce{CrO3}: $84$&\\
  2000\, K & \ce{H2O}: $19$ &\ce{O}: $0.47$&\ce{HF}: $760$&\ce{Fe(OH)2}: $77$& 31.704\\
  0.01\, bar &\ce{Na}: $12$&\ce{NaF}: $0.31$&\ce{CO}: $690$&\ce{Fe}: $71$&\\
   &\textbf{\ce{SO2}: 6.3}&\ce{KCl}: $0.21$&\ce{CrO2}: $520$&\ce{Mn}: $53$&\\
   &\ce{CO2}: $3.7$&\ce{NaO}: $0.13$&\ce{H}: $460$&\ce{K+}: $30$&\\
   &\ce{NaOH}: $3.2$&\ce{KF}: $0.13$&\ce{FeO}: $340$&\textbf{\ce{SO3}: 26}&\\
   &\ce{K}: $1.1$&&\ce{NO}: $170$&\ce{HCl}: $23$&\\
   &\ce{OH}: $1.1$&&\ce{N2}: $150$&\ce{PO}: $23$&\\
   &\ce{PO2}: $1.1$&&\textbf{\ce{SO}: 140}&\ce{MnO}: $23$&\\
   &\ce{KOH}: $1.0$&&\ce{KO}: $130$&\ce{SiO}: $17$&\\
   &&&&\ce{SiO2}: $14$&\\
   &&&&\ce{TiO2}: $13$&\\
   &&&&&\\
 \hline
   &&&&&\\
      BSE & \ce{H2O}: $47$&\ce{NaOH}: $0.50$ &\ce{HCl}: $740$&\ce{AlF2O}: $79$&\\
  2000\, K & \ce{O2}: $28$ &\ce{OH}: $0.49$&\ce{NaF}: $560$&\ce{Cl}: $56$& 31.252\\
  1.0\, bar &\textbf{\ce{SO2}: 13}&\ce{KCl}: $0.44$&\ce{Na}: $430$&\textbf{\ce{Na2SO4}: 50}&\\
   &\ce{CO2}: $8.0$&\ce{HF}: $0.21$&\textbf{\ce{SO3}: 410}&\ce{K}: $42$&\\
   &\ce{NaCl}: $1.3$&\ce{KOH}: $0.16$&\ce{N2}: $400$&\textbf{\ce{SO}: 39}&\\
   &&&\ce{O}: $350$&\ce{NaO}: $36$&\\
   &&&\ce{H2}: $260$&\ce{H}: $26$&\\
   &&&\ce{KF}: $230$&\ce{CrO3}: $18$&\\
   &&&\ce{NO}: $210$&\ce{PO2}: $15$&\\
   &&&\ce{CO}: $200$&\ce{HO2}: $14$&\\
   &&&\ce{Fe(OH)2}: $190$&\ce{CrO2}: $14$&\\
 \hline
   &&&&&\\
      BSE & \ce{H2O}: $66$&\ce{HCl}: $0.80$ &\ce{N2}: $690$&\ce{(KCl)2}: $95$&\\
  2000\, K & \textbf{\ce{SO2}: 19} &\ce{KCl}: $0.40$&\ce{(NaCl)2}: $680$&\ce{NaF}: $86$& $30.724$\\
  100\, bar &\ce{CO2}: $11$&\ce{HF}: $0.38$&\ce{NaOH}: $600$&\textbf{\ce{Na2SO4}: 70}&\\
   &\ce{NaCl}: $1.2$&\ce{O2}: $0.28$&\textbf{\ce{SO3}: 580}&\ce{FeCl2}: $58$&\\
   &&&\ce{OH}: $580$&\textbf{\ce{SO}: 54}&\\
   &&&\ce{H2}: $360$&\ce{Cl}: $51$&\\
   &&&\ce{CO}: $280$&\ce{KF}: $35$&\\
   &&&\ce{Fe(OH)2}: $260$&\ce{NO}: $28$&\\
   &&&\ce{KOH}: $190$&&\\
   &&&\ce{AlF2O}: $180$&&\\
 \hline
   &&&&&\\
      BSE & \ce{H2O}: $83$&\ce{HCl}: $0.58$ &\ce{N2}: $900$&\ce{NaCl}: $61$&\\
  1000\, K & \ce{CO2}: $14$ &\textbf{\ce{H2S}: 0.56}&&\textbf{\ce{COS}: 21}& $21.709$\\
  10\, bar &\ce{H2}: $1.3$&\ce{HF}: $0.40$&&\textbf{\ce{SO2}: 11}&\\
   &&\ce{CO}: $0.16$&&&\\
 \hline
   &&&&&\\
      BSE & \ce{O2}: $55$&\ce{KCl}: $0.54$ &\ce{CO}: $700$&\textbf{\ce{K2SO4}: 98}&\\
  2500\, K & \ce{H2O}: $23$ &\ce{NaCl}: $0.46$&\ce{CrO2}: $660$&\ce{Mg(OH)2}: $89$& $32.969$\\
  10\, bar &\textbf{\ce{SO2}: 7.5}&\ce{Na}: $0.42$&\ce{H2}: $580$&\ce{N2}: $83$&\\
   &\ce{CO2}: $4.4$&\ce{KF}: $0.36$&\ce{Fe(OH)2}: $520$&\textbf{\ce{Na2SO4}: 81}&\\
   &\ce{KOH}: $2.8$&\ce{O}: $0.34$&\ce{KO}: $490$&\ce{FeO}: $78$&\\
   &\ce{NaOH}: $2.7$&\ce{NaF}: $0.27$&\ce{AlF2O}: $480$&\ce{MnO}: $40$&\\
   &\ce{OH}: $1.6$&\ce{HF}: $0.22$&\ce{NO}: $400$&\ce{Cl}: $32$&\\
   &&\ce{K}: $0.17$&\textbf{\ce{SO3}: 340}&\ce{Ca(OH)2}: $21$&\\
   &&\ce{NaO}: $0.12$&\ce{PO2}: $330$&\ce{K+}: $16$&\\
   &&&\ce{CrO3}: $300$&\ce{Mn}: $14$&\\
   &&&\ce{H}: $190$&\ce{Cl-}: $11$&\\
   &&&\textbf{\ce{SO}: 180}&&\\
   &&&\ce{HCl}: $140$&&\\
   &&&\ce{HO2}: $130$&&\\
 \hline
    &&&&&\\
      solar & \ce{H2}: $99.8$&&\ce{CO}: $530$&\ce{Mg}: $75$&\\
  2000\, K &&&\ce{H}: $510$&\ce{N2}: $66$& $2.0449$\\
  10\, bar &&&\ce{H2O}: $380$&\ce{SiO}: $60$&\\
   &&&&\ce{Fe}: $37$&\\
   &&&&\textbf{\ce{H2S}: $21$}&\\
   &&&&&\\
 \hline
\hline
\end{tabular}}
\end{center}
\vspace*{-2mm}
%{\footnotesize ffz}
\end{table*}

\begin{table*}
\begin{center}
\caption{Condensate families occuring in Fig.~\ref{one_model}}
\vspace*{-1mm}
\label{condensates}
\vspace*{-1mm}\hspace*{-2mm}
\resizebox{16cm}{!}{
\begin{tabular}{|c|c|c|} 
   \hline
   &&\\[-2.2ex]
    \bf Ca-Al-Ti& \bf Silicates & \bf Feldspar\\
   &&\\[-2.2ex]
   \hline
   &&\\[-2.2ex]
  \ce{Ca2MgSi2O7[s]} {\sl (akermanite)}&\ce{SiO2[l]} {\sl (liquid quartz)} & \ce{KAlSi3O8[s]} {\sl (microcline)}\\
  \ce{Mg2TiO4[l]} {\sl (liquid qandilit)}&\ce{Mg2SiO4[l/S]} {\sl (liquid and solid fosterite)}&\ce{CaAl2Si2O8[s]} {\sl (anorthite)}\\
  \ce{MgAl2O4[l/s]} {\sl (liquid and solid spinel)}&\ce{MgSiO3[l/s]} {\sl (liquid and solid enstatite)}&\ce{NaAlSi3O8[s]} {\sl (albite)}\\
  \ce{MgTi2O5[l]} {\sl (liquid Mg-dititanate)}&\ce{Ca2SiO4[s]} {\sl (larnite)}&\\
  \ce{MnTiO3[s]} {\sl (pyrophanite)}&\ce{Fe2SiO4[s]} {\sl (fayalite)}&\\
  \ce{Fe2TiO4[s]} {\sl (ulvospinel)}&\ce{Mn2SiO4[s]} {\sl (tephroite)}&\\
  \ce{FeAl2O4[s]} {\sl (hercynite)}&\ce{KAlSiO4[s]} {\sl (kalsilite)}&\\
  \ce{FeTiO3[s]} {\sl (ilmenite)}&\ce{NaAlSiO4[s]} {\sl (nepheline)}&\\
  \ce{Mn3Al2Si3O12[s]} {\sl (spessartine)}&\ce{MnSiO3[s]} {\sl (pyroxmangite)}&\\
  &\ce{Na2SiO3[l]} {\sl (liquid NA-metasilicate)}&\\
  &\ce{KAlSi2O6[s]} {\sl (leucite)}&\\
  &\ce{CaMgSi2O6[s]} {\sl (diopside)}&\\
  \hline
  \hline
  &&\\[-2.2ex]
  \bf Phyllosilicates& \bf Metal-oxides & \bf Iron-oxides\\
  &&\\[-2.2ex]
  \hline
  &&\\[-2.2ex]
 \ce{KMg3AlSi3O12H2[s]} {\sl (phlogopite)}&\ce{MgFe2O4[s]} {\sl (magnesioferrite)}&\ce{Fe3O4[s]} {\sl (magnetite)}\\
 \ce{NaMg3AlSi3O12H2[s]} {\sl (sodaphlogopite)}&\ce{MgCr2O4[s]} {\sl (picrochromite)}&\ce{FeO[l/s]} {\sl (liquid and solid ferropericlase)}\\
 &\ce{MnO[s]} {\sl (manganosite)}&\\
 \hline
 \hline
 &&\\[-2.2ex]
 \bf Sulphide &\bf P-compounds& \bf Other\\
 &&\\[-2.2ex]
 \hline
 &&\\[-2.2ex]
 \ce{FeS[s]} {\sl (troilite)}&\ce{Ca5P3O13H[s]} {\sl (hydroxyapatite)}&\ce{NaCl[s]} {\sl (halite)}\\
 &\ce{Ca5P3O12F[s]} {\sl (fluorapatite)}&\ce{KMg3AlSi3O10F2[s]} {\sl (fluorphlogopite)}\\
 \hline
\end{tabular}}
\end{center}
\vspace*{-2mm}
\hspace*{25mm} $^{(1)}$: The mass densities of these condensates have been co-added before plotting in Fig.~\ref{one_model}
\end{table*}

\section{Transmission spectra}
\label{A4}

\subsection{\ce{SO2} and \ce{H2S}}

\begin{figure*}[!t]
\centering
\begin{tabular}{cc}
\hspace*{-7mm}
\includegraphics[width=82mm,height=62mm,trim=4 0 0 18,clip]{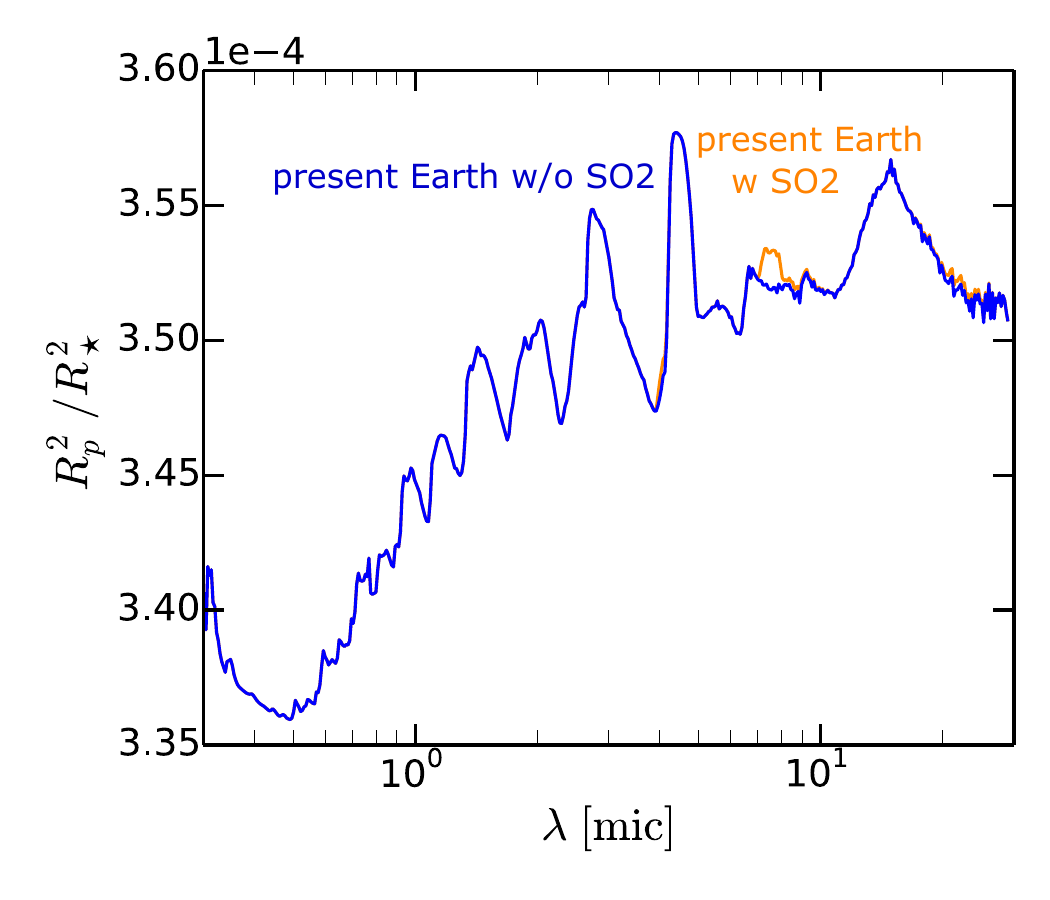} &
\hspace*{-7mm}
\includegraphics[width=82mm,height=62mm,trim=4 0 0 18,clip]{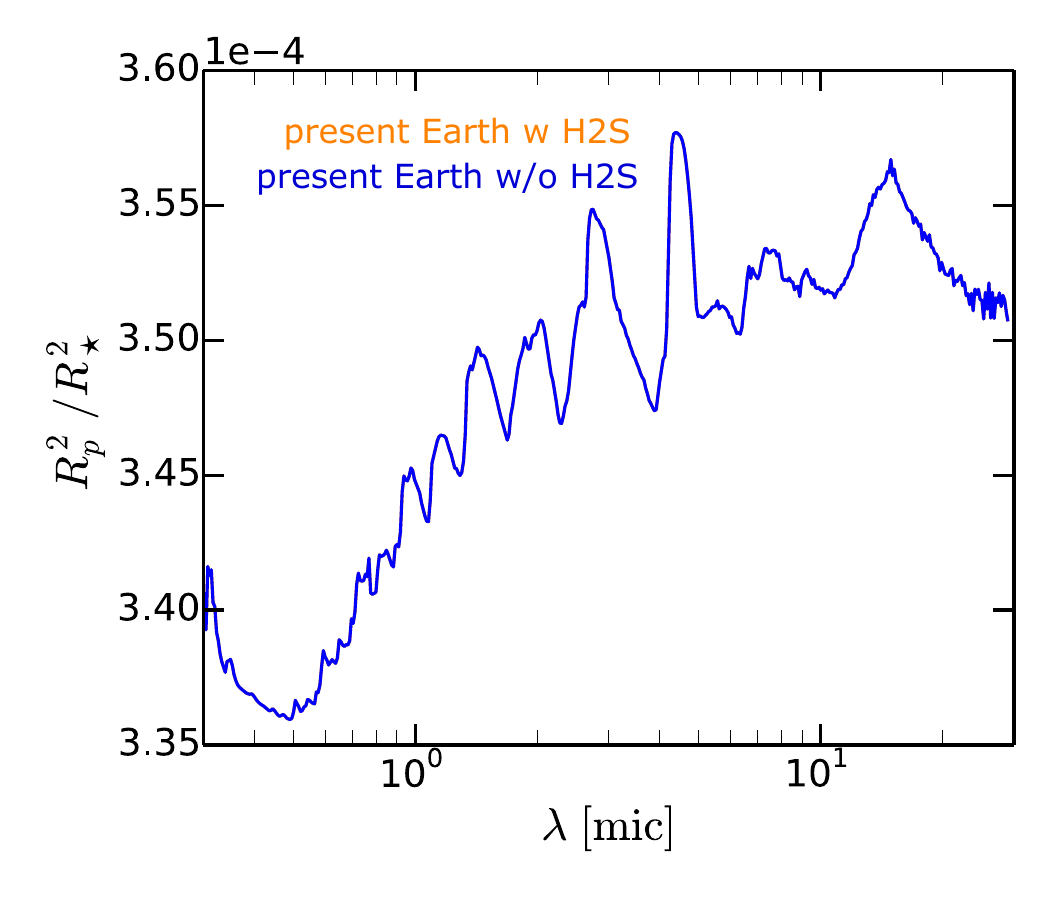} \\[-8mm]
\hspace*{-5.5mm}
\includegraphics[width=80.5mm,height=62mm,trim=0 0 0 18,clip]{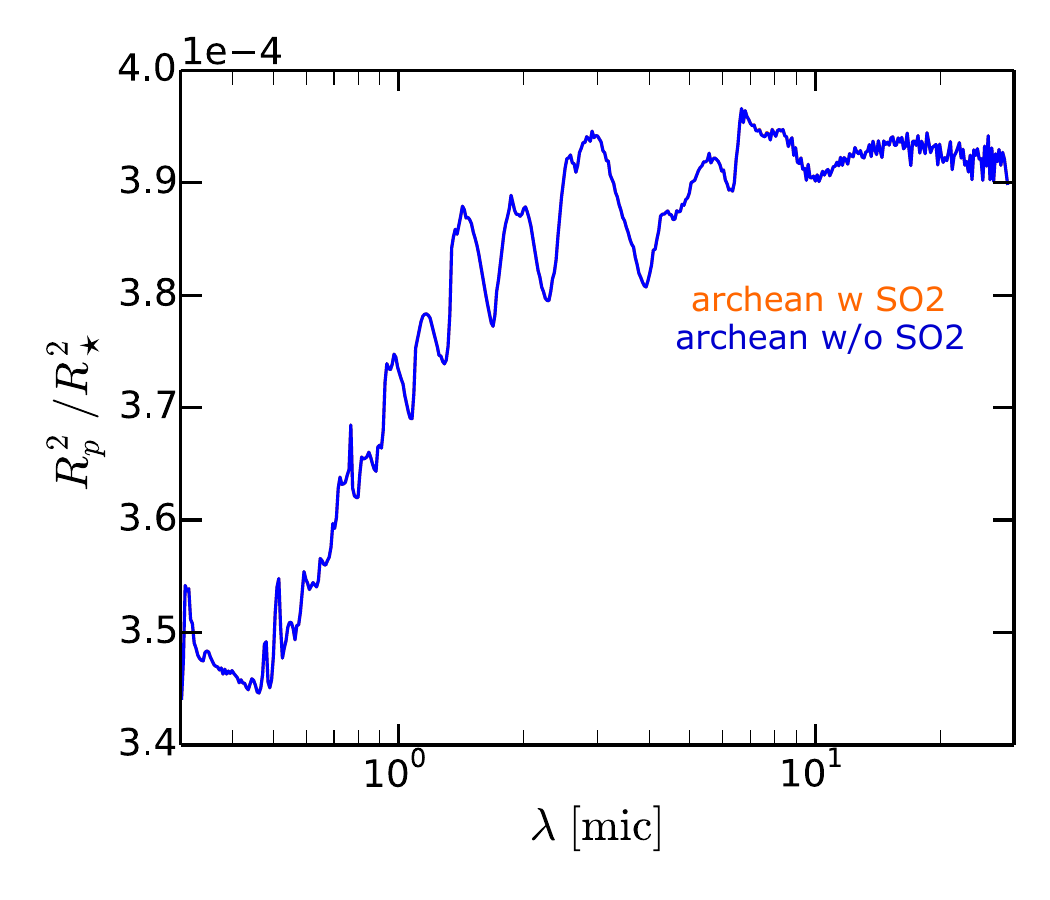} &
\hspace*{-5.5mm}
\includegraphics[width=80.5mm,height=62mm,trim=0 0 0 18,clip]{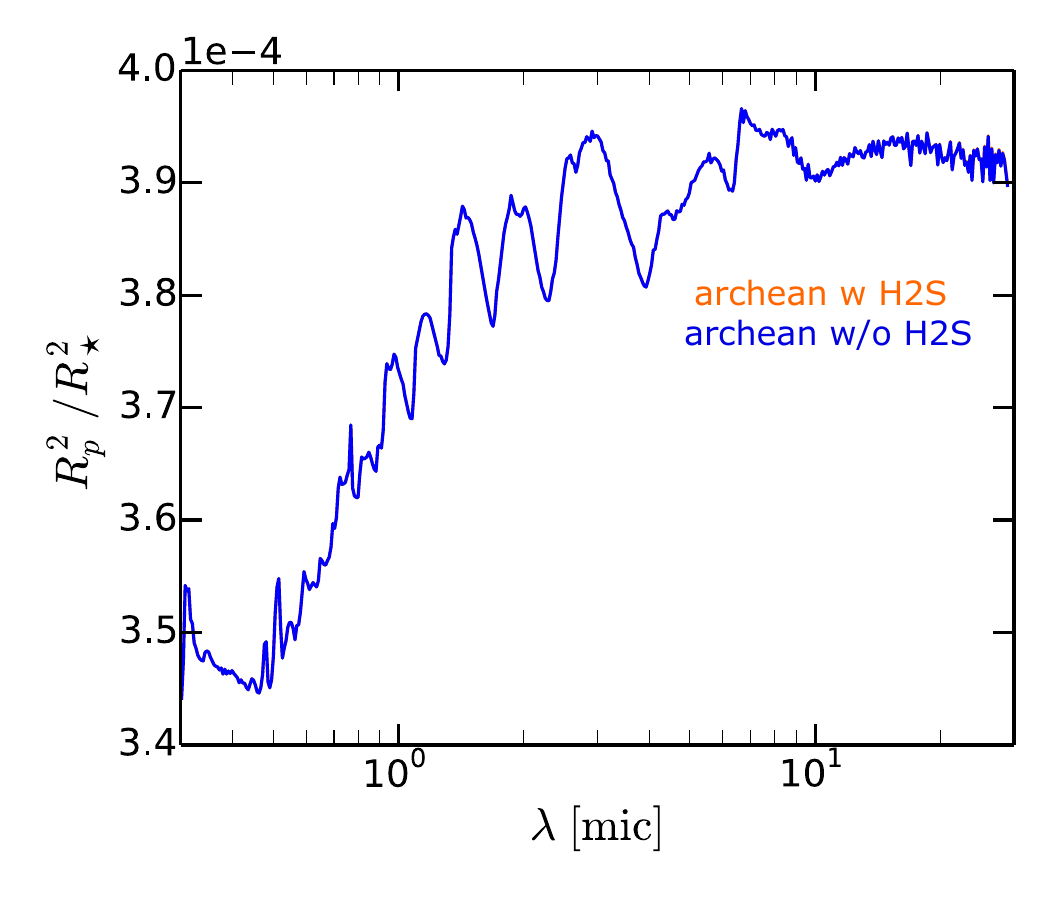} \\[-8mm]
\hspace*{-7mm}
\includegraphics[width=82mm,height=62mm,trim=4 0 0 18,clip]{SO2_BSE_2000K.pdf} &
\hspace*{-7mm}
\includegraphics[width=82mm,height=62mm,trim=4 0 0 18,clip]{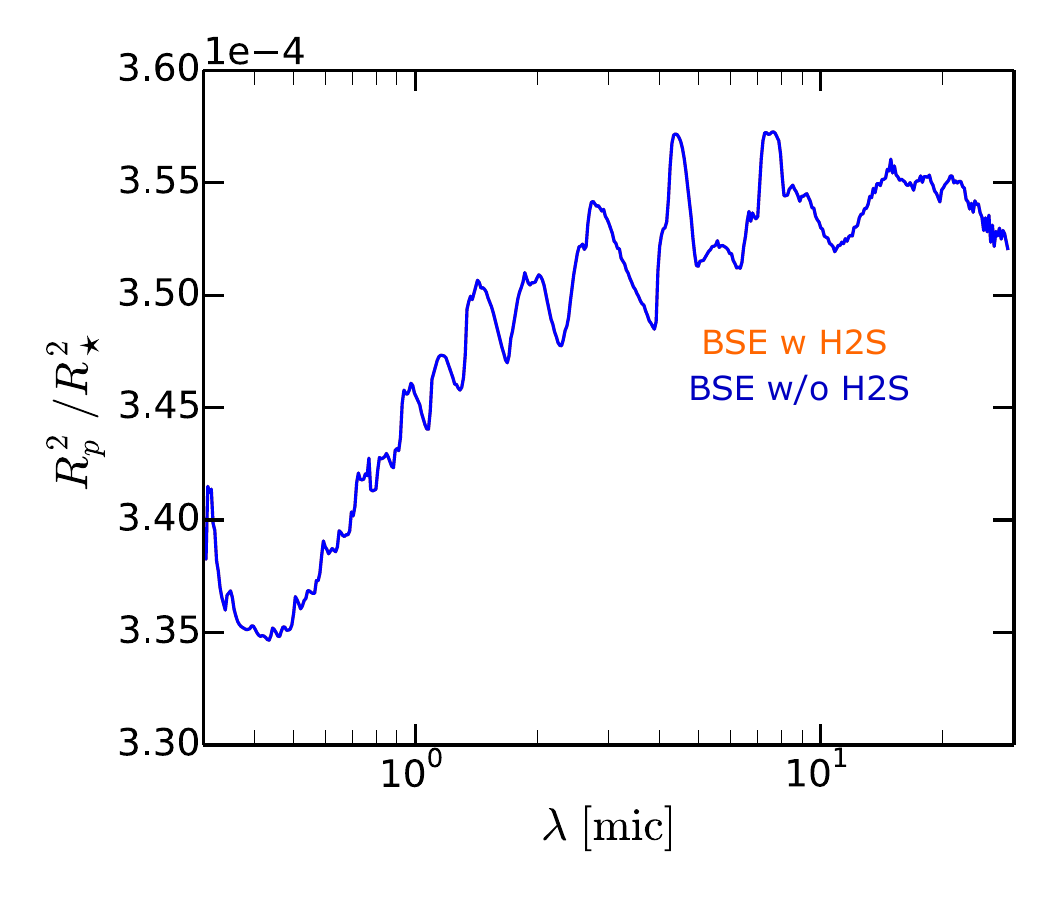} \\[-4mm]
\end{tabular}
\caption{Impact of \ce{SO2} and \ce{H2S} on the transmission spectra generated with ARCiS for isothermal atmospheres with $T\!=\!2000\,$K and $p_{\rm surf}\!=\!10\,$bar. The model input parameters are listed in Table~\ref{ARCIS_parms}.
On the left, the blue line shows the spectrum when the \ce{SO2} line
opacity is omitted. On the right, the blue line shows the spectrum
when the \ce{H2S} line opacity is omitted.}
%The mean molecular weight of the respective molecules is still accounted for and the scale of the y-axis is adapted in each row.}
\label{Fig_extrasolar}
\end{figure*}

\begin{figure*}[!t]
\centering
\begin{tabular}{cc}
\hspace*{-7mm}
\includegraphics[width=82mm,height=62mm,trim=4 0 0 18]{SO2_CC_ink.pdf} &
\hspace*{-7mm}
\includegraphics[width=82mm,height=62mm,trim=4 0 0 18]{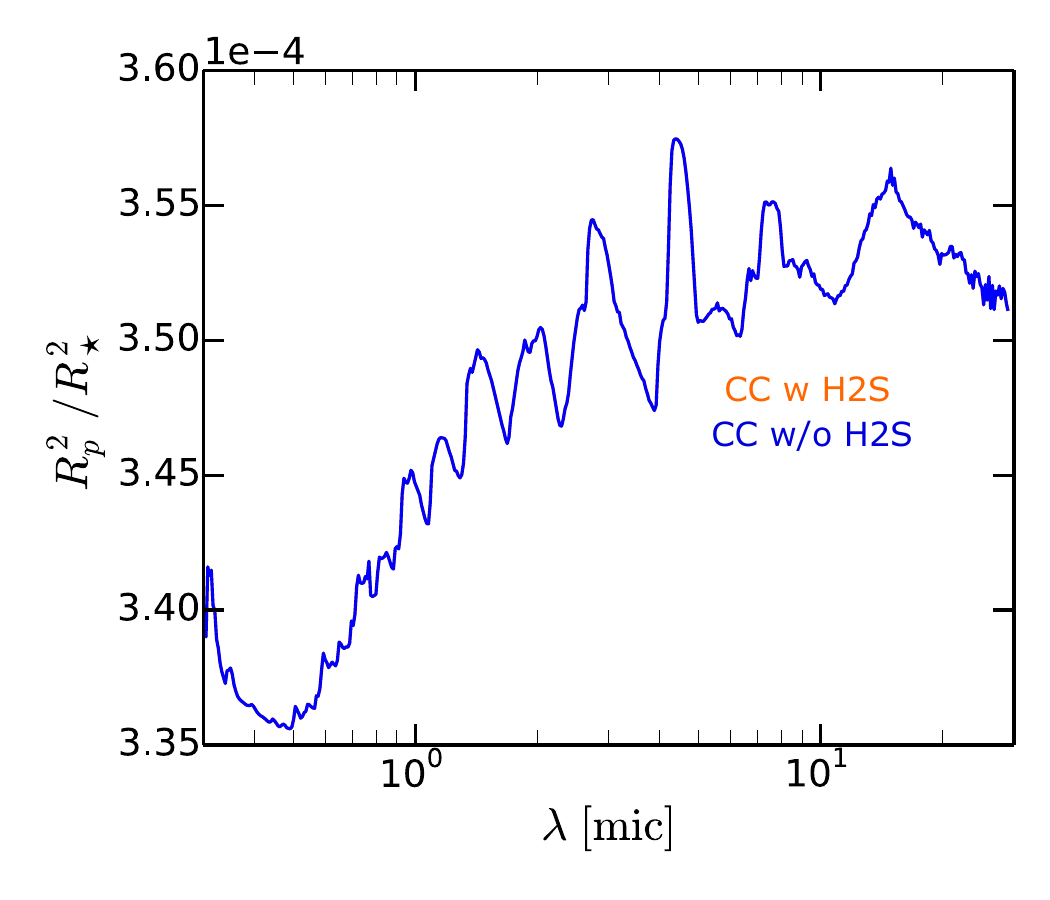} \\[-6mm]
\hspace*{-7mm}
\includegraphics[width=82mm,height=62mm,trim=4 0 0 18]{SO2_MORB_ink.pdf} &
\hspace*{-7mm}
\includegraphics[width=82mm,height=62mm,trim=4 0 0 18]{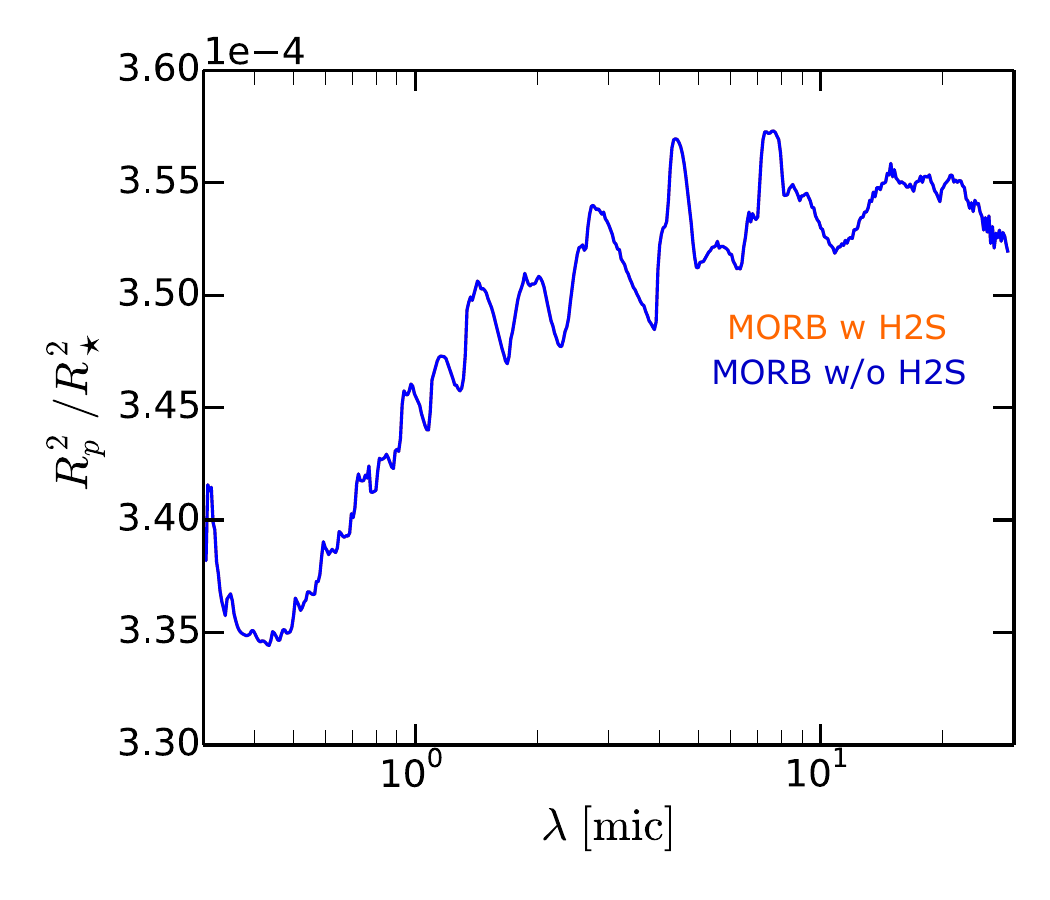} \\[-6mm]
\hspace*{-7mm}
\includegraphics[width=82mm,height=62mm,trim=4 0 0 18]{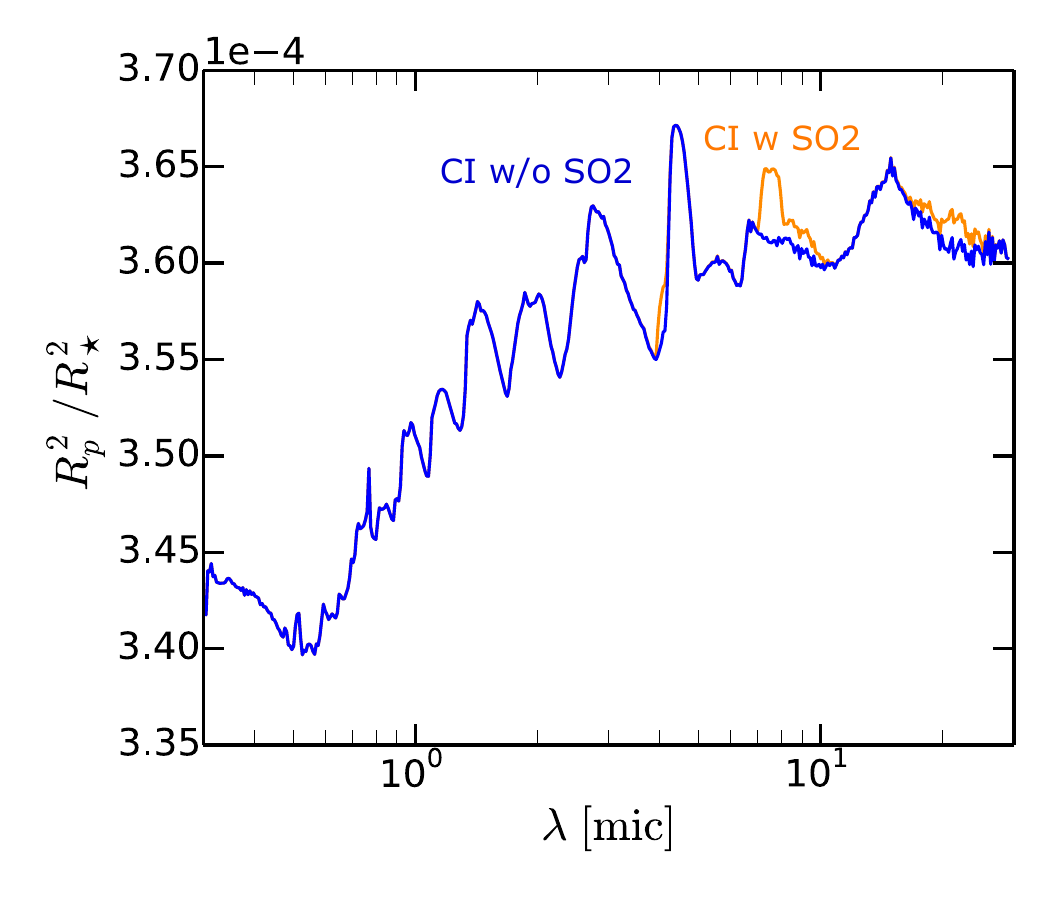} &
\hspace*{-7mm}
\includegraphics[width=82mm,height=62mm,trim=4 0 0 18]{H2S_CI_2000K.pdf} \\[-4mm]
\end{tabular}
\caption{Same as Fig.~\ref{Fig_extrasolar}, but for different element compositions.}
\label{Fig_ssystem}
\end{figure*}

\begin{figure*}[!ht]
\centering
\begin{tabular}{cc}
\hspace*{-7mm}
\includegraphics[width=82mm,height=62mm,trim=4 0 0 18,clip]{SO2_PWDBSE_ink.pdf} &
\hspace*{-7mm}
\includegraphics[width=82mm,height=62mm,trim=4 0 0 18,clip]{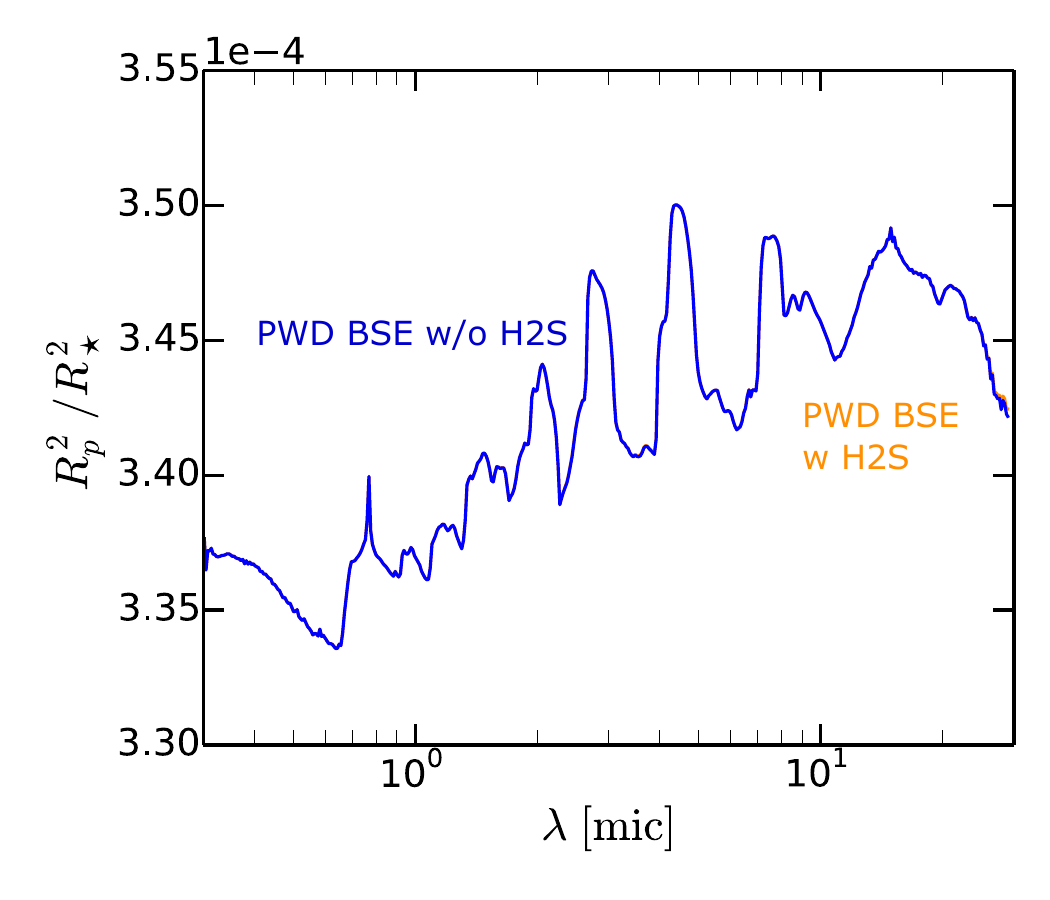} \\[-9mm]
\hspace*{-7mm}
\includegraphics[width=82mm,height=62mm,trim=4 0 0 18,clip]{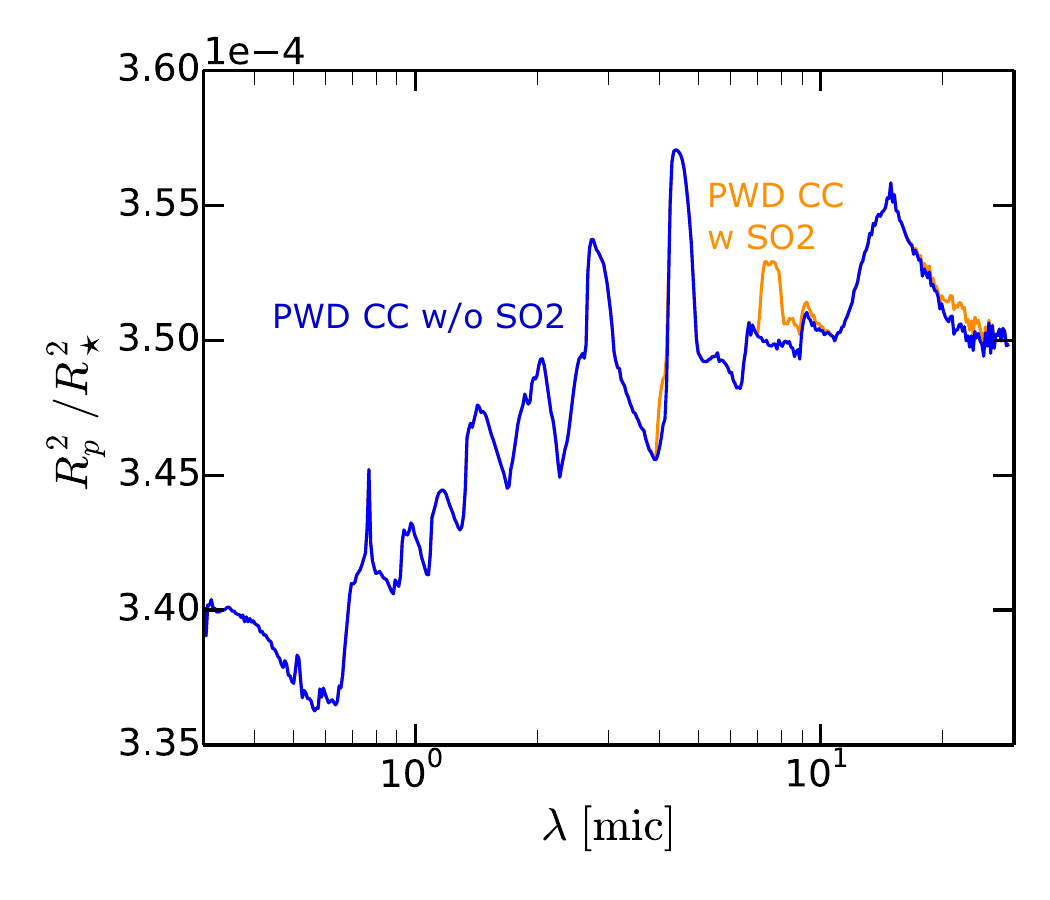} &
\hspace*{-7mm}
\includegraphics[width=82mm,height=62mm,trim=4 0 0 18,clip]{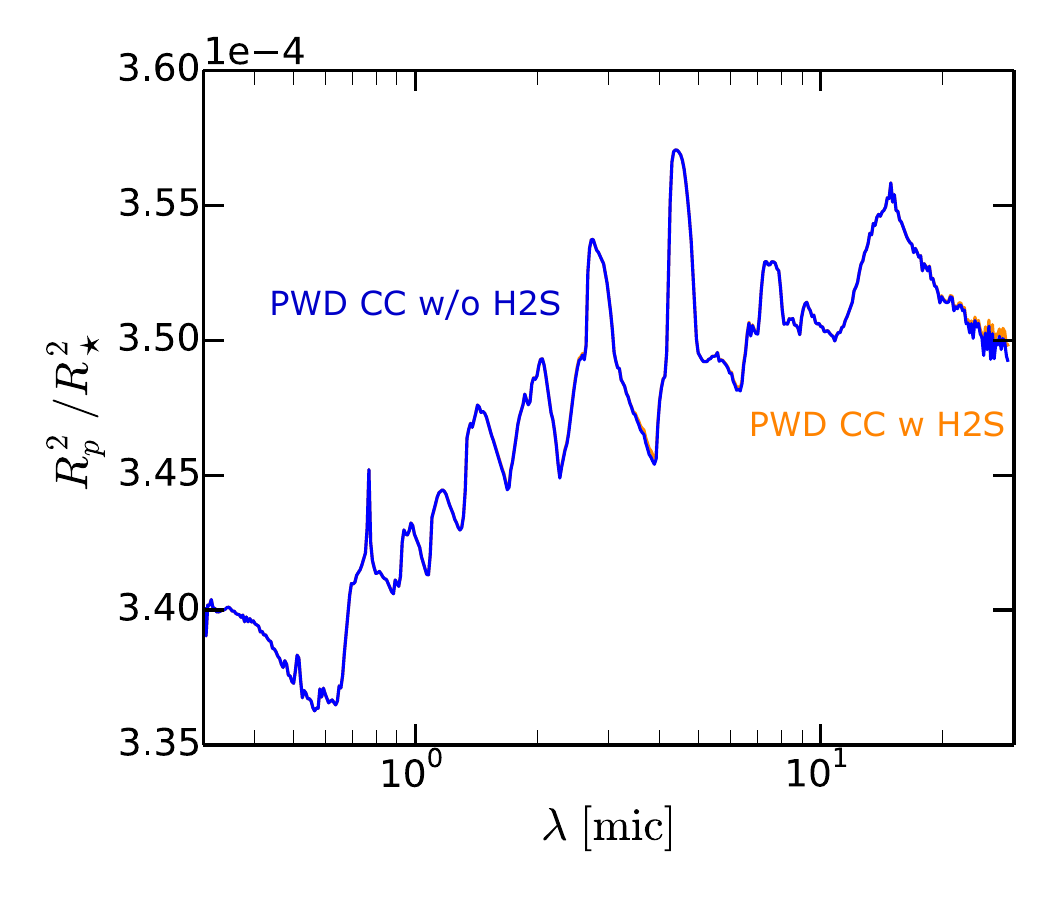} \\[-9mm]
\hspace*{-5.5mm}
\includegraphics[width=80.5mm,height=62mm,trim=0 0 0 18,clip]{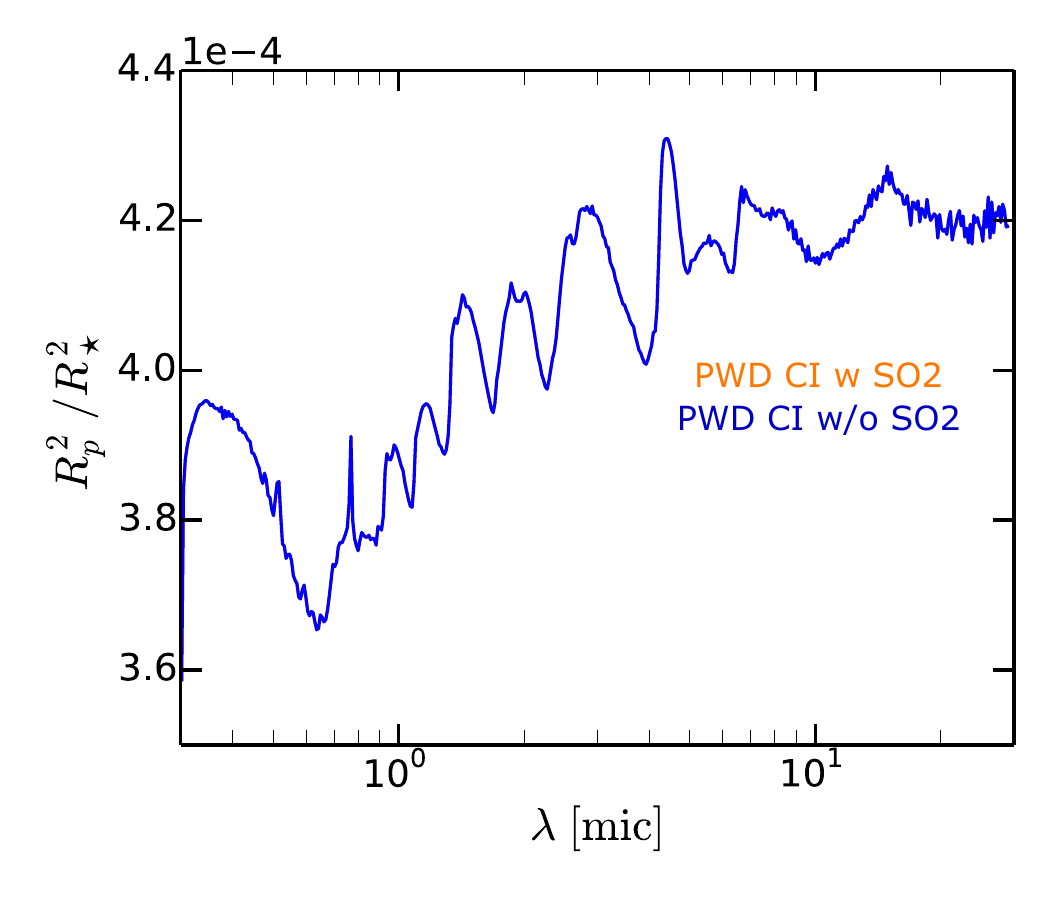} &
\hspace*{-5.5mm}
\includegraphics[width=80.5mm,height=62mm,trim=0 0 0 18,clip]{H2S_PWDCI_2000K.pdf} \\[-4mm]
\end{tabular}
\caption{Same as Fig.~\ref{Fig_extrasolar} but for element abundances derived from polluted white dwarfs.}
\label{FigPWDspectra}
\end{figure*}

\begin{figure*}[!ht]
\centering
\begin{tabular}{cc}
\hspace*{-4mm}
\includegraphics[width=80mm,trim=16 10 11 10,clip]{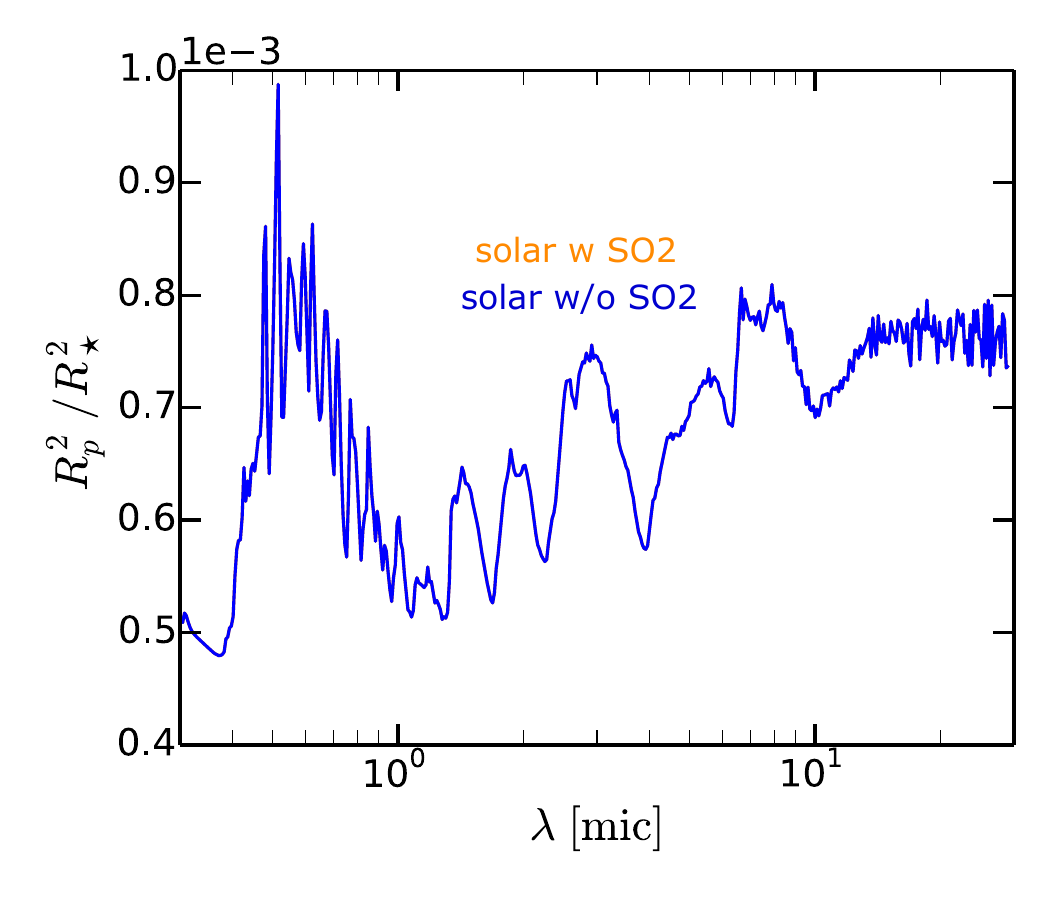} &
\hspace*{-4mm}
\includegraphics[width=80mm,trim=16 10 11 10,clip]{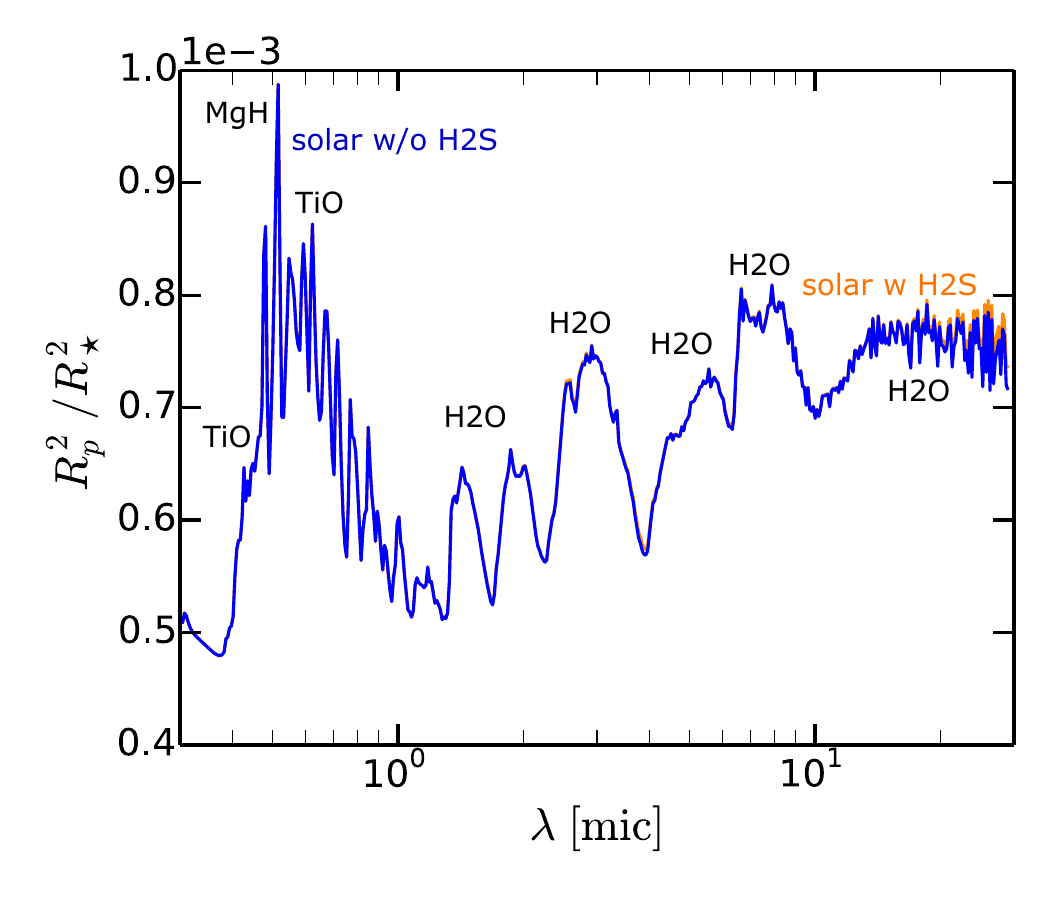} \\[-2mm]
\end{tabular}
\caption{Same as Fig.~\ref{Fig_extrasolar} but for solar element abundances, only showing the effect of \ce{SO2}, and \ce{H2S}. Both do not give rise to any strong spectral features. The towering absorption features at optical wavelengths are mainly due to TiO in between $0.38\!-\!1.03\,\mu$m with a sharp MgH feature between $0.47\!-\!0.61\,\mu$m.}
\vspace*{-2mm}
\label{Fig_solar}
\end{figure*}

This section shows the transmission spectra generated with ARCiS
for all the element compositions which we selected: present Earth,
Archean Earth, solar, BSE, MORB, CC, CI, PWD-BSE, PWD-CC and
PWD-CI. Fig. ~\ref{Fig_extrasolar}, Fig. ~\ref{Fig_ssystem},
Fig. ~\ref{FigPWDspectra} and Fig. ~\ref{Fig_solar} put the spectra
of specific compositions discussed in the main text (BSE and PWD-BSE
and MORB for \ce{SO2} and CI and PWD-CI for \ce{H2S}) into a wider
context by comparing them to the spectra of atmospheres resulting
from different types of possible rocky compositions. The figures suggest that \ce{H2S} is indeed difficult to detect in warm rocky planet atmospheres.
%Solar element abundances of rocky crusts do neither show \ce{SO2} nor \ce{H2S} absorption features from their atmospheres, see Fig. ~\ref{Fig_solar}.

\subsection{PS}

\begin{figure*}[!ht]
\centering
\begin{tabular}{cc}
\hspace*{-4mm}
\includegraphics[width=80mm,height=62mm,trim=16 10 11 10,clip]{PS_PWDCI_ink.pdf} &
\hspace*{-4mm}
\includegraphics[width=82mm,height=62mm,trim=16 10 11 10,clip]{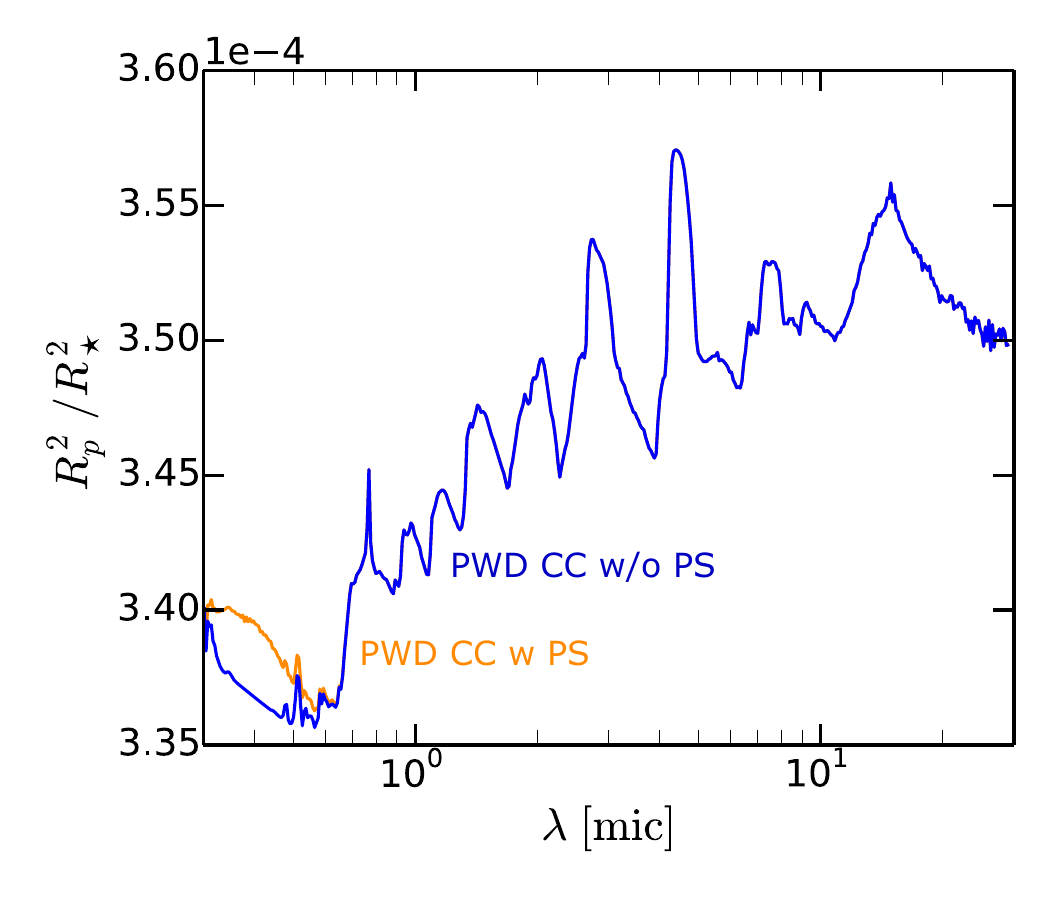}\\[-7.5mm]
\hspace*{-6mm}
\includegraphics[width=82mm,height=62mm,trim=16 10 11 10,clip]{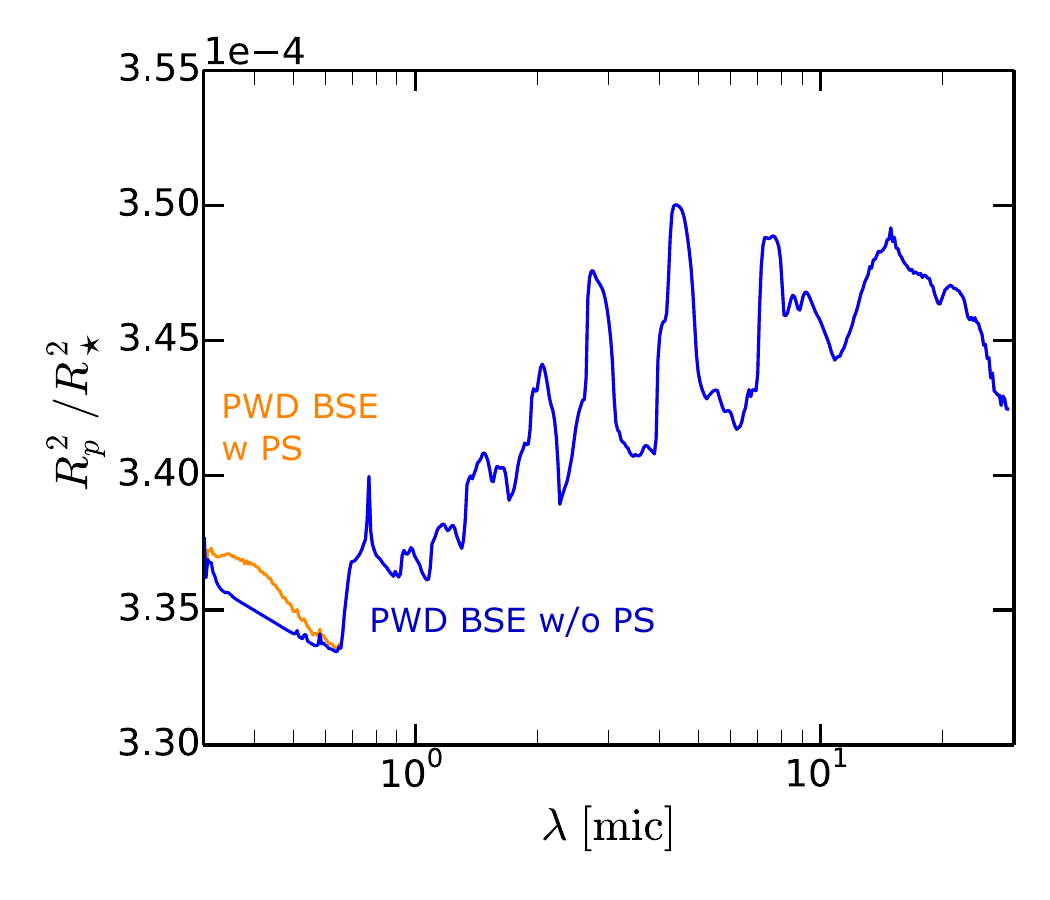}&
\hspace*{-2mm}
\includegraphics[width=80mm,height=62mm,trim=16 10 11 10,clip]{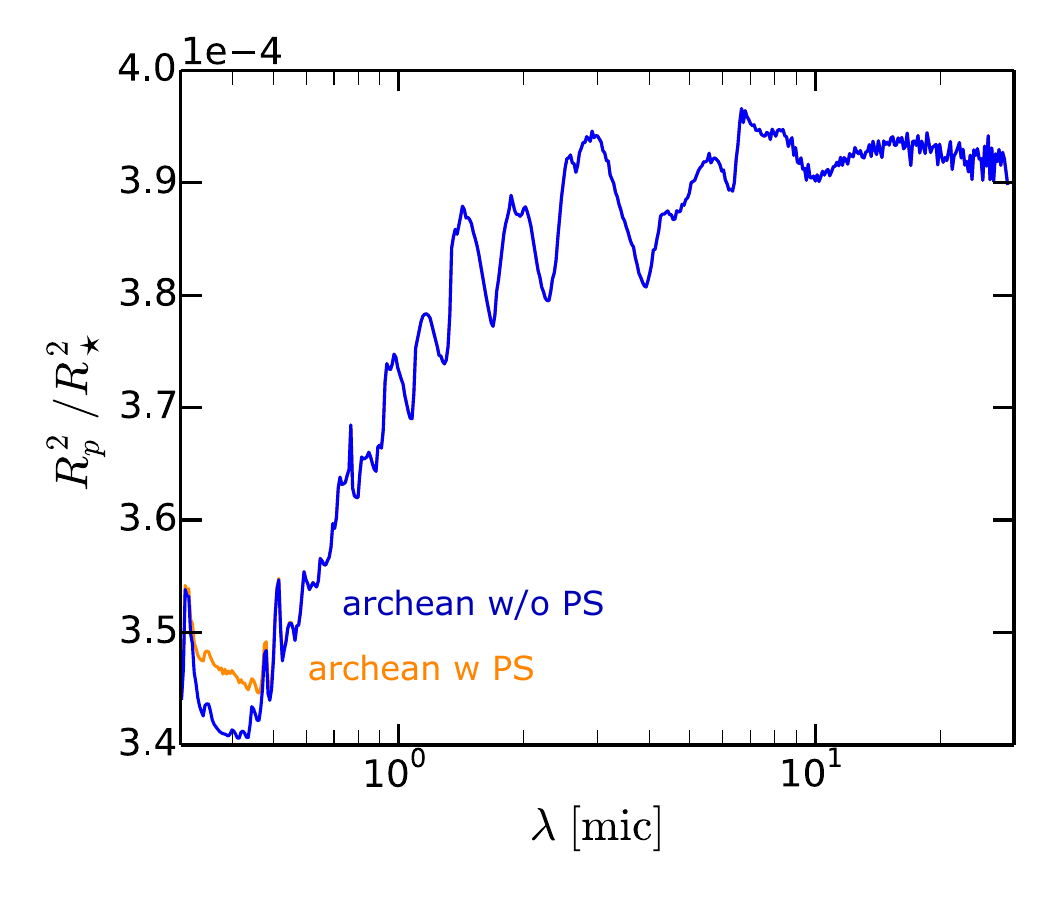}\\[-7.5mm]
\hspace*{-6mm}
\includegraphics[width=82mm,height=62mm,trim=16 10 11 10,clip]{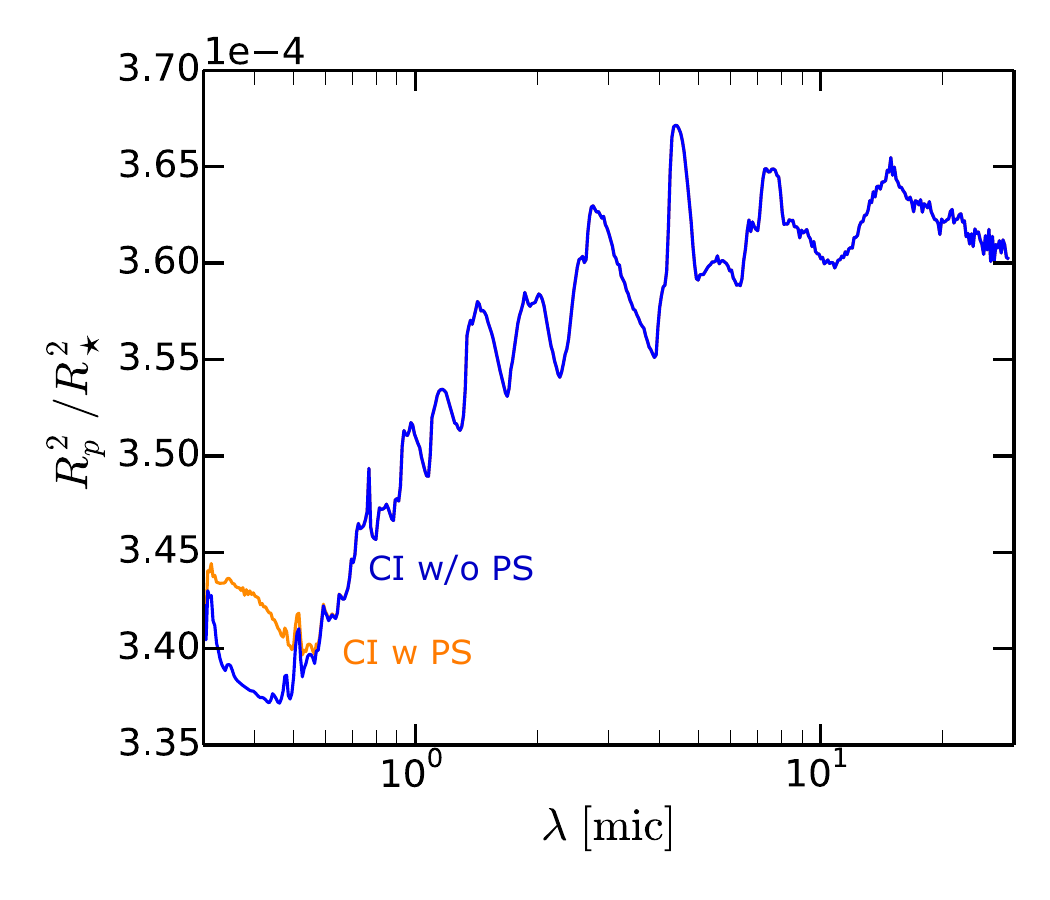}&
\hspace*{-2mm}
\includegraphics[width=80mm,height=62mm,trim=16 10 11 10,clip]{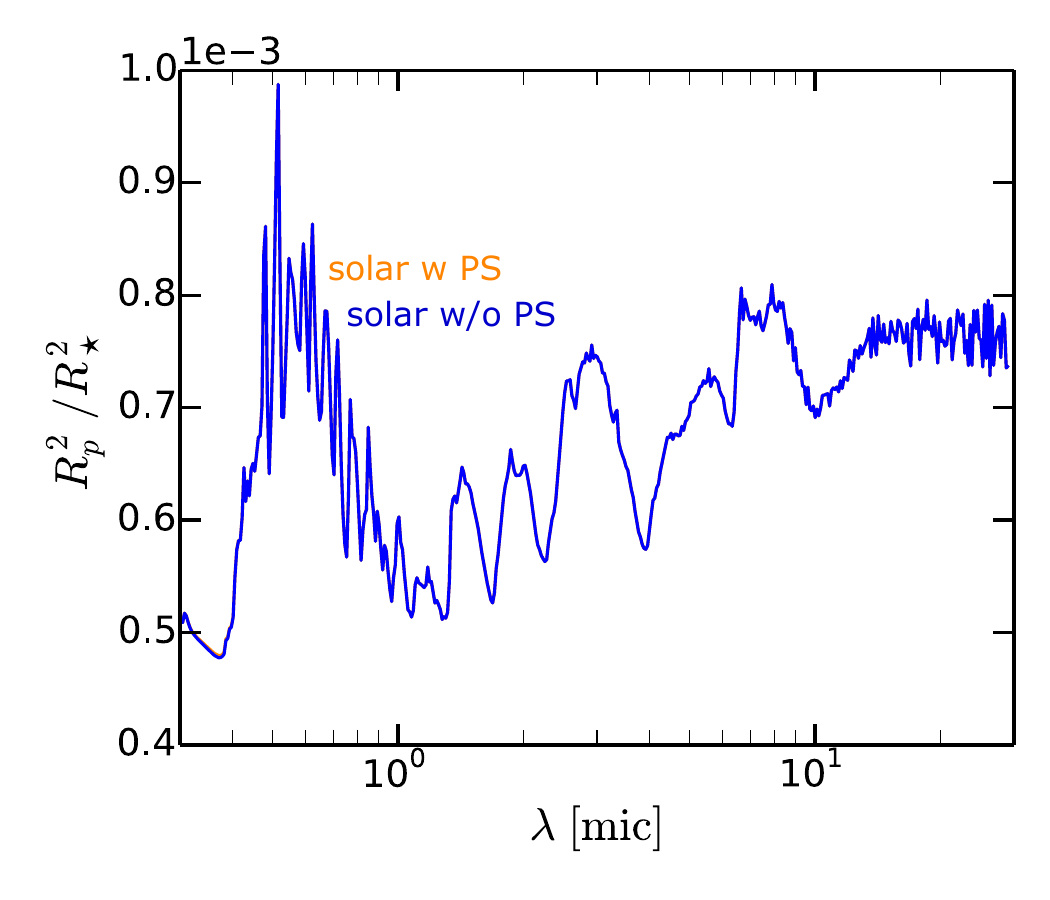}\\[-3mm]
\end{tabular}
\caption{Transmission spectra obtained with ARCiS for Polluted White Dwarf element abundances, the archean Earth, CI chondrite and solar compositions at 2000\,K and 10\,bar. The orange curves show the complete results including PS, whilst the blue curves are the spectra obtained when the PS opacity has been set to zero. The ARCiS model parameter settings are listed in Table~\ref{ARCIS_parms}.}
\label{PS_effect}
\end{figure*}

{ Fig. ~\ref{PS_effect} compares the PS absorption features of the Polluted White Dwarf compositions. Whilst for the PWD-CI composition the absorption feature around $0.3-0.8\,\mu$m reaches 40\,ppm, the feature strength reaches 5\,ppm at most in the other compositions. Apart form the low oxygen abundance in the PWD-CI model, it is to note, that the scale height in this composition is particularly large, which could be one of the causes for this result. No PS feature can be observed in the solar like composition at all.}% The UV wavelengths there are masked by other absorption features

\end{document}